\documentclass{wsmod}
\usepackage[super,compress]{cite}
\usepackage[dvips,breaklinks]{hyperref}
\hypersetup{colorlinks,urlcolor=black,citecolor=black,linkcolor=black,filecolor=black}
\usepackage{breakurl}
\usepackage{graphicx}

\def\pt{\ensuremath{p_{\mathrm{T}}}} % Subscript roman not italic (EE)
\def\HT{\ensuremath{H_{\mathrm{T}}}} % Subscript roman not italic (EE)
\newcommand{\Eslash}{\mbox{$E \kern-0.6em\slash$}}
\newcommand{\pslash}{\mbox{$p \kern-0.5em\slash$}}
\newcommand{\met}{\mbox{\ensuremath{\Eslash_{\kern-0.3emT}\!}}}
\newcommand{\metvec}{\mbox{\ensuremath{\vec{\Eslash_{\kern-0.3emT}\!}}}}
\newcommand{\misspt}{\mbox{\ensuremath{\pslash_{\kern-0.1emT}\!}}}
\newcommand{\missptvec}{\mbox{\ensuremath{\vec{\pslash_{\kern-0.1emT}\!}}}}

\begin{document}
\markboth{Thomas R. Junk and Aurelio Juste}
{Higgs Boson Physics}

\catchline{}{}{}{}{}
%
%%%%%%%%%%%%%%%%%%%%%%%%%%%%%%%%%%%%%%%%%%%%%%%%%%%%%%%%%%%%%%%%%%%%

\title{REVIEW OF PHYSICS RESULTS FROM THE TEVATRON: \\
HIGGS BOSON PHYSICS
%\footnote{For the title, try not to use more than
%3 lines. Typeset the title in 10 pt roman, uppercase and
%boldface.}
}

\author{Thomas R. Junk%\footnote{}
}

\address{Fermi National Accelerator Laboratory\\
Batavia, Illinois 60510, United States of America\\
trj@fnal.gov}

\author{Aurelio Juste}

\address{
Instituci\'{o} Catalana de Recerca i Estudis Avan\c{c}ats (ICREA) and\\
Institut de F\'isica d'Altes Energies (IFAE)\\
Edifici Cn, Facultat de Ciencies\\
Universitat Aut\`onoma de Barcelona\\
E-08193 Bellaterra (Barcelona), Spain\\
juste@ifae.es}

\maketitle

\begin{history}
% version submitted to IJMPA {\center 17 September, 2014}
{\center 12 February, 2015}
%\received{Day Month Year}
%\revised{Day Month Year}
\end{history}

\begin{abstract}
We review the techniques and results of the searches for the Higgs boson performed
by the two Tevatron collaborations, CDF and D\O.  The Higgs boson predicted by the Standard Model
was sought in the mass range 90~GeV$<m_H<200$~GeV in all main production modes at the Tevatron:
gluon-gluon fusion, $WH$ and $ZH$ associated production, vector boson fusion, and $t\bar{t}H$ production, and in five main decay modes:
$H\rightarrow b{\bar{b}}$, $H\rightarrow \tau^+\tau^-$, $H\to WW^{(*)}$,
$H\to ZZ^{(*)}$, and $H\rightarrow\gamma\gamma$.  An excess of events was seen in the
$H\rightarrow b{\bar{b}}$ searches consistent with a Standard Model Higgs boson with a mass in the range
115~GeV$<m_H<$135~GeV.  Assuming a Higgs boson mass of $m_H=125$~GeV, studies 
of Higgs boson properties were performed, including measurements of the product of the 
cross section times the branching ratio in various 
production and decay modes, constraints on Higgs boson couplings to fermions and vector bosons, and tests of spin and parity.
We also summarize the results of searches for supersymmetric Higgs bosons, and Higgs bosons in other extensions of the Standard Model.

\end{abstract}

\ccode{PACS numbers: 13.85.Rm, 14.80.Bn, 14.80.Da, 14.80.Ec, 14.80.Fd}

\tableofcontents

\section{Introduction}
The recent observation of the Higgs boson by the ATLAS and CMS
Collaborations at the Large Hadron
Collider~\cite{Aad:2012tfa,Chatrchyan:2012ufa} closes a long chapter in
experimental particle physics and begins a new one in which the
properties of the Higgs boson are used to test for new physical
phenomena.  In 1964 the existence of a massive scalar boson became a 
key testable prediction of the Higgs mechanism~\cite{Higgs:1964pj,Englert:1964et,Guralnik:1964eu,Higgs:1966ev},
which is the simplest description of how the observed masses of the
$W$ and $Z$ gauge bosons, as well as those of the fermions, are
consistent with the SU(2)$_L\times$U(1)$_Y$ gauge symmetry.  This
symmetry, when broken by the Higgs mechanism to the U(1)$_{\rm{EM}}$
symmetry of quantum electrodynamics, provides the basis of the
Standard Model (SM)~\cite{Glashow:1961tr,Weinberg:1967tq,Salam:1968rm}, a very successful framework that
predicts, or at least accommodates, all particle physics measurements
made to date.  The mysteries of dark matter, dark energy, and a
quantum description of gravity remain beyond the scope of the SM, 
though the Higgs bosons produced in the laboratory can
be a window to testing alternate hypotheses motivated by these
unexplained phenomena.  For many years, the non-observation of exotic
particles, Higgs bosons among them, have constrained many possible
models of new physics.

The search for Higgs bosons was a central component of the Run~II physics program
at the Tevatron.  Early estimates of the sensitivity~\cite{Carena:2000yx,Babukhadia:2003zu} 
indicated that tests of the presence or absence of the SM Higgs boson were achievable, even though
these estimates were uncertain due to the level of precision of the available signal cross section predictions 
as well as the rudimentary estimates and handling of backgrounds rates, signal efficiencies, and systematic uncertainties.
Models of exotic Higgs boson production provided motivation to search for Higgs bosons even with smaller data sets.
Many of the upgrades to the Tevatron and the two detectors, CDF and D\O, described elsewhere in this review,
were motivated by the Higgs boson physics program, though these upgrades also had positive impacts on
the broad physics objectives of the two collaborations.

With the full Tevatron Run~II data set, CDF and D\O\ combined their search results together and in July 2012 obtained the first evidence
for a particle produced in association with vector bosons and which decays to $b{\bar{b}}$, consistent with the
expectation for the SM Higgs boson~\cite{Aaltonen:2012qt}.  Measurements of the cross sections times decay branching ratios in 
different production and decay modes, as well as tests of couplings and spin and
parity, were performed~\cite{Aaltonen:2013kxa,Abazov:2014doa,Aaltonen:2015tsa,Aaltonen:2015mka}.  
No significant deviations from the predictions
for the SM Higgs boson with a mass near 125~GeV were seen.  Because the Tevatron searches were most sensitive to processes in which
the Higgs boson decays to fermion pairs, they are naturally complementary with the LHC searches, which are most sensitive to 
decays of the Higgs boson to pairs of bosons ($\gamma\gamma$, $ZZ^{(*)}$, and $WW^{(*)}$).
This article describes the components of the Tevatron searches for the
Higgs boson and their interpretation, starting with the models under
test, and proceeding with the experimental equipment, analysis tools,
and results.

\section{Higgs Boson Theory and Phenomenology}	
\label{sec:sec2}
The simplest implementation of the Higgs mechanism is that used by the
SM.  A doublet of self-interacting complex scalar fields
is introduced that, by virtue of the opposite sign of the quadratic
and quartic terms in the Higgs potential, acquires a vacuum expectation value at the minimum
of the potential, which has a three-dimensional degeneracy.
This degeneracy would result in three massless Goldstone bosons, which
are not observed.  Instead, the three degrees of freedom appear as the
longitudinal polarization components of the $W^+$, $W^-$, and $Z$
bosons, endowing these particles with their masses.  The fourth degree of freedom has finite-mass excitations
corresponding to a neutral scalar boson $H$ with a mass $m_H$ that is
not predicted by the theory.  Together with the gauge interactions of
the SM, the Higgs mechanism completes the model by
allowing for both fermion and gauge boson masses while preserving
renormalizability~\cite{'tHooft:1972fi}.

\subsection{Standard Model Higgs boson production}

At tree level in the SM, the Higgs boson couples to a species of
fermion with a strength proportional to that fermion's mass, and to a
species of boson with a strength proportional to the square of that
boson's mass.  This feature, along with the kinematic availability of
each final state, determines the decay branching ratios of the SM
Higgs boson as a function of its mass.  The dominant decay modes are
to the heaviest particles kinematically available, with a preference
for decays to massive bosons.  The couplings of the Higgs boson to
SM particles tend to be smaller than electroweak and strong
couplings, leading to the challenge of searching for rare Higgs boson
processes among much more copious backgrounds.

At one loop and higher, the Higgs boson couples to the massless gauge
bosons $g$ and $\gamma$, even though the tree-level couplings vanish.
The $Hgg$ coupling is dominated by a top-quark loop, although the
$b$-quark loop also contributes a non-negligible amount.  The presence
of additional gluons radiated by the gluons coupling to the Higgs
boson increases both the $gg\rightarrow H$ production cross section at
hadron colliders, and the decay branching ratio for $H\rightarrow gg$.
Gluon radiation also modifies the branching ratios of the Higgs boson
to quarks.  Because of the small couplings of the Higgs boson to the
$u$ and $d$ quarks, the primary constituents of the proton, the
$gg\rightarrow H$ production mechanism is the dominant process at both
the Tevatron and the LHC.  The sub-dominant processes are production
in association with a vector boson ($q\bar{q}\rightarrow WH, ZH$, referred to as $VH$), as well as vector
boson fusion (VBF) ($q\bar{q}\rightarrow q'\bar{q}'H$), and Yukawa radiation from a
top quark pair ($t\bar{t}H$) or a $b$-quark pair ($b\bar{b}H$), the latter
of which can be dominant in extensions of the SM with enhanced couplings to
down-type quarks.  More rare modes include production in association
with a single top quark and the production of a pair of Higgs bosons.

\begin{figure}[t]
\centerline{\includegraphics[width=0.47\columnwidth]{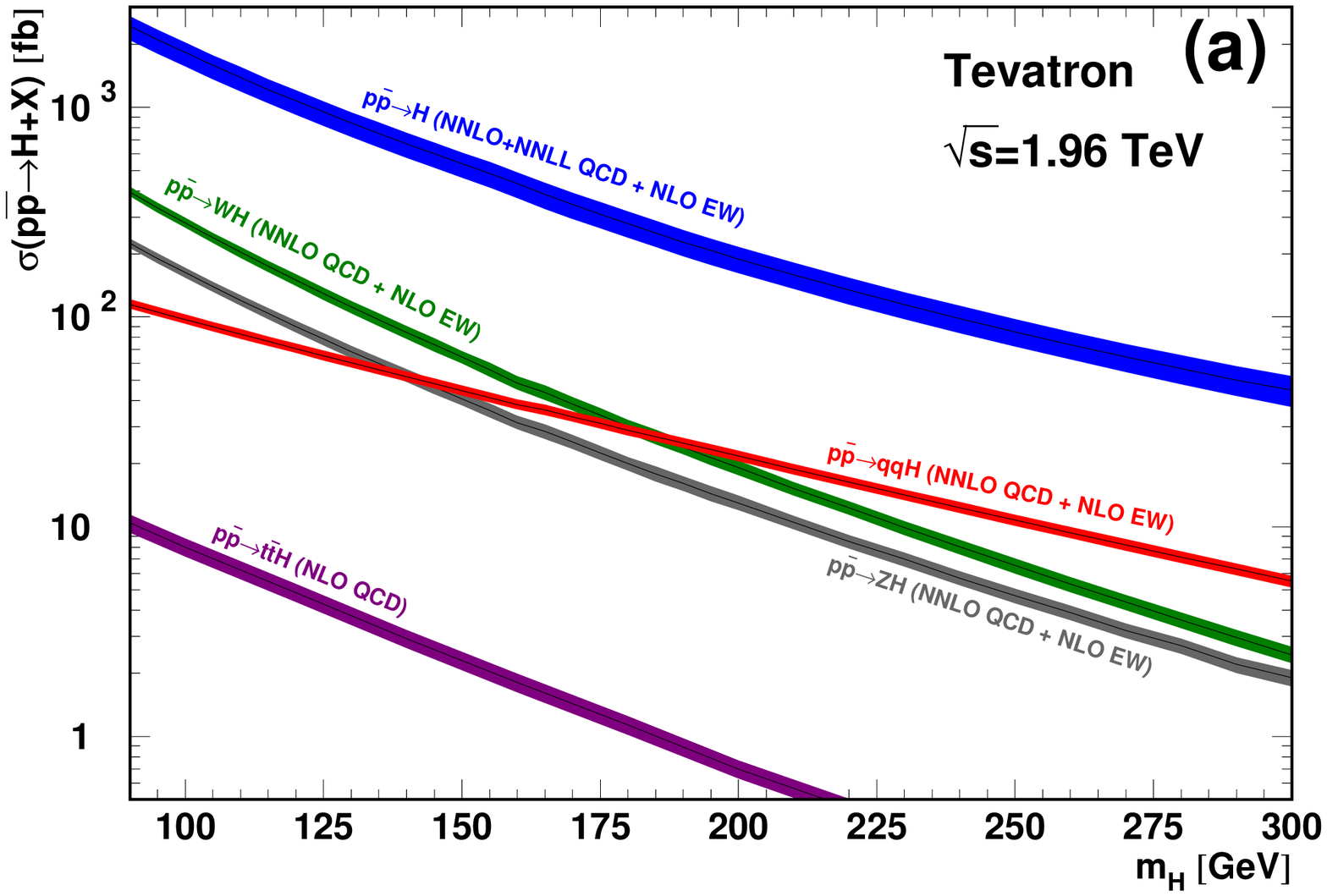}\includegraphics[width=0.47\columnwidth]{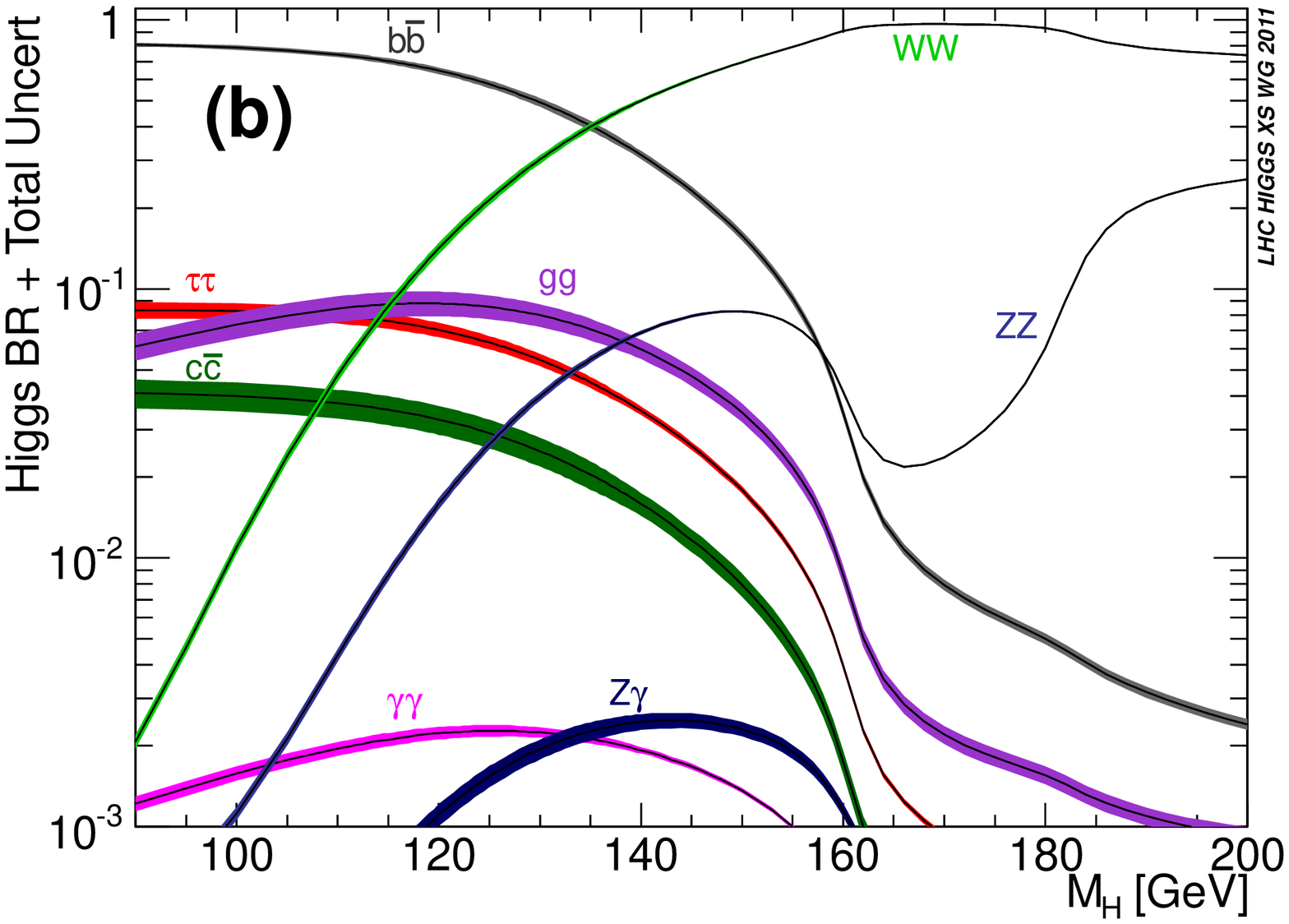}}
\caption{(a) Production cross sections for the gluon-fusion process (labeled $p{\bar{p}}\rightarrow H$), the $WH$, $ZH$,
VBF (labeled $p{\bar{p}}\rightarrow qqH$), and $t{\bar{t}}H$
processes. From Ref.~\citen{Beringer:1900zz}.  (b) Decay branching fractions
for the SM Higgs boson as functions of its mass. From Ref.~\citen{Dittmaier:2011ti}.
\label{fig:smxsbr}}
\end{figure}

The SM predictions of the production rates in $p{\bar{p}}$ collisions
are shown as functions of $m_H$ in Fig.~\ref{fig:smxsbr}(a).  The
$gg\rightarrow H$ production rate is computed at
next-to-next-to-leading order (NNLO) with next-to-next-to-leading-logarithmic (NNLL) soft-gluon
summation accuracy 
in QCD (referred to as NNLO+NNLL)~\cite{Harlander:2002wh,Anastasiou:2002yz,Ravindran:2003um,Kramer:1996iq,Catani:2001ic,Harlander:2001is,Catani:2003zt,Aglietti:2004nj,deFlorian:2009hc,Anastasiou:2008tj}, including the effects of mixed
QCD-electroweak corrections and the running $b$-quark mass.
Higher-order corrections are very important in this process due to the
strong coupling of gluons to additional particles: the next-to-leading order (NLO) $k$-factor
is approximately 2.0, and at NNLO, there is an additional factor of
$\sim 1.5$.  A partial calculation at next-to-next-to-next-to-leading order (NNNLO) in 
QCD~\cite{Moch:2005ky} provides
some confidence that the corrections from further terms in the series
become smaller, and are adequately covered by the factorization and
renormalization scale uncertainties customarily assigned.  The parton distribution functions (PDFs)
used in the cross section calculations used for the Tevatron results are the MSTW2008 set~\cite{Martin:2009iq}
and the recommended uncertainties~\cite{Alekhin:2011sk,Botje:2011sn}.
The differential spectrum of $gg\rightarrow H$ production is complex and
has experimental consequences.  The production rates for $gg\rightarrow H+1$~jet
and $gg\rightarrow H+\geq 2$~jets have been calculated at NLO 
in QCD~\cite{Anastasiou:2008ik,Campbell:2010cz}, and the $p_T$ spectrum at
NLO+NNLL~\cite{Bozzi:2003jy,Bozzi:2005wk,deFlorian:2011xf}.

The theoretical uncertainty on the total production rate for $gg\rightarrow H$ is 
approximately $\pm 10$\%~\cite{deFlorian:2009hc,Anastasiou:2008tj}, although the uncertainty
on the production rate for $gg\rightarrow H+$jets is significantly larger -- it is $\pm 23$\%
for the $\geq 2$~jets category.  The CDF and D\O\ Collaborations follow the procedure
of Ref.~\citen{Stewart:2011cf} in order to account for the correlations (positive and negative) 
between the predictions of the rates in the exclusive observable jet categories used to analyze the data.
The impact of the factorization and renormalization scale uncertainties on the PDF uncertainties are
considered correlated with the factorization and renomalization scale uncertainties and are added linearly
with those.  The remaining components of the PDF uncertainties are considered uncorrelated.

For the associated production cross sections, $p{\bar{p}}\rightarrow WH+X$ and $p{\bar{p}}\rightarrow ZH+X$, the
CDF and D\O\ Collaborations
use the calculations of Ref.~\citen{Baglio:2010um}, which are performed at NNLO precision in QCD and
NLO precision in electroweak corrections.  A similar calculation is available in Ref.~\citen{Ferrera:2011bk}.
The theoretical uncertainties in these predictions are approximately $\pm 8$\% at $m_H$=125~GeV, mostly
due to the PDF uncertainties.

The VBF cross section is computed at NNLO in QCD~\cite{Bolzoni:2010xr}, and the
electroweak corrections are computed with the {\sc hawk} program~\cite{Ciccolini:2007ec}.
The $t{\bar{t}}H$ production cross sections are computed at NLO in 
QCD~\cite{Beenakker:2002nc,Beenakker:2001rj,Reina:2001sf}, although this last cross section
calculation was computed using the CTEQ6M PDF set~\cite{Pumplin:2002vw}.

\subsection{Standard Model Higgs boson decay}

The decay branching ratios to pairs of particles are shown in
Fig.~\ref{fig:smxsbr}(b)~\cite{Dittmaier:2011ti}.  Refinements to these calculations
are provided in Refs.~\citen{Dittmaier:2012vm} and~\citen{Heinemeyer:2013tqa}. These are
obtained by using the {\sc hdecay} program~\cite{Djouadi:325078} to compute the partial widths for
all decay modes except the four-fermion final states resulting from the
$H\rightarrow WW^{(*)}$ and $H\rightarrow ZZ^{(*)}$ decay modes, which
interfere quantum mechanically.  These latter contributions are computed
using {\sc prophecy4f}~\cite{Bredenstein:1053995}.
The partial widths are then summed
and the fractions of the total widths are the resulting branching
ratios.  Uncertainties in the branching ratios are assessed from
uncertainties in the masses of the final state particles, specifically
the $b$ and $c$ quark masses, and the factorization and
renormalization scale uncertainties used to estimate the effects of missing higher
order terms in the calculation.  At $m_H=125$~GeV, the SM predictions
for the branching ratios are approximately~\cite{Dittmaier:2011ti}:
${\cal B}(H\rightarrow b{\bar{b}})\approx 57.8\%$, 
${\cal B}(H\rightarrow WW^{(*)})\approx 21.6\%$, 
${\cal B}(H\rightarrow \tau^+\tau^-)\approx 6.37\%$, 
${\cal B}(H\rightarrow \gamma\gamma)\approx 0.23\%$,
${\cal B}(H\rightarrow gg)\approx 8.56\%$, 
${\cal B}(H\rightarrow ZZ^{(*)})\approx 0.23\%$, and
${\cal B}(H\rightarrow Z\gamma)\approx 0.16\%$.

\subsection{Precision electroweak constraints and direct searches}

Searches by the four LEP collaborations excluded
$m_H<114.4$~GeV at the 95\% confidence level (C.L.), assuming SM properties of the Higgs boson, taking
advantage of the associated production mode, $e^+e^- \to ZH$~\cite{Barate:2003sz,Abbiendi:2002qp}.  LEP also placed strong
limits on the production of additional Higgs bosons predicted by the minimal supersymmetric standard model (MSSM)~\cite{Nilles:1983ge,Haber:1984rc}, 
which can provide visible signatures even when the coupling
strengths of the $ZH$ and $Zh$ processes are suppressed~\cite{Schael:2006cr}.  LEP
sought a great variety of Higgs boson decay final states explicitly,
such as $H\rightarrow\gamma\gamma$~\cite{LHWG:Photons},
$H\rightarrow$~hadrons~\cite{Searches:2001aa} and $H\rightarrow$~invisible~\cite{Searches:2001ab}, in
addition to the SM and MSSM searches.  The limit on the mass of the SM
Higgs boson from LEP provides a lower bound in the region of interest for
the Tevatron searches, although those searches were extended down to $m_H=90$~GeV, largely to
validate the modeling of the lower-mass Higgs boson searches, and to make a point of comparison
with the corresponding measurements of non-resonant diboson production, $WZ$ and $ZZ$, described
in Sec.~\ref{sec:vhsearch}.

Precision electroweak measurements also provide constraints on $m_H$
assuming SM relations~\cite{ALEPH:2005ab}.  Combined with the
precision measurements of the top quark mass $m_t$~\cite{Aaltonen:2012ra} and
the $W$ boson mass $M_W$~\cite{Group:2012gb}, an upper bound of
$m_H<152$~GeV is obtained at the 95\% C.L.  Adding in the constraint from 
direct LEP searches raises the upper bound to $m_H<171$~GeV at the 95\% C.L.

\subsection{Higgs bosons in extensions of the Standard Model}
\label{sec:bsmpheno}

The observation of a SM-like Higgs boson at the LHC has established
the existence of a state of the electroweak symmetry breaking sector,
but it has not proven that the minimal SM Higgs mechanism provides a
full description of it.  In fact, many extensions of the SM postulate
the existence of an extended scalar Higgs sector.  For instance,
introducing a second Higgs doublet, such as in Two Higgs Doublet
Models (2HDM)~\cite{Gunion:2002zf,Branco:2011iw}, leads to five
physical Higgs bosons: a light and a heavy CP-even Higgs boson ($h$
and $H$), a CP-odd Higgs boson ($A$), and a pair of charged bosons
($H^\pm$). The MSSM is, at leading order, an
example of a Type-II 2HDM, where up- and down-type quarks couple to
separate Higgs doublets.  CP violation in the MSSM Higgs
sector~\cite{Carena:2002bb,Carena:2001fw} would relax the production
selection rules and enlarge the possible space of parameters to
search.  The next-to-minimal MSSM (NMSSM)~\cite{Chang:2008cw} further
extends the MSSM to include an additional CP-even and CP-odd neutral
Higgs bosons. In the NMSSM, the lightest CP-odd Higgs boson, $a$, can
be very light, even below the $b\bar{b}$ threshold. Alternatively,
Higgs triplet models extend the SM by adding a complex SU(2)$_L$
triplet scalar field, predicting a pair of doubly-charged Higgs bosons
($H^{\pm\pm}$), in addition to the five Higgs bosons present in 2HDMs.
Finally, some extensions of the SM predict the existence of massive
metastable particles that can only decay to SM particles through
diagrams containing a new high-mass force carrier or a loop of very
massive particles. These scenarios, referred to as ``hidden-valley"
(HV) models~\cite{Strassler:2006im}, can involve a HV scalar particle
that can mix with the SM Higgs boson, so that the latter could decay with
substantial branching ratio to HV
particles~\cite{Strassler:2006ri,Falkowski:2010cm,Falkowski:2010gv}.

Additional particle content may be present to augment the SM's three
generations of fermions and its set of gauge bosons.  A model that
consists of the SM with one additional generation of fermions is
referred to as SM4~\cite{Holdom:2009rf}.  The presence of two very
heavy quarks would lead to an enhancement in the $ggH$ coupling by a
factor of approximately three, since each heavy quark $Q$ would contribute
as much as the top quark in the SM.  The contributions to the $ggH$
amplitude are nearly independent of the masses of the new heavy quarks
as the suppression factors from the propagators cancel the enhancement
from the $QQH$
couplings~\cite{Arik:2005ed,Kribs:2007nz,Anastasiou:2010bt}.  The
$gg\rightarrow H$ production cross section therefore rises by a factor
of approximately nine relative to the SM prediction in the range of
$m_H$ the Tevatron is sensitive to, and the partial decay width
of the Higgs boson to gluons also increases by a factor of nine.  Even
with a higher decay rate to a pair of gluons, the $H\to WW^{(*)}$ decay
mode continues to dominate for $m_H>135$ GeV.

%\section{The Tevatron and the CDF and D\O\ Detectors}
%\label{sec:tevatron}
%\input{sectevatron}

\section{Analysis Tools}
\label{sec:sec3}
The searches for the SM Higgs boson and Higgs bosons in exotic models
are especially challenging due to the small signal production cross sections and
the large background rates.  The analyses reported here make extensive use of nearly every capability
of the CDF and D\O\ detectors.  As data were collected, the experience gained in refining the tools
was used to improve the sensitivity of the Higgs boson searches, so that over time the sensitivity
increased significantly faster than expected from simple luminosity scaling.

\subsection{Particle identification}
\label{sec:particleid}

Searches for $WH\rightarrow\ell\nu b{\bar{b}}$, $ZH\rightarrow \ell^+\ell^- b{\bar{b}}$, and
$H\rightarrow WW^{(*)}\rightarrow\ell^+\nu\ell^-\bar\nu$ rely heavily on identifying leptons
with high efficiency and low rates of backgrounds from misidentified jets.  Typically, analyses are designed 
to select electrons and muons, as the detectors are optimized to separate these from hadronic backgrounds.
Tracks in the CDF COT and the D\O\ fiber tracker are associated with EM showers in the calorimeters to
identify electrons and with track segments in the surrounding muon chambers to identify muons.
Activity in the hadronic calorimeters inconsistent with electrons or muon signatures is used to veto
hadrons that otherwise might pass the lepton identification selections, and the spatial distribution
of the energy in the electromagnetic calorimeters is also used as a discriminating variable, helping
to reduce the background from $\pi^0 \to \gamma\gamma$ decays misidentified as electrons.  Lepton candidates are categorized in terms of their
quality -- how many selection requirements they pass, and whether they are detected in the central
portion of the detectors or the forward portions, or travel through uninstrumented materials.  Isolated
tracks are also counted as lepton candidates in some analyses; these channels are analyzed separately
from the others so as not to dilute the purity of higher-quality lepton selections.

The lepton identification efficiencies and energy resolutions are calibrated using $Z\rightarrow e^+e^-$
and $Z\rightarrow\mu^+\mu^-$ samples in the data.  Lepton triggers are based on identifying one lepton
at a time, and so tag-and-probe methods are used to calibrate the trigger efficiencies:
in $Z\to \ell^+\ell^-$ events in which one lepton satisfies the trigger requirements, the other is used to probe the trigger efficiency
with minimal bias.  Similar methods are used to calibrate the efficiencies of lepton identification
requirements.  Lepton energy scales are calibrated with leptonically-decaying
$Z$, $J/\psi$, and $\Upsilon$ events.

\subsection{Jet identification and energy measurement}
\label{sec:jes}

Energy deposits in the electromagnetic and hadronic calorimeters are grouped into jets using
a cone-based algorithm with a radius $\Delta R=0.4$ (CDF) and 0.5 (D\O), where
$\Delta R=\sqrt{(\Delta\eta)^2+(\Delta\phi)^2}$~\footnote{Both CDF and D\O\ use
right-handed coordinate systems, with the $z$~axis directed along the proton beam.
The azimuthal angle $\phi$ around the beam axis is defined with respect to a horizontal ray running outwards
from the center of the Tevatron, and radii are measured with respect to the beam
axis.   The polar angle $\theta$ is defined
with respect to the proton beam direction, and the pseudorapidity $\eta$ is
defined to be $\eta=-\ln\left[\tan(\theta/2)\right]$.}.  In many analyses, the charged particle momenta
are measured for tracks within these jets and are combined with the calorimetric measurements in order
to improve the energy resolution, a key ingredient to the searches for $H\rightarrow b{\bar{b}}$.
The jet energy scale is likewise important for the same searches.  Not all of the energy of the hadrons
in jets is measured by the calorimeters -- some of it is absorbed in nuclear interactions, some of it
leaks out the back of the calorimeters, and some of it falls outside of the jet cones.
Data samples, such as dijets, photons
recoiling against jets or $Z\rightarrow e^+e^-,\mu^+\mu^-$ recoiling against jets, are used to
calibrate the response of the calorimeters and the jet algorithms to hadronic jets with known transverse
momenta~\cite{Bhatti:2005ai,Abazov:2013hda}.  Typical resolutions for jet 
energies are of order 8\%, with higher-$E_T$ jets being better measured
than lower-$E_T$ jets.  Jets originating from gluons at the hard-scatter vertex tend to be wider than
jets originating from quarks, resulting in a different energy scale due to the jet identification
and energy clustering algorithms~\cite{Aaltonen:2013ioi,Aaltonen:2014mdq}.  
Even though quark and gluon jets may be well simulated by parton-shower Monte Carlos (MC) such as {\sc pythia},
the application of a calorimetric energy correction factor derived on a data sample
with a given fraction of quark jets to a sample with a different fraction is not correct; instead, two
separate correction factors must be derived and applied separately to quark and gluon jets in the MC samples.

Jets containing $B$~hadrons suffer from additional jet energy biases compared with light-flavored jets, due
to the high masses of $B$  hadrons which disperse the decay particles outside of the jet cones and also because
semileptonic decays of $B$ and $D$ hadrons produce neutrinos whose momenta are not measured.  Algorithms are
devised to improve the jet energy scale and resolution specifically for $b$~jets~\cite{Aaltonen:2011bp}.
 
\subsection{{\it b}-Tagging}
\label{sec:btagging}

Searches for $WH\rightarrow Wb{\bar{b}}$ and $ZH\rightarrow Zb{\bar{b}}$
have large backgrounds from vector bosons producted in association with jets.  The vast majority of these jets
are light-flavored, and so separation of $b$~jets from light-flavored jets provides a significant
improvement in the signal-to-background ratio in these searches.  Multivariate Algorithms (MVAs) are designed to key on the
large mass ($\approx 5$~GeV) and long lifetimes ($\approx 1.4$~ps) of $B$ hadrons.  Charged tracks
from $B$ (and subsequent $D$) hadron decay tend to have significantly larger impact parameters with
respect to the beam axis than tracks created promptly at the primary vertex, whose impact parameters
are dominated by resolution effects and multiple scattering.  Displaced vertices are identified topologically
and their properties, such as the invariant masses of the contributing tracks, the decay length, and
the presence of leptons are all used to separate $b$~jets from light-flavored 
jets~\cite{Freeman:2012uf,Acosta:2005am,Abulencia:2006kv,Abazov:2010ab,Collaboration:2012hv,Abazov:2013gaa}.  
Typical performances achieved for $b$~tagging are 50\% efficiency for $b$ jets from top quark decay
with a 0.5\% mistag rate of light-flavored jets in the same momentum range, for a typical tight operating point
corresponding to a requirement on the MVA score.  A benefit of using a continuous variable to rank jets as being more or less $b$-like 
is that multiple operating points of the tagger can be used within the analyses.  A typical loose requirement
yields a $b$-tagging efficiency of 80\% with a mistag rate of 10\%, although some of these jets are also tagged by the
tighter requirements.  Analyses are constructed out of the exclusive subsets of tagged events. The $b$-tagging efficiencies
and mistag rates are calibrated with data control samples such as $t{\bar{t}}$ decays, $W/Z+1$~jet events
(where the flavor composition is measured with other taggers), and multijet events. 

\subsection{Missing energy}

Because the PDFs are broad, events can be boosted along the beam axis by an unknown amount.
Therefore, unlike an $e^+e^-$ collider, which typically has a known total three-momentum and energy of the interactions, only
the sum of the transverse momenta is possible to constrain at a hadron collider.  The $WH\rightarrow\ell\nu b{\bar{b}}$ channel
and the $H\rightarrow WW^{(*)}\rightarrow\ell^+\nu\ell^-{\bar{\nu}}$ and the $WH\rightarrow WWW$ channels seek leptonically-decaying
$W$ bosons, and the accompanying high-momentum neutrinos are not reconstructed.  The $ZH\rightarrow\nu{\bar{\nu}}b{\bar{b}}$ channel has two
high-momentum neutrinos in each signal event.  The presence of a lepton and missing transverse momentum, or merely missing transverse
momentum by itself, are powerful discriminant variables for reducing backgrounds and selecting Higgs boson events.  Since the recoiling
system is often hadronic, the calorimetry is used to sum the visible energy in an event, and the angle from the primary vertex
is used to compute the transverse projections of the calorimeter energies.  The negative vector sum of these transverse energies is
denoted $\metvec$, and its magnitude is $\met$.

Because $\met$ is an inference of unmeasured momenta from a sum of measurements that are subject to physical, detector, and 
reconstruction effects, its value is often rather different from the sum of the neutrino momenta it approximates.  Jet energies
are corrected for the jet energy scale as described in Sec.~\ref{sec:jes}, although individual jet mismeasurement constitutes the main cause
for $\met$ mismeasurement.  Frequently the difference between $\metvec$ and $\missptvec$, where $\missptvec$ is the missing
momentum using the tracks measured by the tracking detectors, is used to help identify events
with mismeasured $\met$.

%redundant.
%In the $ZH\rightarrow\ell^+\ell^-b{\bar{b}}$ search, there is no true missing 
%transverse energy, and any measured missing transverse energy is due to mismeasured jets. 
%Projections of the mismeasured missing transverse energy along the measured jet axes provide
%guidance of how to adjust the measured jet energies to improve the resolution on $m_{jj}$.   In the CDF
%version of this analysis~\cite{Aaltonen:2012id}, a neural network is trained including the measured missing
%transverse energy and the jet energies and directions in order to optimize the improvement in the jet
%energy resolution.

\subsection{Top quark identification and reconstruction}

The $t{\bar{t}}$ production cross section at the Tevatron is approximately 7~pb, significantly larger than the
Higgs boson production cross section.  Its decays, to $W^+bW^-\bar{b}$, can mimic the signal in all of the main
search channels: $WH\rightarrow\ell\nu b{\bar{b}}$, $H\rightarrow WW^{(*)}$, $ZH\rightarrow\ell^+\ell^- b{\bar{b}}$,
$WH+ZH\rightarrow\met b{\bar{b}}$, and others, usually as a result of one or more of the decay products of one or
both top quarks falling outside of the detector acceptance or being misreconstructed.  The highest-purity $t\bar{t}$ samples involve 
more jets than are normally required of the Higgs boson searches, and thus e.g. rejecting
events with four or more jets is effective at reducing the $t{\bar{t}}$ background in the $WH\rightarrow \ell\nu b{\bar{b}}$ searches.
In the $H\rightarrow WW^{(*)}\rightarrow\ell^+\nu\ell^-{\bar{\nu}}$ search, the signal is not expected to contain
$b$ quarks, and so the $b$-tag requirement is inverted on reconstructed jets within the acceptance of the silicon
detectors in order to reduce the $t{\bar{t}}$ background.  Full reconstruction of top quarks is rarely needed in order
to reject events in which they may be present, particularly in cases in which particles are missing or mismeasured.

Single top quark production has a final state that is the same as that of the $WH\rightarrow\ell\nu b{\bar{b}}$ search, and
it has a cross section of approximately 3~pb.  Fortunately, the kinematics of single top quark production are quite
striking.  Variables such as $m_{jj}$, $m_{\ell\nu b}$ and $q\times\eta$, where $q$ is the charge of the lepton and $\eta$ is
the pseudo-rapidity of the non-b-tagged jet~\cite{Yuan:1989tc} are quite powerful in separating single top quark production
from Higgs boson production.

\subsection{Multivariate analyses}

The small predicted signal cross sections and the large non-resonant
backgrounds to Higgs boson production require that all possible
methods be used in order to distinguish signal-like events from
background-like ones.  The usual distinguishing feature -- the
invariant mass of reconstructed candidates, which ought to produce a
localized excess in its distribution at the mass of the Higgs boson --
is not a powerful enough variable to perform the searches only with it.
In $H\rightarrow b{\bar{b}}$ searches, the dijet mass distribution is wide
enough and the expected signal small enough that a noticeable excess
would not be seen on top of the background.  In the 
$H\rightarrow WW^{(*)}\rightarrow\ell^+\nu\ell^-\bar\nu$ searches, the invariant mass
of the Higgs boson cannot be reconstructed with good resolution due to
the missing neutrino momenta.  Other variables, such as the transverse
momentum of the dijet system, the missing transverse energy, or the
angle between the two leptons in the 
$H\rightarrow WW^{(*)}\rightarrow\ell^+\nu\ell^-\bar\nu$ searches, help provide
separation between the signal and the background

The relatively large systematic uncertainties in the background
predictions would wash out a small potential signal if events were
merely counted after applying selection cuts -- it is impossible to
discover or exclude a signal that is smaller than the uncertainty on
the background.  Furthermore, if an analysis were to simply select
events and count them, different event selection requirements would
need to be chosen in order to optimize the analysis for setting
limits, making a discovery, and measuring the signal rate.
Multivariate analysis techniques provide solutions to these challenges
by scoring events according to how signal-like (or background-like)
their measured properties are.  Higgs boson searches at the Tevatron
typically make use of neural
networks~\cite{Hocker:2007ht,Neal:1996:BLN:525544,Feindt:2006pm},
boosted decision trees~\cite{Friedman98additivelogistic}, and
matrix-element techniques~\cite{Abazov:2006bd,Aaltonen:2010jr}.  Some
analyses use several MVA discriminant functions sequentially in order
to separate the signal from more than one distinct source of
background contamination.  Typically, signal and background MC 
samples are used to train MVA classifier functions, with event
samples that are statistically independent from those used to predict
the signal and background rates in the subsequent statistical
analyses.  In some analyses and for some discriminants, data events in
background-enriched control samples (in which the signal contribution
is expected to be negligible) are used in the background training
samples.

The MVA discriminants are functions of reconstructed quantities for
each event and their distributions are used in the statistical
analyses, as described in Sec.~\ref{sec:statistics}.  Events
falling in low signal-to-background portions of the MVA discriminant
outputs serve as sideband constraints for the backgrounds, while
events in the high signal-to-background regions provide the most
powerful tests of the presence or absence of a Higgs boson, and
measure its production rate.  In the most sensitive Higgs boson searches,
the signal predictions in the highest-score bins are much larger than 
the corresponding post-fit background uncertainties. 
Multiple MVA functions are used in order to separate background
contributions from each other in order to reduce the total uncertainty
on the background contributions by providing measurements of each
component.

In order to optimize the sensitivity of the searches, separate MVA
functions are trained at each hypothesized value of $m_H$, typically
on a grid between 90~GeV and 200~GeV, in steps of
5~GeV.  The input variables to the MVA selections are also
optimized for each $m_H$ value in some analyses.  These differences
give rise to some statistical fluctuations in the observed cross
sections and limits as functions of $m_H$ even though the sensitivity
is expected to be a smooth function if all searches are optimized at
each $m_H$.

\subsection{Statistical methods}
\label{sec:statistics}

The statistical methods used to extract results from the Higgs boson
searches at the Tevatron are described in
Refs.~\citen{Beringer:1900zz} and~\citen{Aaltonen:2013kxa}.  Both Bayesian and
Modified Frequentist methods are used, and their results are compared
to check that the conclusions reached do not depend significantly
on the choice of statistical method.  The methods are chosen to make
maximum use of the separation power of the MVA techniques, while at
the same time incorporating the effects of systematic uncertainties in
the rates, shapes, and independent bin-by-bin uncertainties that arise
from limited MC sample (or data control sample) statistics.
The inclusion of uncertainties on the shapes of the distributions of complex
MVA discriminant variables allays concerns that unmodeled shape distortions can
give spurious results if only the rates of contributing signal and background
contributions are allowed to vary.  Shape distortions due to systematic uncertainties
are estimated by holding the discriminant function fixed and varying the uncertain
parameters in the modeling and producing alternate distributions for the variable
in question.   Examples are provided below.

Both the Bayesian and the Modified Frequentist techniques rely on a
binned likelihood function of the data, the model parameters, and the
nuisance parameters. In most analyses the model parameters are $m_H$
and $\mu$, a signal strength modifier which scales the SM predictions
in all combined channels together.  Each independent source of
systematic uncertainty is assigned a nuisance parameter, and
correlated systematic uncertainties are decomposed by their sources in
order to assign independent parameters.  The predictions of the yields
in each bin for the expected signal and the backgrounds are itemized
by process and they depend on the model parameters and the nuisance
parameters.  Some nuisance parameters, such as the integrated
luminosity, scale the predicted yields in each bin of all processes
affected by them.  In the case of the luminosity, this consists of all
processes using theoretical predictions and MC models.  Other
parameters, such as the jet energy scale in the detector simulation and the QCD factorization and renormalization
scale parameters in the event generators affect both the total rates
of processes (due to the fraction of events passing the event selection requirements) 
and also the shapes of the predicted distributions of kinematic
variables.  Correlations are included by parameterizing all bins and
all channels' predictions that are sensitive to a particular
systematic effect by the same nuisance parameter.  Each process in
each bin is also subject to a random, independent 
uncertainty due to MC (or data from a control sample)
statistics and given a separate nuisance parameter.  In the searches
presented here, sufficient MC samples have been simulated in
order to render negligible the effects of limited MC statistics.

In the Bayesian method, the nuisance parameters are integrated over (``marginalized''):
\begin{equation}
\label{eq:lprime}
L^\prime=\int L({\rm{data}}|{\vec{\bf{\theta}}},{\vec{\bf{\nu}}})\pi({\vec{\bf{\nu}}})d{\vec{\bf{\nu}}}
\end{equation}
where $L({\rm{data}}|{\vec{\bf{\theta}}},{\vec{\bf{\nu}}})$ is the likelihood function of the data, 
${\vec{\bf{\theta}}}$ are the model parameters $m_H$ and $\mu$, and ${\vec{\nu}}$ are the nuisance parameters.
Typically nuisance parameters are given Gaussian priors $\pi({\vec{\bf{\nu}}})$, 
truncated so that no prediction of any signal or background is negative, although more sophisticated priors are also possible.  
Studies have shown that in practical applications, the RMS widths
of the prior distributions is the most important feature controlling the impact of a particular systematic uncertainty on the
results.

The 95\% credibility level (C.L.) upper limit on the rate $\mu$ of a process ($\mu_{\rm{limit}}$) is given by
\begin{equation}
\label{eq:bayeslimits}
0.95 = \frac{\int_0^{\mu_{\rm{limit}}} L^\prime({\rm{data}}|\mu)\pi(\mu)d\mu}{\int_0^\infty L^\prime({\rm{data}}|\mu)\pi(\mu)d\mu}
\end{equation}
where the prior probability distribution for $\mu$ is taken to be uniform~\footnote{There is a formal divergence in limits
computed with truncated Gaussian priors and uniform priors on $\mu$, although they hardly appear in practice as the integration
ranges are typically chosen to be very large instead of infinite, and the sampling of the signal rates near zero is not
infinitely fine.}.  Markov Chain MC techniques~\cite{Metropolis:1953:ESC,Hastings70} are used to compute the integrals of 
Eqs.~\ref{eq:lprime} and~\ref{eq:bayeslimits} efficiently.  The sensitivity of the search is quantified by the expected
limit, which is computed as the median limit in a sample of simulated datasets with only background processes contributing,
sampling over the systematic uncertainties.  Expected sensitivity calculations also include 68\% and 95\% probability intervals
for the limits, computed with the same simulated pseudo-datasets.
To measure cross sections, the maximum of the
posterior probability density of $\mu$, $L^\prime({\rm{data}}|\mu)\pi(\mu)$, is found, and the
uncertainty is quoted using the shortest interval containing 68\% of the integral of the posterior density.

In the Modified Frequentist method, two $p$-values are computed using a log-likelihood ratio (LLR) as the test statistic:
\begin{equation}
{\rm{LLR}} = \frac{L({\rm{data}}|\mu,{\hat{\hat{\vec{\bf{\nu}}}}})}{L({\rm{data}}|\mu=0,{\hat{\vec{\bf{\nu}}}})}.
\end{equation}
Two maximum-likelihood fits are performed to the data, allowing the nuisance parameters $\vec{\bf{\nu}}$ to float.
One fit assumes that a signal is present with strength $\mu$, and the best-fit nuisance parameters in this case are
denoted ${\hat{\hat{\vec{\bf{\nu}}}}}$, and the other fit is performed assuming a signal is absent ($\mu=0$),
and the corresponding best-fit nuisance parameters are denoted ${\hat{\vec{\bf{\nu}}}}$.  The two $p$-values are 
\begin{equation}
{\rm CL_{s+b}} = p({\rm{LLR}}\ge {\rm{LLR_{obs}}} | \mu)
\end{equation}
and
\begin{equation}
{\rm CL_{b}} = p({\rm{LLR}}\ge {\rm{LLR_{obs}}} | \mu=0).
\end{equation}
The impact of systematic uncertainties on the $p$-values is included by sampling the values of the nuisance parameters
within their prior distributions in the process of generating pseudo-datasets in the calculation of the $p$-values.
The $p$-value
\begin{equation}
1-{\rm CL_{b}} = p({\rm{LLR}}\le {\rm{LLR_{obs}}} | \mu=0)
\end{equation}
is used to discover a new process.  If it is small, then the ability of the null hypothesis to explain the data is
small.  Small $p$-values are reported in units of Gaussian significance $z$ using the integral of one side of a Gaussian
distribution:
\begin{equation}
p = (1-{\rm{erf}}(z/\sqrt{2}))/2.
\end{equation}
A significance $z$ of 3 is the customary threshold for claiming evidence, and a significance of $5$ is the
threshold for claiming observation or discovery.  These $p$-values are computed separately for 
each value of $m_H$ and are called ``local'' $p$-values.
The ``Look-Elsewhere Effect" (LEE)~\cite{Lyons:1900zz,Dunn:1961my}, also called the multiple-tests effect, is taken into
account by studying the distribution of the smallest $1-{\rm{CL_{b}}}$ over a sample of simulated background-only
datasets and the ``global'' $p$-value is the probability of obtaining a specific value of the smallest
local $p$-value or smaller.  The sensitivity of the search at a specific mass $m_H$
is quantified by the median expected local $p$-value assuming a signal is truly present at that mass. 

Limits on $\mu$ are obtained using the CL$_{\rm{s}}$ technique in addition to the Bayesian limits described above.  The ratio
CL$_{\rm{s}}$=CL$_{\rm{s+b}}$/CL$_{\rm{b}}$ is computed as a function of the signal strength modifier $\mu$, and the upper limit
on $\mu$ is defined to be that which yields CL$_{\rm{s}}=0.05$.  The median expected upper limit on $\mu$ and the 68\% and 95\%
intervals of the distribution of the upper limit on $\mu$ are quoted to illustrate the sensitivity of the search and quantify
the expected distribution of possible outcomes.

Combined Tevatron Higgs search results use the Bayesian technique to quote limits and cross section
measurements, and the Frequentist $1-{\rm{CL_b}}$ $p$-value to quantify the significance of a signal.  The values
of LLR are displayed along with their expected distributions to also quantify the data's preference for either
the signal-plus-background or the background-only predictions.

\section{Searches for the Standard Model Higgs Boson}
\label{sec:sec4}

An early search for the SM Higgs boson~\cite{Abe:1998na} was performed by the CDF Collaboration 
using 91 pb$^{-1}$ of data recorded at $\sqrt{s}=1.8$ TeV during Run~I of the Tevatron Collider.
This search considered the associated production mode of a Higgs boson with a hadronically-decaying 
$W$ or $Z$ boson, with $H\to b\bar{b}$. The low available integrated luminosity and the small total 
selection efficiency achieved of $\approx 1$--2\%, mainly driven by the limited trigger and double $b$-tagging
efficiency, resulted in a cross section limit that was about two orders of magnitude larger than the
SM prediction for a Higgs boson with mass in the 70--140 GeV range.

A broad and competitive program of searches for the SM Higgs boson had to wait until
Run~II, exploiting much improved detectors and reconstruction algorithms, as well as a factor
of $\approx 100$ times larger integrated luminosity.
Eventually, with 10~fb$^{-1}$ of data analyzed per experiment at $\sqrt{s}=1.96$ TeV, the combination
of searches by the CDF and D\O\ Collaborations was expected to achieve 95\% C.L. exclusion
sensitivity to a Higgs boson with mass in the range between 90 GeV and 185 GeV.
In the following sections we review the search strategies followed by the Tevatron experiments
to achieve this goal, as well as discuss the main characteristics and results of the 
search channels considered. Then we summarize the final results on the SM Higgs boson from
the combination of all available searches, which constitute one of the main legacies of the
Tevatron physics program.

\subsection{Search strategies}

The main search modes at the Tevatron in the low $m_H$ region ($\approx 90$--120 GeV)
involve the associated production of a $W$ and $Z$ with a Higgs boson,
with the $W$ and $Z$ boson decaying leptonically and $H\to b\bar{b}$. 
At higher mass ($\approx 130$--185 GeV), the main search mode
is gluon-gluon fusion ($gg \to H$), with $H\to WW^{(*)}$, again involving leptonic $W$ boson decays. 
For $m_H\sim 125$ GeV, searches for $H\to b\bar{b}$ and $H\to WW^{(*)}$ have comparable sensitivity.
Although the above represent the main search channels, other combinations of production and decay modes 
have also been considered in order to further improve the sensitivity as well as the model-independence of the search. In particular,
in the low mass region, decay modes such as $H\to ZZ^{(*)}$, $\tau^+\tau^-$ and
$\gamma\gamma$ have also been exploited.

Just considering the main search channels,  $\approx 40$--70 Higgs boson events (assuming $m_H$ in the
range of 110--160 GeV) are expected 
to be produced per experiment and per fb$^{-1}$, adding up to an expected sample of about 1000 
Higgs bosons produced over the complete Run II data set. Selecting and identifying the signal candidate 
events from the overwhelming background represents a major challenge. This resulted in an aggressive
program of improvements by the CDF and D\O\ Collaborations leading to the development of some of the most 
sophisticated analyses up until now. For every Higgs boson search, the basic strategy involves: (a) selection of the
candidate sample, trying to maintain the highest possible acceptance; (b) classification of events into separate 
categories with different signal-to-background ratio; (c) validation of background predictions in dedicated
data control samples; (d) for each category, construction of a variable that is a function of the
measured quantities for each event that has the most discrimination between the
signal and background, typically involving the use of an MVA discriminant; and
(e) test of hypothesis involving the combination of all event categories, including 
in situ constraints on the dominant systematic uncertainties using high-statistics data control samples.

\subsection{Signal and background modeling}

The modeling of Higgs boson production is performed via leading-order
(LO) MC simulations provided by {\sc pythia}~\cite{Sjostrand:2006za} 
using the LO CTE5L or
CTEQ6L1~\cite{Lai:1999wy,Pumplin:2002vw} PDF sets.
While this provides a sufficiently accurate
model of the kinematics of Higgs boson production for most processes,
further accuracy is sought in modeling the $\pt$ spectrum of Higgs bosons
produced in the gluon-gluon fusion process.  Monte Carlo signal events in
this process are reweighted in order match the prediction of the $\pt$ distribution
predicted at NLO+NNLL accuracy by the {\sc hqt} program~\cite{Bozzi:2005wk}.
The decay of the Higgs boson is modeled by {\sc pythia} with
branching ratios predictions from Ref.~\citen{Dittmaier:2012vm}. All MC
samples are normalized to the highest-order (NLO or higher) cross
section calculation available for the corresponding production process
(see Sec.~\ref{sec:sec2}).
 
Higgs boson searches at the Tevatron are affected by large backgrounds
that can be categorized as ``physics" and ``instrumental"
backgrounds. The optimized event selections used in the Higgs boson
searches often result in the former dominating over the latter.  The
main physics backgrounds involve the production of a vector boson
produced in association with jets ($W/Z$+jets), single and pair
production of top quarks, and diboson ($WW$, $WZ$, $ZZ$)
production. Backgrounds from $W/Z$+jets are typically simulated using
MC matrix element generators such as {\sc alpgen}~\cite{Mangano:2002ea}, allowing the simulation of high parton
multiplicities at LO. This includes the generation of samples with
extra heavy-flavor quarks, such as $W/Zb\bar{b}$+jets and
$W/Zc\bar{c}$+jets. These samples are interfaced with {\sc pythia} for
further showering and hadronization, and implement the MLM parton-jet
matching algorithm~\cite{Mangano:2001xp} to avoid double-counting of
radiation between the matrix-element calculation and the parton
shower.  Backgrounds from top quark pair production are modeled using
{\sc pythia} (CDF) or {\sc alpgen}+{\sc pythia} (D\O), while
backgrounds from single top quark production are modeled using the
{\sc madevent}~\cite{Maltoni:2002qb} (CDF) or {\sc singletop}~\cite{Boos:2006af} (D\O) 
event generators, both interfaced
to {\sc pythia}. Finally, diboson production is modeled with {\sc pythia}. 
The corresponding MC samples are normalized to higher-order
theoretical cross sections (typically at NLO or higher). In the case
of $W/Z$+jets events, imperfections in the modeling of the vector
boson $\pt$ or jet kinematics, or the heavy-flavor content, are
corrected using data control samples.

Instrumental backgrounds are estimated either entirely from data or by
applying data-driven corrections to dedicated MC samples. Examples of
instrumental backgrounds include QCD multijet production with jets
misidentified as isolated leptons and/or jet energy mismeasurements
generating spurious $\met$. Other examples include $W$+$\gamma$/jets
production with photons and jets misidentified as leptons or
$Z/\gamma^*(\to \ell^+\ell^-)$+jets production with fake $\met$ because
of jet energy or lepton momentum mismeasurements. Some searches
requiring same-charge leptons are also sensitive to lepton charge
mismeasurements, which the simulation programs often do not predict
accurately enough and need to be corrected based on measurements in
data control samples.

All MC samples are processed through {\sc geant}~\cite{geant} simulations of the
detectors and reconstructed using the same software as used for 
collision data. The effects from electronic noise and additional
proton-antiproton interactions are included either via the simulation
(CDF) or by overlaying data events from randomly selected beam
crossings on the MC events, in both cases attempting to
reproduce the instantaneous luminosity spectrum of the analyzed
dataset. Averaging over the entire Run II data sample, approximately
two additional proton-antiproton interactions per event were present,
which did not result in a significant degradation of the performance
of reconstruction algorithms.  This is contrast with the average of
$\approx 20$ proton-proton interactions per crossing recorded by the
ATLAS and CMS experiments during the run in 2012 at $\sqrt{s}=8$ TeV.
 
\subsection{Searches for $H\to b\bar{b}$}
\label{sec:sec4_hbb}

Searches for $H\to b\bar{b}$ at the Tevatron dominate the sensitivity
in the low $m_H$ region ($\approx 90$--120 GeV), capitalizing on the
$VH$ production modes with leptonic $W$ and $Z$ decays, which
facilitate event triggering and identification.  The main search
channels exploited are $WH\to \ell\nu b\bar{b}$, $ZH\to\ell^+\ell^-
b\bar{b}$ and $ZH\to\nu\nu b\bar{b}$.  While identifying a data sample
enriched in $W/Z$+jets is straightforward using $W$ and $Z$ decays
into electrons or muons, it is more challenging in the case of
$Z(\to \nu\nu)$+jets owing to the large background from QCD multijet
production in such jets+$\met$ signature. Nevertheless, sophisticated
techniques have been developed that allow the effective suppression of the
QCD background, making $ZH\to\nu\nu b\bar{b}$ one of the most
sensitive search channels in the $H\to b\bar{b}$ decay mode.

The main backgrounds to $VH(\to b\bar{b})$ searches are $W/Z$+jets and
$t\bar{t}$ production. Smaller backgrounds originate from single top,
QCD multijet and diboson production. A key experimental handle to
suppress the background is the requirement of having at least one
$b$-tagged jet. Over the years, sophisticated MVA $b$-tagging
algorithms have been developed by the CDF and D\O\ Collaborations (see
Sec.~\ref{sec:btagging}).  The $b$-tagging information can most optimally
be used by categorizing events according to the purity and number of
$b$-tagged jets. Samples with more stringent $b$-tagging requirements
are dominated by $V$+heavy-flavor jets, and in particular $Vb\bar{b}$,
which constitutes an irreducible background.  The main discriminating
variable between the $VH$ signal and the backgrounds is the dijet
invariant mass distribution, which shows a resonant structure around
the Higgs boson mass for signal, while it has a smoothly-falling spectrum for
background. Therefore, significant efforts have been undertaken to
improve the dijet mass resolution (see
e.g. Ref.~\citen{Aaltonen:2011bp}). The final step for these searches
is to combine a number of kinematic variables into MVA
discriminants in order to maximize the sensitivity. A crucial
validation of the overall search strategy is provided by the precise
measurement of small-cross section backgrounds with the same signature
as the signal, such as single top quark and diboson production.  Lower-sensitivity 
searches have also been carried out in the $VH$ and VBF
production modes with fully-hadronic final states, as well as in the
$t\bar{t}H$ production mode.  A summary of the main features and
results for all $H\to b\bar{b}$ searches is provided below.

%%%%%%%%%%%%%%%%%%%%%
\begin{figure}[t]
\begin{center}
\begin{tabular}{cc}
\includegraphics[height=5.5cm, width=0.49\textwidth]{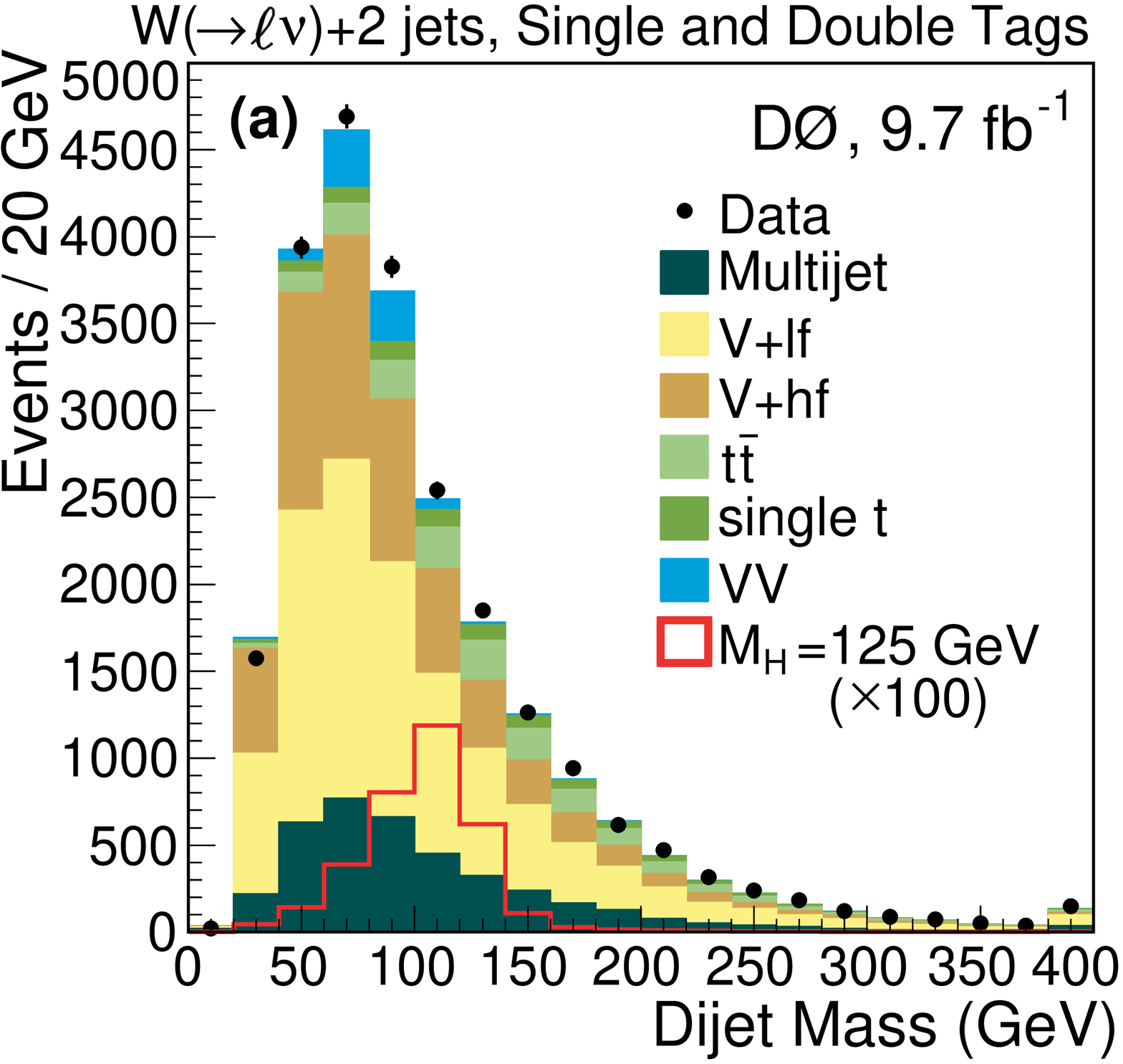} &
\includegraphics[height=5.5cm, width=0.49\textwidth]{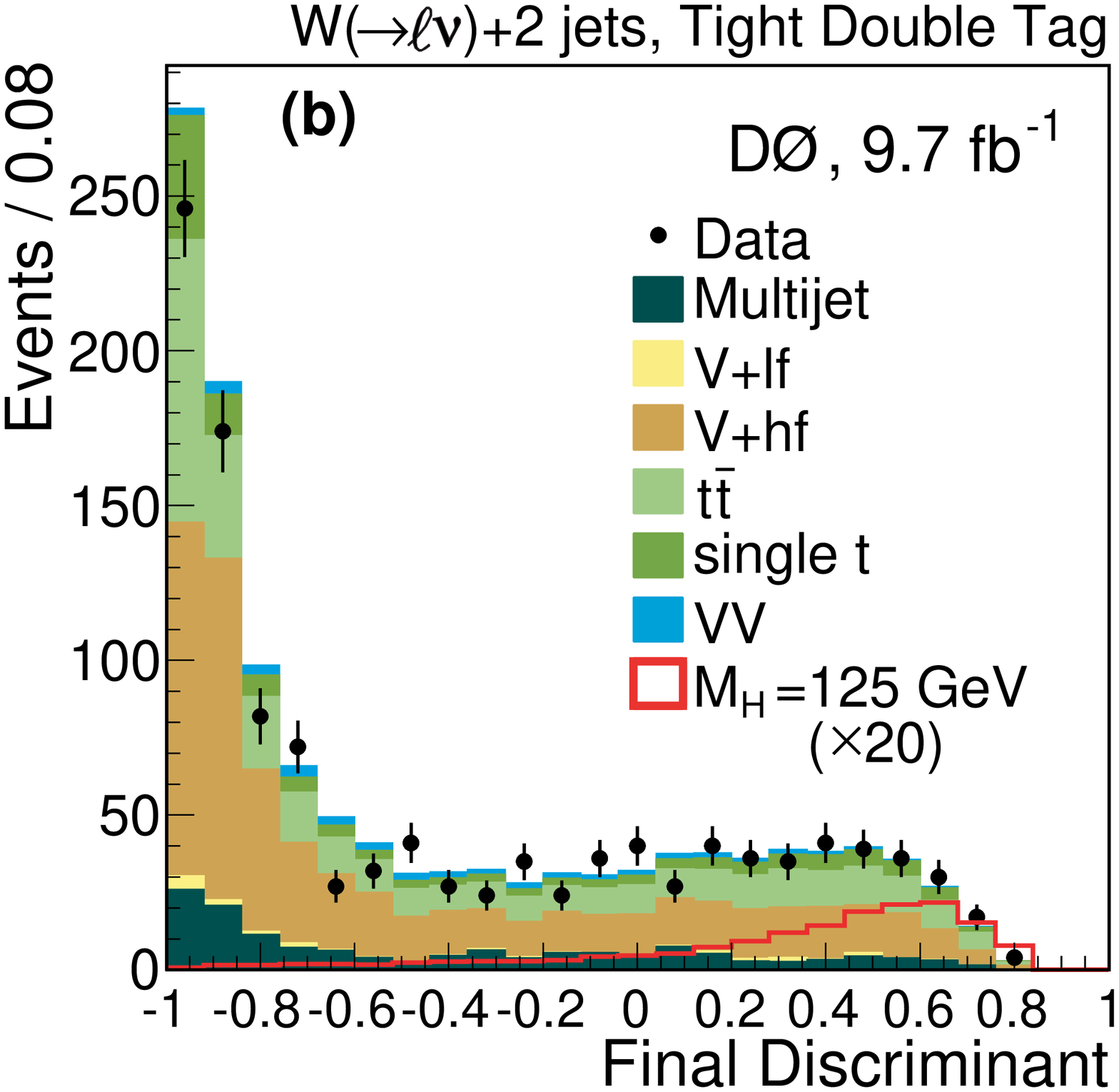} \\
\end{tabular}
\end{center}
\caption{(a) Distribution of the dijet invariant mass for all $b$-tag categories combined in the 2-jet channel of the D\O\ $WH\to \ell\nu b\bar{b}$ search.
The data (points with error bars) are compared to the background prediction, 
broken down into its individual components. Also shown is the expected contribution from a SM Higgs 
boson with $m_H = 125$~GeV scaled by a factor of 100. 
(b) Distribution of the final BDT distribution for the tight double $b$-tag category in the 2-jet channel of the D\O\ $WH\to \ell\nu b\bar{b}$ search.
The data (points with error bars) are compared to the background prediction, 
broken down into its individual components. Also shown is the expected contribution from a SM Higgs 
boson with $m_H = 125$~GeV scaled by a factor of 20. 
From Ref.~\citen{Abazov:2012zla}.
\label{fig:hbb_1}}
\end{figure}
%%%%%%%%%%%%%%%%%%%%%

\subsubsection{$WH\to \ell \nu b\bar{b}$}

The CDF and D\O\ Collaborations have performed searches for $WH\to \ell \nu b\bar{b}$ using the full
Run II dataset~\cite{Aaltonen:2012ic,Abazov:2012zla,Abazov:2013mjc}.   These searches have much in common
with the single top quark searches and subsequent observations and measurements made previously\cite{Abazov:2008kt,Abazov:2009ii,
Aaltonen:2010jr,Aaltonen:2009jj}.  Specifically, since the final state contains the same particle content, the backgrounds
to both analyses are from the same processes, though the signals have different kinematic properties.  The single top quark
signal has a higher production cross section and more distinct kinematic properties; the fact that the top quark mass
was known precisely also helped.  The techniques for search and discovery such as background estimation, cut and MVA optimization,
and systematic uncertainty estimation were tested,  improved, and validated first in 
the search for single top quarks and then refined for the Higgs boson analyses.  Interestingly, many of
the improvements made in the Higgs boson searches, such as the more sophisticated $b$-taggers,
were then propagated back into the final 
single top results~\cite{Aaltonen:2014ura,CDF:2014uma,Aaltonen:2014qja,Abazov:2013qka,Abazov:2011pt,Abazov:2011pm,Abazov:2011rz}.

Candidate events are selected
requiring a single isolated lepton ($e$ or $\mu$), large $\met$ and two or three jets in the event, at least
one of which is required to be $b$-tagged. Lepton selections are kept as loose as possible in order to
maximize acceptance, requiring the development of sophisticated techniques to suppress the QCD multijet 
background, leaving a sample dominated by background events containing real leptonic $W$ decays.
In both the CDF and D\O\ analyses, events are categorized into different channels depending on the
jet multiplicity (2 or 3 jets) and the number and purity (``loose'' (L) or ``tight'' (T)) 
of $b$-tagged jets. As a result, 
the CDF analysis considers five $b$-tagging categories (TT, TL, LL, T, L) for the 2-jet sample and
two categories (TT, TL) for the 3-jet sample, while the D\O\ analysis considers four categories (TT, TL, LL, T)
and two categories (LL, T), respectively. Channels with two $b$-tagged jets are enriched in $Wb\bar{b}$,
$t\bar{t}$ and single top backgrounds, while channels with one $b$-tagged jet are dominated by 
$W$+light or charm jets, and contain also sizable QCD multijet contributions. 
For each of the analysis channels, optimized MVA discriminants are trained against the corresponding backgrounds, 
considering a number of kinematic distributions, in addition to the dijet mass. Figure~\ref{fig:hbb_1} shows examples
of the inclusive dijet mass distribution (summed over all analysis channels) and the final MVA discriminant in the
most sensitive channel of the D\O\ search.
The observed (expected) cross section limits at $m_H=125$~GeV for the CDF and D\O\ analyses are 
4.9 (2.8) and 5.2 (4.7) times the SM prediction, respectively.

\subsubsection{$ZH\to \ell^+ \ell^- b\bar{b}$}

The CDF and D\O\ Collaborations have performed searches for $ZH\to \ell^+ \ell^- b\bar{b}$ using the full
Run II dataset~\cite{Aaltonen:2012id,Abazov:2012tka,Abazov:2013alg}. Candidate events are selected
requiring two opposite-sign (OS) same-flavor isolated leptons ($e^+e^-$ or $\mu^+\mu^-$) and two or three jets 
in the event, at least one of which is required to be $b$-tagged. Lepton selections are kept as efficient as possible 
in order to maximize acceptance, since after the requirement that the two leptons form a $Z$ boson candidate,
background from misidentified leptons is negligible. In any case, events are categorized according to the
quality of the identified leptons. Similarly to the $WH$ analysis, different event categories are defined based
on jet and $b$-tag multiplicities, and on $b$-tag purity requirements. The CDF analysis considers four $b$-tagging 
categories (TT, TL, LL, T) for both the 2-jet and 3-jet samples, while the D\O\ analysis considers two categories (TL, T)
for both the 2-jet and 3-jet samples. The absence of $\met$ in the event allows for improved invariant mass
resolution by imposing event-wide transverse momentum constraints: in the case of the CDF analysis corrections to the jet energies are performed 
via a dedicated NN relating the measured jet energies and directions to the $\met$ vector on an event-by-event basis;
in the case of the D\O\ analysis an improved measurement of the jet energy is obtained from a kinematic fit 
imposing constraints on the dilepton mass to be consistent with the $Z$ boson mass and that the missing transverse momentum of the leptons-plus-jets 
system should be consistent with zero. A sophisticated MVA strategy is followed whereby different MVA discriminants are trained to separate the signal from the
different backgrounds ($Z$+jets, $t\bar{t}$  and diboson), one at a time.
The observed (expected) cross section limits at $m_H=125$~GeV for the CDF and D\O\ analyses are 
7.1 (3.9) and 7.1 (5.1) times the SM prediction, respectively.

\subsubsection{$WH,ZH\to \met \, b\bar{b}$}

The CDF and D\O\ Collaborations have performed searches for $ZH\to \nu\nu b\bar{b}$ using the full
Run II dataset~\cite{Aaltonen:2012ii,Aaltonen:2013js,Collaboration:2012hv}. Candidate events are 
selected requiring at least two jets (CDF) or exactly two jets (D\O), no identified leptons and significant 
$\met$ not aligned with the jet directions.
Dedicated triggers using $\met$ are used with or without accompanying jets. About half of the signal events
in this channel originate from $WH\to \ell \nu b\bar{b}$ with the charged lepton not identified, hence the
name $WH,ZH\to \met\; b\bar{b}$ given to this search. As the previous $VH(\to b\bar{b})$
searches, different event categories are defined based on $b$-tag multiplicity and purity requirements: 
the CDF analysis considers three $b$-tagging  categories (TT, TL, T), while the D\O\ analysis considers 
two categories which are defined by requirements on 
the sum of the $b$-tagging output variables for the two taggable jets in the event.  The large QCD multijet background
with spurious $\met$ is effectively suppressed via MVA discriminants that exploit information of the $\met$
as measured by the calorimeter and by the tracker, including the correlation between their directions in the transverse
plane and with respect to the directions of the jets. As a result, after final selection the QCD multijet background can be
made comparable or significantly smaller than the physics background, dominated by $V$+heavy-flavor and $t\bar{t}$.
For each of the analyzed samples,  MVA discriminants are trained between the signal and all backgrounds.
The observed (expected) cross section limits at $m_H=125$~GeV for the CDF and D\O\ analyses are 
3.1 (3.3) and 4.3 (3.9) times the SM prediction, respectively.

\subsubsection{$VH(\to b\bar{b})$ search results and validation}
\label{sec:vhsearch}

The CDF and D\O\ Collaborations have performed individual combinations
of their $VH(\to b\bar{b})$ search results. The CDF combined
search~\cite{Aaltonen:2012if} excludes a SM Higgs boson with a mass in
the range of 90--96~GeV and finds a broad excess with smallest local
$p$-value at a mass of 125 GeV corresponding to a significance of 2.7
s.d. In the case of the D\O\ combination~\cite{Abazov:2012qya}, the
excluded mass range is 100--102 GeV and the smallest local $p$-value
is found at a mass of 135 GeV and corresponds to a significance of 1.7
s.d. The combined result from both
experiments~\cite{Aaltonen:2012qt} reached a maximum local
significance of 3.3 s.d. at a mass of 135 GeV, becoming 3.1 s.d. after
taking into account the LEE, thus representing the first
evidence for the presence of a particle produced in association with a
$W$ or $Z$ boson and decaying to $b\bar{b}$. Since then, the CDF
$WH,ZH\to \met \, b\bar{b}$ result was updated~\cite{Aaltonen:2013js}
to use a more powerful MVA $b$-tagging algorithm along with changes in
the kinematic selections, resulting in a statistical fluctuation that
slightly reduced the significance of the excess. The measured combined
cross section times branching ratio at $m_H=125$ GeV from the updated
combination~\cite{Aaltonen:2013kxa} is
$(\sigma_{WH}+\sigma_{ZH})\times {\cal B}(H\to
b\bar{b})=0.19^{+0.08}_{-0.09}\;{\rm (stat+syst)}$~pb, about 1.5 times
larger than the SM prediction at the same mass.  Results from the
combination of $H\to b\bar{b}$ searches with the rest of search
channels are provided in Sec.~\ref{sec:sec4_results}.

%%%%%%%%%%%%%%%%%%%%%
\begin{figure}[t]
\begin{center}
\begin{tabular}{cc}
\includegraphics[height=5.5cm, width=0.49\textwidth]{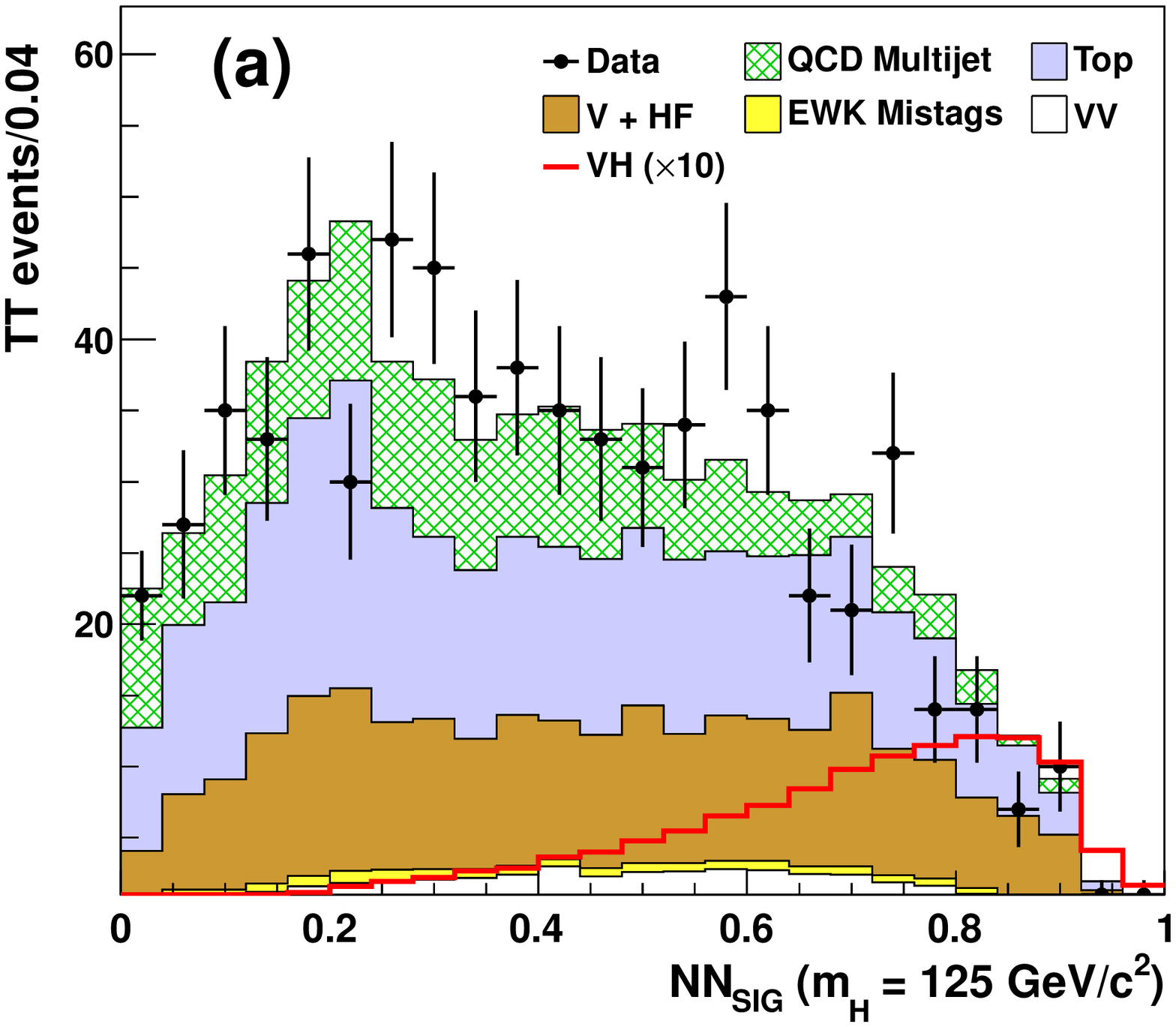} &
\includegraphics[height=5.2cm, width=0.49\textwidth]{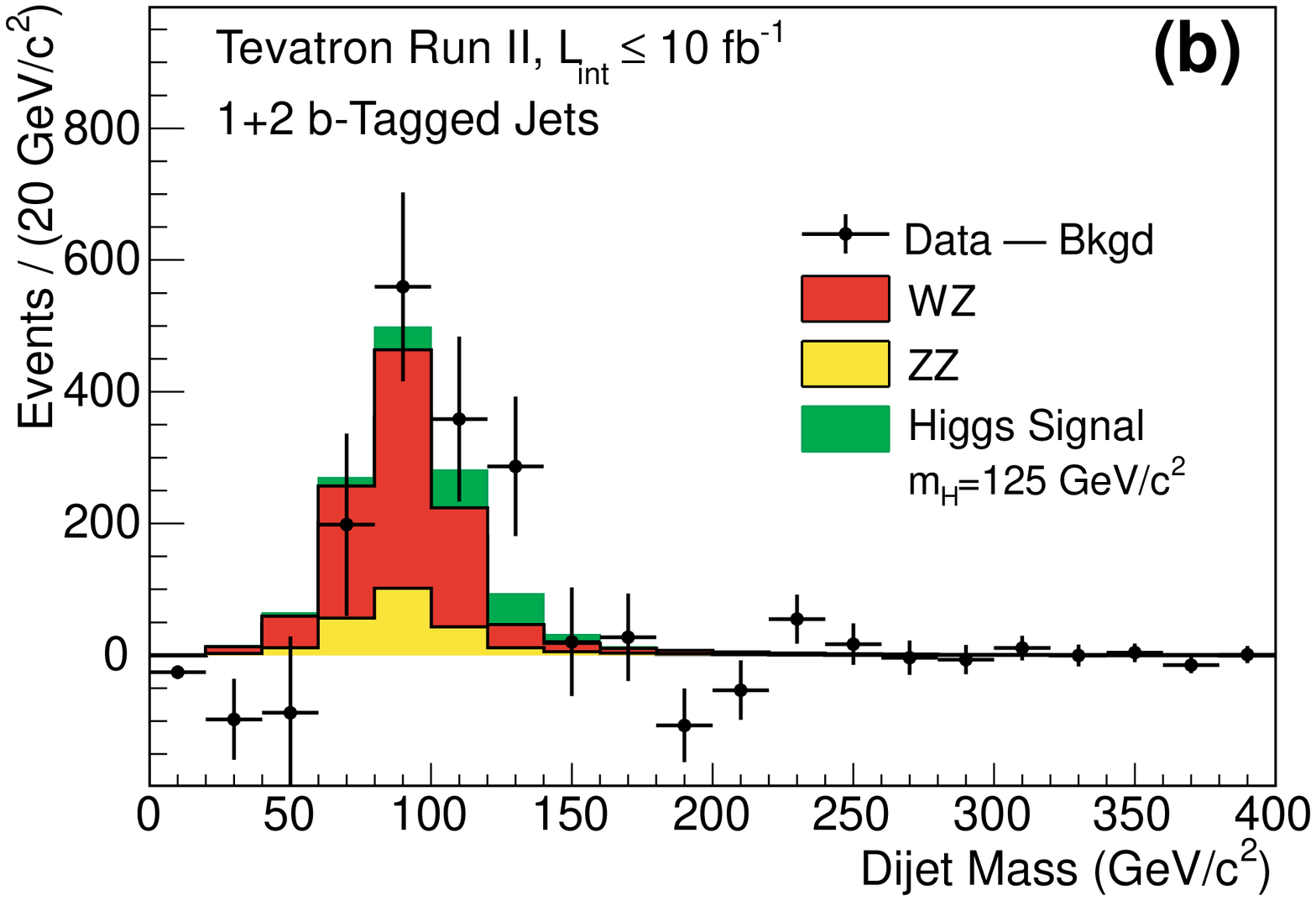} \\
\end{tabular}
\end{center}
\caption{(a) Distribution of the final NN discriminant for the TT category of the CDF $WH,ZH\to \met\; b\bar{b}$ search (see text for details).
The data (points with error bars) are compared to the background prediction, 
broken down into its individual components. Also shown is the expected contribution from a SM Higgs 
boson with $m_H = 125$~GeV scaled by a factor of 10. 
Adapted from Ref.~\citen{Aaltonen:2013js}.
(b) Background-subtracted distribution of the reconstructed dijet mass, summed over all CDF and D\O\ 
channels contributing to the $VZ$ analysis. The fitted $VZ$ and the expected
SM Higgs (assuming $m_H = 125$~GeV) contributions are shown with filled histograms.
From Ref.~\citen{Aaltonen:2013kxa}.
\label{fig:hbb_2}}
\end{figure}
%%%%%%%%%%%%%%%%%%%%%

The sophisticated analysis techniques and methodology used in the
$VH(\to b\bar{b})$ searches are validated by measuring the cross
section for $VZ$ production, with the $Z$ boson decaying into
heavy-flavor jets. This process has the same signature as the signals
of interest, including the feature of a resonance in the $b\bar{b}$
invariant mass spectrum. While the SM prediction for the cross section
for $VZ(\to b\bar{b})$ is about six times larger than for the Higgs
boson signal, this process is affected by larger background from
$V$+jets owing to the lower invariant mass of the $b\bar{b}$ system compared to the $VH$ signal. 
Exactly the same analyses as for the Higgs boson search are used for this measurement,
with the only difference being that MVA discriminants are
trained considering $VZ$ as the signal of interest, and potential
contributions from Higgs boson production are not considered. The
measured cross section from the combination of CDF and D\O\
analyses~\cite{Aaltonen:2013kxa} is $\sigma_{VZ}=3.0 \pm 0.6\;
({\rm stat.})\pm 0.7\;({\rm syst.})$~pb, in good agreement with SM
prediction of $4.4\pm 0.3$~pb~\cite{Campbell:1999ah}.  Individual
measurements by the CDF and D\O\
Collaborations~\cite{Aaltonen:2013ipa,Abazov:2013gmz} are also found
to be consistent with the SM prediction. Figure~\ref{fig:hbb_2}(b)
shows the combined background-subtracted dijet invariant mass
distribution, clearly showing an excess compatible in yield and shape
with that expected from $VZ$.

\subsubsection{$VH,q\bar{q}H \to jj b\bar{b}$}

The CDF Collaboration has performed a search for $H\to b\bar{b}$ in
the fully-hadronic final state using 9.45 fb$^{-1}$ of
data~\cite{Aaltonen:2012ji}.  This search focuses on the $VH$ and VBF
production modes resulting in a signature consisting of four or five
jets, at least two of which are $b$-tagged. Two different $b$-tagging
algorithms with different efficiency and purity are employed, and
different analysis channels are defined based on the algorithms
contributing to each $b$-tagged jet. The main background originates
from QCD multijet production and is modeled directly from
data. Multivariate discriminant variables are constructed in each of
the analyzed channels to separate the signal from the
background.  The observed (expected) cross section limit at
$m_H=125$~GeV is 9.0 (11.0) times the SM prediction.

\subsubsection{$t\bar{t}H \to t\bar{t} b\bar{b}$}

The CDF Collaboration has performed a search for $t\bar{t}H\to t\bar{t}b\bar{b}$ in the lepton-plus-jets final state 
using 9.45 fb$^{-1}$ of data~\cite{Collaboration:2012bk}. Events are selected
requiring one electron or muon, large $\met$ and at least four jets.
Similarly to the $VH,q\bar{q}H \to jj b\bar{b}$ search, two different
$b$-tagging algorithms with different efficiency and purity are
employed.  Events are categorized into different channels depending on
their jet multiplicity (4, 5 and $\geq 6$ jets), the number of
$b$-tags and the algorithms contributing to each $b$-tagged
jet. Multivariate discriminant variables are constructed in each of
the analyzed channels to separate the signal from the
dominant $t\bar{t}$+jets background.  The observed (expected) cross
section limit at $m_H=125$~GeV is 20.5 (12.6) times the SM prediction.
The D\O\ Collaboration has also performed a search for $t\bar{t}H\to t\bar{t}b\bar{b}$
using 2.1~fb$^{-1}$ of data~\cite{d0tthprel}, with an observed (expected) limit of 84.8 (64.2) 
times the SM prediction; this result is not included in the combination.

\subsection{Searches for $H\to \tau^+ \tau^-$}

The CDF and D\O\ Collaborations have performed a number of searches
involving hadronic $\tau$ leptons ($\tau_h$) attempting to probe the
$H\to\tau^+\tau^-$ decay mode. Searches for $H+X \to \ell \tau_h$+jets
are sensitive to the main Higgs boson production mechanisms,
$gg\rightarrow H$, $VH$ and VBF, and to both $H\to WW^{(*)}$ and
$H\to \tau^+\tau^-$ decays.  The D\O\ search~\cite{Abazov:2012ee} uses
9.7 fb$^{-1}$ of data and requires an electron or muon, a $\tau_h$ and at
least two jets, while the CDF search~\cite{Aaltonen:2012jh}, which uses
6 fb$^{-1}$ of data, considers in addition events with exactly one jet. The
dominant backgrounds in these searches originate from $W/Z$+jets,
$t\bar{t}$ and QCD multijet backgrounds. Multivariate techniques are
employed to separate signal from background by making use of a large
number of kinematic variables. In the case of the D\O\ analysis, a BDT
trained to distinguish between $H\to\tau^+\tau^-$ and $H\to WW^{(*)}$
signals is used to define $\tau\tau$- and $WW$-dominated samples, which
are analyzed separately for decay mode-specific limits, or in
combination assuming the SM prediction for the ratio of
$H\to\tau^+\tau^-$ and $H\to WW^{(*)}$ branching ratios. The most
restrictive limit for the $H\to\tau^+\tau^-$ decay mode is obtained by
the D\O\ $\tau\tau$-specific discriminant, yielding an observed
(expected) cross section limit at $m_H=125$~GeV of 12.8 (10.4) times
the SM prediction. The observed (expected) cross section limit obtained by the CDF
Collaboration is 16.4 (16.9) times the SM prediction, but it does not
correspond specifically to the $H\to \tau^+\tau^-$ decay mode, as it 
also includes a non-negligible contribution from $H\to WW^{(*)}$ decays.  Additional
sensitivity to the $H\to\tau^+\tau^-$ decay mode is achieved via searches probing the $VH$ production
mechanism together with $H\to \tau^+\tau^+$, leading to trilepton
final states ($ee\mu$, $e\mu\mu$ and $\mu\tau_h\tau_h$) involving both
leptonic and hadronic $\tau$
decays~\cite{Aaltonen:2013iia,Abazov:2013eha}.

\subsection{Searches for $H\to WW^{(*)}$}
\label{sec:sec4_hww}

While searches for $H\to WW^{(*)}$ typically reach their highest
sensitivity at the maximum of $\sigma_H \times {\cal B}(H\to WW^{(*)})$ as a function of $m_H$, 
around $m_H=165$~GeV, these searches significantly
contribute to the combined sensitivity down to $m_H\sim 125$ GeV. A number of 
searches have been developed targeting different Higgs boson
production and decay modes, resulting in very different topologies,
depending on the number of charged leptons, whether they have the same or
opposite charge, and the number of additional jets considered. The
most sensitive topology involves a pair of OS charged
leptons and no additional jets, aiming at isolating the 
$gg \to H \to WW^{(*)} \to \ell^+\nu \ell^- {\bar\nu}$ signal. Searches for OS dileptons
are dominated by $e^+e^-$, $\mu^+\mu^-$ and $e^\pm\mu^\mp$ but final
states with one $\tau_h$, $e\tau_h$ and $\mu\tau_h$, have also
been considered. Other searches require two same-sign (SS) charged
leptons or three leptons (trileptons), being primarily sensitive to
the $VH$ production mode, with $H \to WW^{(*)}$.  More challenging
searches exploiting the semileptonic decay mode of the $WW^{(*)}$
pair, $H \to WW^{(*)} \to \ell \nu q\bar{q}'$, have also been
developed. A summary of the main features and results for these
searches is provided below.

\subsubsection{Opposite-sign dileptons}

\begin{figure}[t]
\begin{center}
\begin{tabular}{cc}
\includegraphics[height=5.5cm, width=0.49\textwidth]{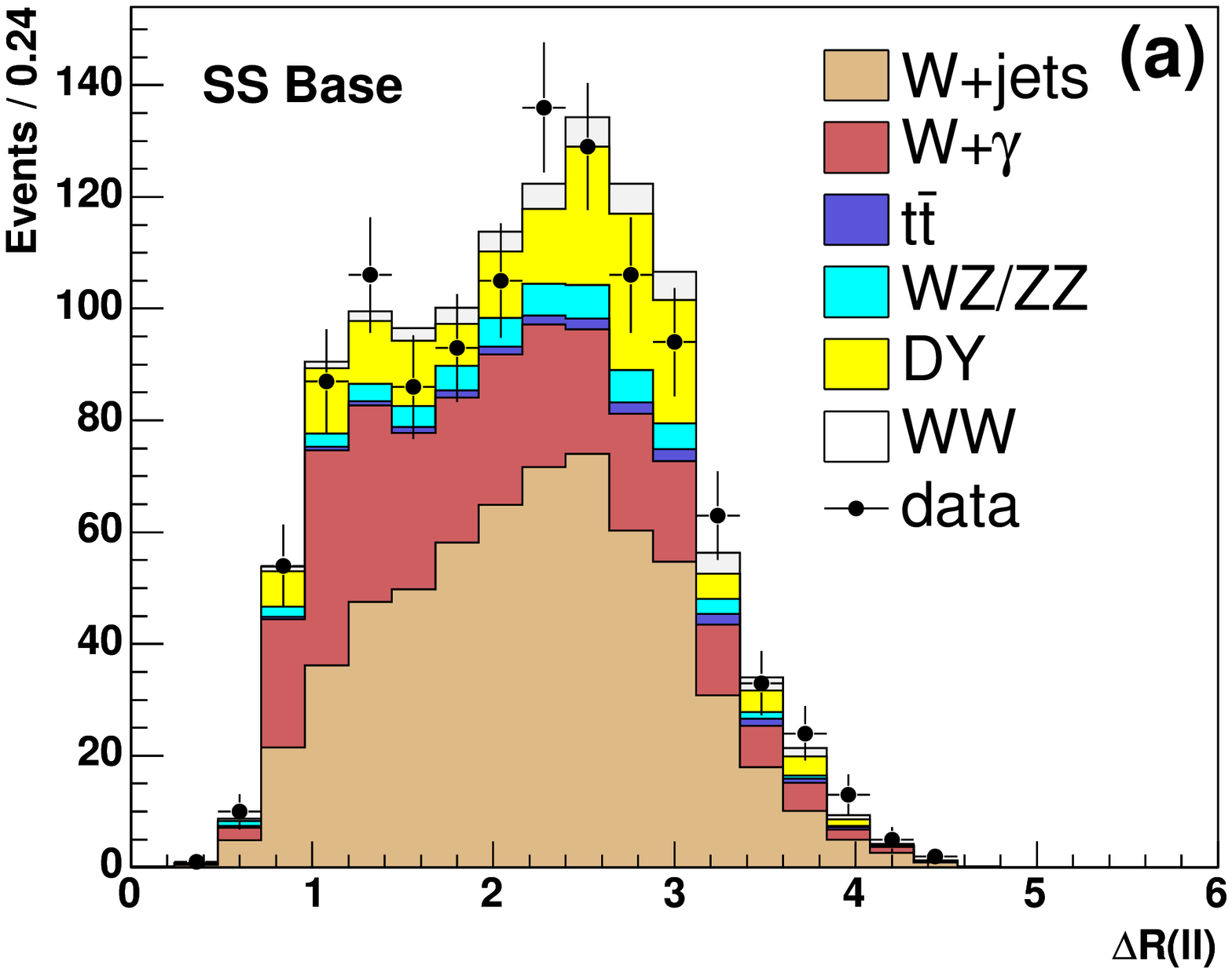} &
\includegraphics[height=5.5cm, width=0.49\textwidth]{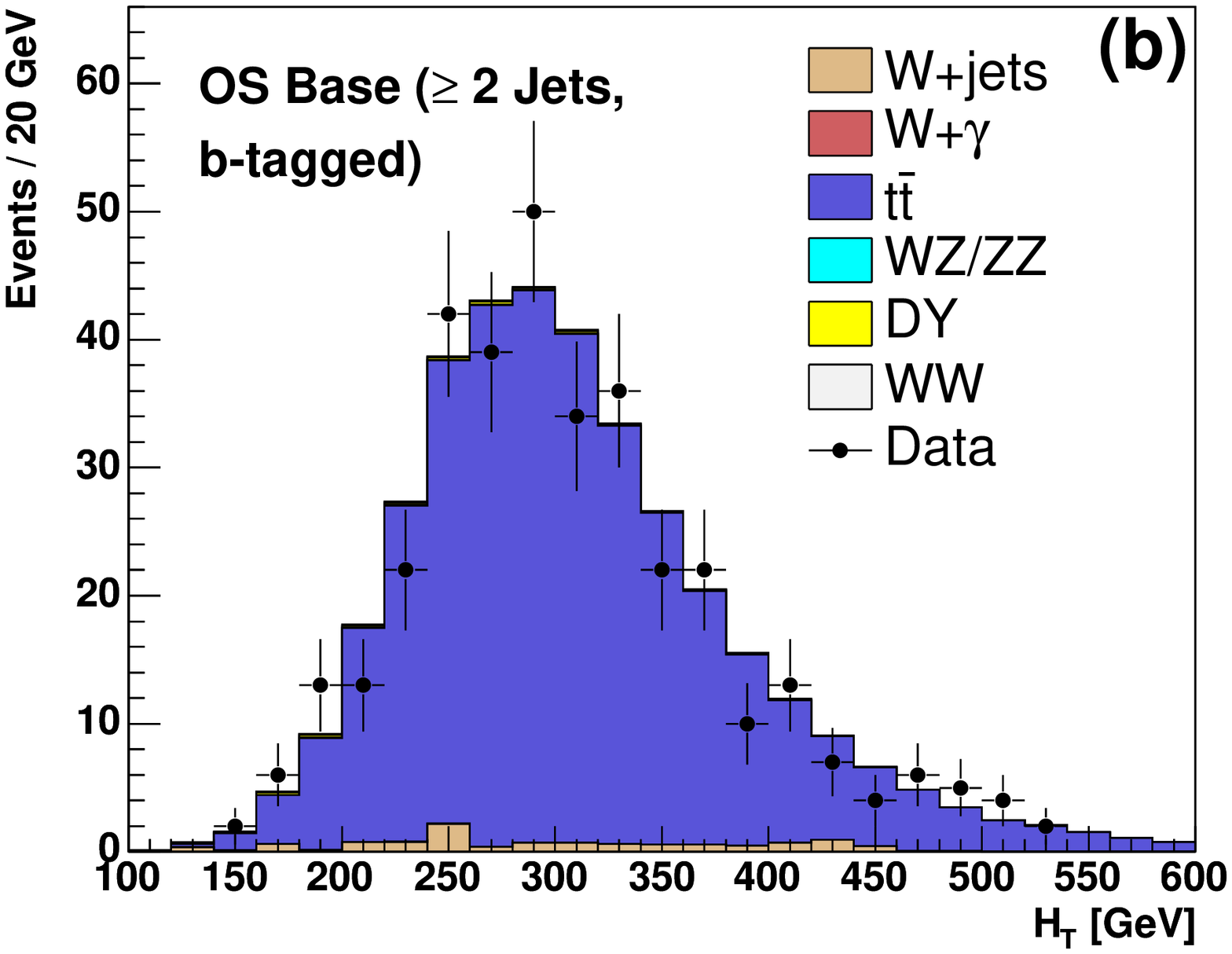} \\
\end{tabular}
\end{center}
\caption{(a) Distribution of the three-dimensional angular separation between the two charged leptons ($\Delta R(\ell^+,\ell^-)$) in the SS dilepton+$0,1$ jets control sample used to validate the modeling of $W$+jets/$\gamma$ backgrounds in the CDF $H\to WW^{(*)}$ search. From Ref.~\citen{Abazov:2013wha}.
(b) Distribution of the scalar sum of the lepton $\pt$, $\met$ and $\pt$ of the jets ($\HT$) in the OS dilepton+$\geq 2$ jets/$\geq 1$ $b$-tags control sample used to validate the modeling of the $t\bar{t}$ background in the CDF $H\to WW^{(*)}$ search.
The data (points with error bars) are compared to the background prediction,  broken down into its individual components.
From Ref.~\citen{Abazov:2013wha}.
\label{fig:hww_control}}
\end{figure}

The CDF and D\O\ Collaborations have searched for $H\to WW^{(*)}$ with both $W$ 
bosons decaying leptonically, giving an
experimentally clean signature of two OS charged
leptons ($e$ or $\mu$) and significant
$\met$~\cite{Aaltonen:2013iia,Abazov:2013wha}.  These searches are
sensitive to the three main Higgs boson production mechanisms,
although $gg\rightarrow H$ dominates the sensitivity.  Since at low
$m_H$ one of the $W$ bosons from the Higgs boson decay is off-shell,
lepton selections optimized down to low $\pt$ have been developed.
After $\met$ requirements, the main backgrounds are non-resonant
$W^+W^-$ with two real leptons, and $W$+jets and $W$+$\gamma$ with a 
jet or photon mimicking the signature of an isolated lepton. Additional contributions,
primarily affecting the same-flavor dilepton channels ($ee$ and
$\mu\mu$) originate from $Z/\gamma^*$+jets with jet energy mismeasurements
causing spurious $\met$. At higher jet multiplicity, the contribution
from dileptonic $t\bar{t}$ events can be substantial even after
vetoing $b$-tagged jets. Finally, smaller background contributions
arise from $WZ$ and $ZZ$ processes. Backgrounds are estimated using a
combination of MC simulations and data-driven techniques. In the case
of backgrounds with real leptons and true $\met$ from neutrinos, the
MC simulation is used. On the other hand, instrumental backgrounds
originating from the misidentification of jets or photons as leptons
($W$+jets/$\gamma$) or mismeasured $\met$ ($Z/\gamma^*$+jets), are
not sufficiently well modeled by the simulation, which is improved by applying 
dedicated data-based corrections. Detailed comparisons between the
data and the background predictions are made in dedicated control
samples enriched in one background at a time (see examples in
Fig.~\ref{fig:hww_control}).

%%%%%%%%%%%%%%%%%%%%%
%%%%%%%%%%%%%%%%%%%%%

\begin{figure}[t]
\begin{center}
\begin{tabular}{cc}
\includegraphics[height=5.5cm,width=0.49\textwidth]{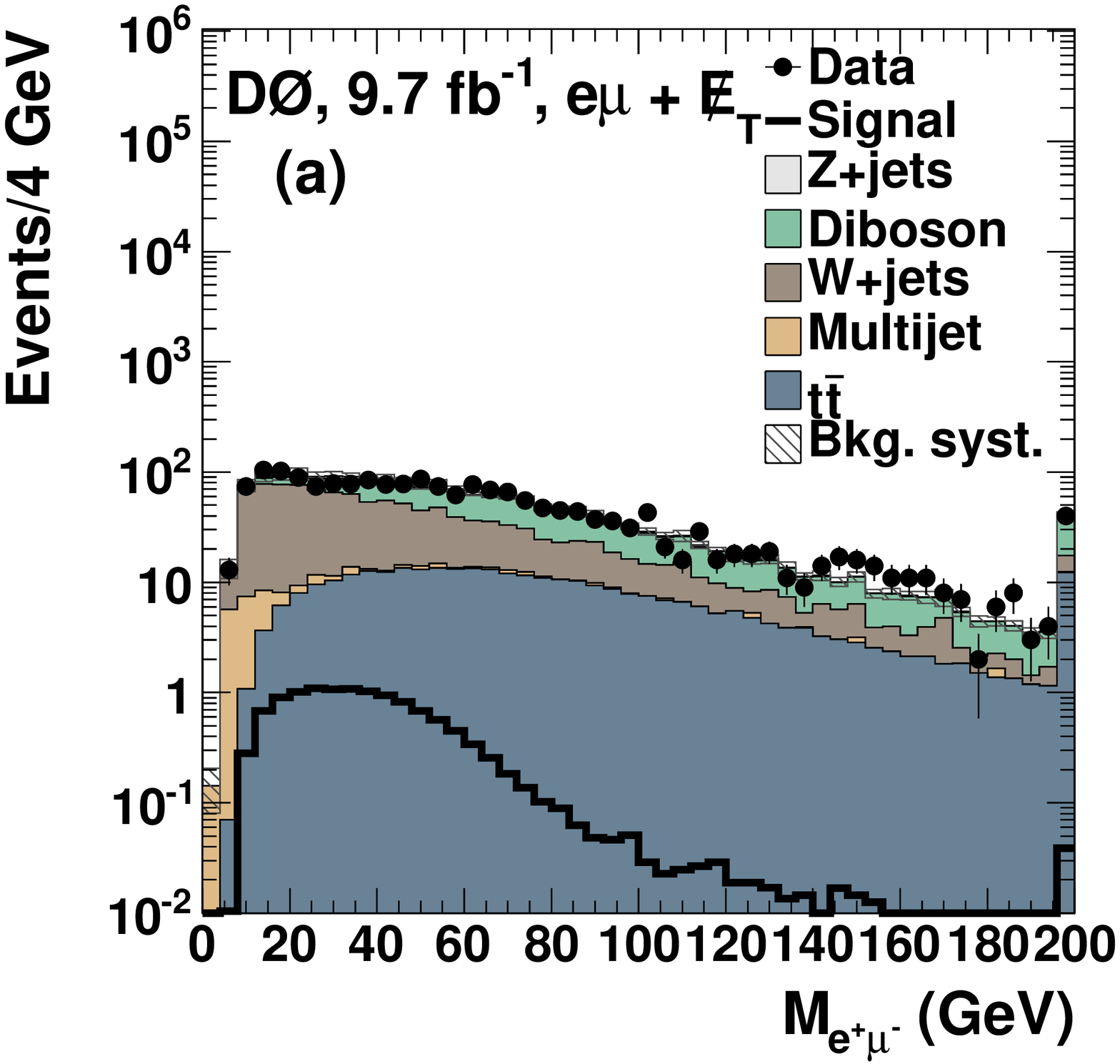} &
\includegraphics[height=6cm, width=0.49\textwidth]{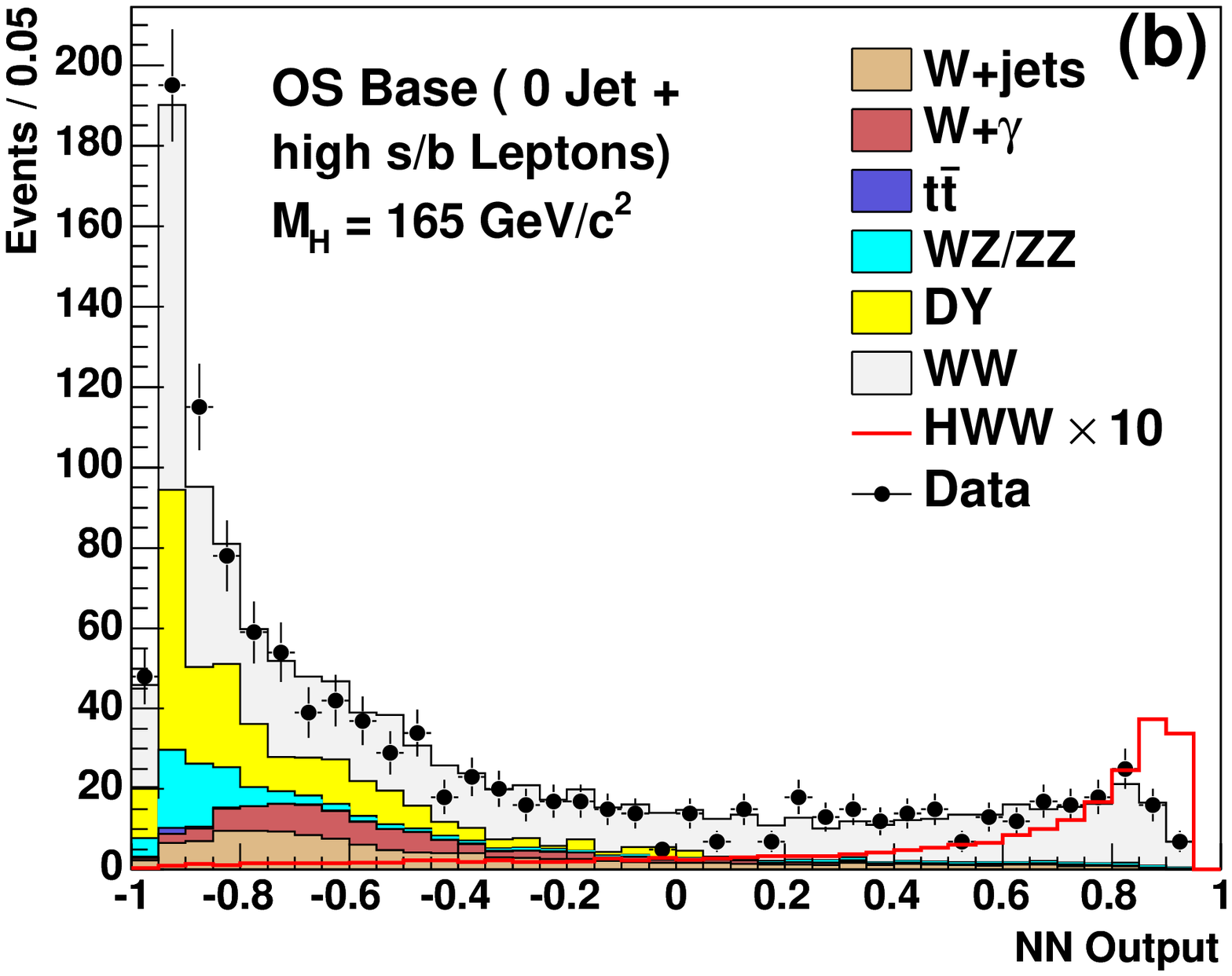} \\
\end{tabular}
\end{center}
\caption{(a) Distribution of the dilepton invariant mass in the $e^{\pm}\mu^{\mp}+\met$ channel from the D\O\ $H\to WW^{(*)}$ search.
The data (points with error bars) are compared to the background prediction, 
broken down into its individual components. Also shown is the expected contribution from a SM Higgs 
boson with  $m_H = 125$~GeV. From Ref.~\citen{Aaltonen:2013iia}. 
(b) Distribution of the neural network output variable in the OS/0-jet/high $s/b$ channel from the CDF $H\to WW^{(*)}$ search.
The data (points with error bars) are compared to the background prediction, 
broken down into its individual components. Also shown is the expected contribution from a SM Higgs 
boson with  $m_H = 125$~GeV scaled by a factor of 10. From Ref.~\citen{Abazov:2013wha}.
\label{fig:hww}}
\end{figure}

In contrast with $H\to b\bar{b}$ searches, in the case of $H\to WW^{(*)}$ 
the presence of two neutrinos in the final state precludes
the reconstruction of the Higgs boson mass, therefore other discriminating
variables against the backgrounds are used.  One of the
most sensitive kinematic variables is the angular separation between
the two charged leptons, either in two dimensions 
($\Delta \phi (\ell^+,\ell^-)$) or three dimensions ($\Delta R(\ell^+,\ell^-)$),
since the spin-zero nature of the Higgs boson causes the leptons to be
produced closer to each other on average than is the case for the background.  For the
same reason, the dilepton invariant mass distribution shows good
discrimination between signal and background (see
Fig.~\ref{fig:hww}(a)).  In order to optimize the search
sensitivity, events are categorized into different analysis channels
with different signal-to-background ratio and background composition, and
optimized MVA discriminants are defined and trained for each of them. Both
CDF and D\O\ categorize events according to the number of jets, with
the 0-jet channel primarily probing the $gg\rightarrow H$ production
mechanism and the 1-jet and 2-jet channels being more sensitive to
$VH$ and VBF production. Categories are also defined based on the
lepton quality (CDF) or lepton flavor (D\O). An example of the MVA
discriminant for the single highest-sensitivity channel in the CDF
analysis, requiring exactly 0 jets and high-purity leptons, is shown
in Fig.~\ref{fig:hww}(b), demonstrating the good separation
between signal and background achieved.  The D\O\ analysis further
categorizes events with exactly 0 or 1 jets according to an MVA
discriminant designed to separate $WW$-like events (including both
$H \to WW^{(*)}$ signal and non-resonant $WW$ background) from non-$WW$ events. The CDF analysis
also considers a separate channel for events with dilepton invariant
mass below 16 GeV. Finally, searches for $H \to WW^{(*)} \to
e\tau_h, \mu\tau_h+\leq 1$ jets have also been
performed~\cite{Aaltonen:2013iia,Abazov:2012ee}.

%%%%%%%%%%%%%%%%%%%%%
%%%%%%%%%%%%%%%%%%%%%

\subsubsection{Same-sign dileptons and trileptons}

The CDF and D\O\ Collaborations have performed a number of searches for $VH$ production in association with  $H\to WW^{(*)}$,
involving two or more leptonic $W$ or $Z$ decays~\cite{Aaltonen:2013iia,Abazov:2013eha}. The resulting signatures include 
SS dileptons-plus-jets, e.g. from $W^\pm H \to W^\pm W^+W^- \to \ell^\pm \nu \ell^\pm \nu jj$, and trileptons, e.g. from 
$W^+ H \to W^+ W^+W^- \to \ell^+ \nu \ell^+ \nu \ell^- \nu$ or $ZH \to Z W^+ W^- \to \ell^+ \ell^- \nu \ell^+ \nu j j$. While SS dilepton
analyses only consider electrons and muons, some trilepton analyses allow for up to one $\tau_h$. These searches are characterized
by small expected signal contributions, but also small backgrounds, dominated by $V$+jets/$\gamma$ with jets or photons misidentified as leptons, and
diboson ($WZ$, $ZZ$) production. Multiple analysis channels are defined, depending on the lepton flavor, the jet multiplicity, and whether
a dilepton pair has mass close to $M_Z$ in the trilepton channels.  MVA
discriminants are constructed for each of them making use of several kinematic
 variables, among which the event $\met$ is found to be particularly 
useful, owing to the presence of multiple neutrinos in the signal as compared to the main background processes.

\subsubsection{Lepton-plus-jets}

The D\O\ Collaboration has performed a search for $H \to WW^{(*)} \to \ell \nu q\bar{q}'$ using the full Run II
dataset~\cite{Abazov:2013mjc}. This search considers events with
exactly one electron or muon, large $\met$ and at least two jets,
requiring either that there are no $b$-tagged jets or at the most
there is exactly one $b$-tagged of the lowest purity that can
originate from a $c$ quark. This ensures a non-overlapping selection
with that used in the $WH \to \ell\nu b\bar{b}$ search. Events are
further categorized according to their lepton flavor ($e$ or $\mu$),
their jet multiplicity (2 jets, 3 jets or $\geq 4$ jets) and the
number of $b$-tagged jets. The selections with 2 or 3 jets are
primarily sensitive to the $gg\rightarrow H$ production mode, while the selection
with $\geq 4$ jets targets the associated production mode,
$VH\to\ell \nu q\bar{q}' q\bar{q}'$. Multivariate discriminants are
trained to separate signal from the overwhelming $W$+jets background.
The best expected sensitivity for these searches is achieved at
$m_H=165$ GeV, reaching 4.0 and 7.3 times the SM Higgs boson cross
section for the 2+3-jet channels and $\geq 4$ jets channel, respectively. The
corresponding observed limits are 2.8 and 8.5 times the SM prediction,
respectively.

\subsubsection{$H\to WW^{(*)}$ search results and validation}

The combination of CDF and D\O\ searches for $H \to WW^{(*)}$ using
only 4.8--5.4 fb$^{-1}$ of data~\cite{Aaltonen:2010yv} reached 95\% C.L. exclusion of 
a SM Higgs boson with mass in the range of 162--166 GeV, the
first exclusion above the LEP limit. After
analyzing the full Run II dataset and substantially improving the
analyses, each experiment has been able to exclude a substantial mass
range: 149--172 GeV and 157--172 GeV in case of the CDF and D\O\ combinations, 
respectively.  The expected sensitivities reached at a
mass of 125 GeV are 3.1 and 3.0 times the SM Higgs boson cross section
for the CDF and D\O\ searches, respectively. More details on the
expected and observed sensitivities, as well as the combination of
searches, are provided in Sec.~\ref{sec:sec4_results}.

Once again, the measurement of diboson cross sections using the same
experimental techniques as for the $H\to WW^{(*)}$ searches, provides
an important validation of the search methodology. Both collaborations
have performed measurements of the $W^+W^-$ cross section in the
$\ell^+\nu\ell^-{\bar\nu}$ final
state~\cite{Abazov:2013wha,Aaltonen:2009aa}, the $ZZ$ cross section in
the $\ell^+\ell^-\nu{\bar\nu}$ final state~\cite{CDF:2011ab,Abazov:2012cj},
and the $WZ$ cross section in the $\ell^+\nu\ell^+\ell^-$ final
state~\cite{Aaltonen:2012vu,Abazov:2012cj}, finding good agreement
with NLO predictions.

\subsection{Searches for $H\to ZZ^{(*)}$}
\label{sec:sec4_hzz}

The CDF and D\O\ Collaborations have performed searches for $H\to
ZZ^{(*)}\to \ell^+\ell^-\ell'^+\ell'^-$ ($\ell, \ell'=e,\mu$) using
the full Run II dataset~\cite{Collaboration:2012pa,Abazov:2013pci}.
While this channel constitutes a discovery mode at the LHC, at the
Tevatron the exceedingly small branching ratio for $H\to ZZ^{(*)}\to 4\ell$, 
coupled with the limited integrated luminosity available,
results in a small expected sensitivity. These searches are characterized
by very small expected signal, but also small backgrounds from
non-resonant production of $(Z/\gamma^*)(Z/\gamma^*)$. In order to
maximize the signal acceptance, these searches select leptons with
$\pt$ down to 10 GeV and relaxed lepton identification criteria. The
four-lepton invariant mass distribution constitutes the most
discriminating variable to separate $gg\to H \to ZZ^{(*)}$ from the
background. In addition, the event $\met$ is employed to increase the
sensitivity to signal contributions from $ZH\to ZW^+W^+\to \ell^+\ell^-\ell^+\nu\ell^-{\bar\nu}$ 
and $ZH\to \ell^+\ell^-\tau^+\tau^-$, particularly at low
$m_H$. Figure~\ref{fig:hzz_hgamgam}(b) shows the four-lepton
invariant mass distribution used by the CDF analysis. No excess
compatible with a Higgs boson signal is found and the
$(Z/\gamma^*)(Z/\gamma^*)$ cross section is measured finding good
agreement with the SM
prediction~\cite{Aaltonen:2014yfa,Abazov:2013pci}.  The best expected
sensitivity is achieved for $m_H$ near 150~GeV and~190 GeV, reaching
approximately 10 times the SM prediction.  The observed (expected)
cross section limits at $m_H=125$~GeV for the CDF and D\O\ analyses
are 29.3 (26.5) and 42.8 (42.3) times the SM prediction, respectively.

\subsection{Searches for $H\to \gamma\gamma$}
\label{sec:sec4_hgamgam}

\begin{figure}[t]
\begin{center}
\begin{tabular}{cc}
\includegraphics[height=5.7cm, width=0.49\textwidth]{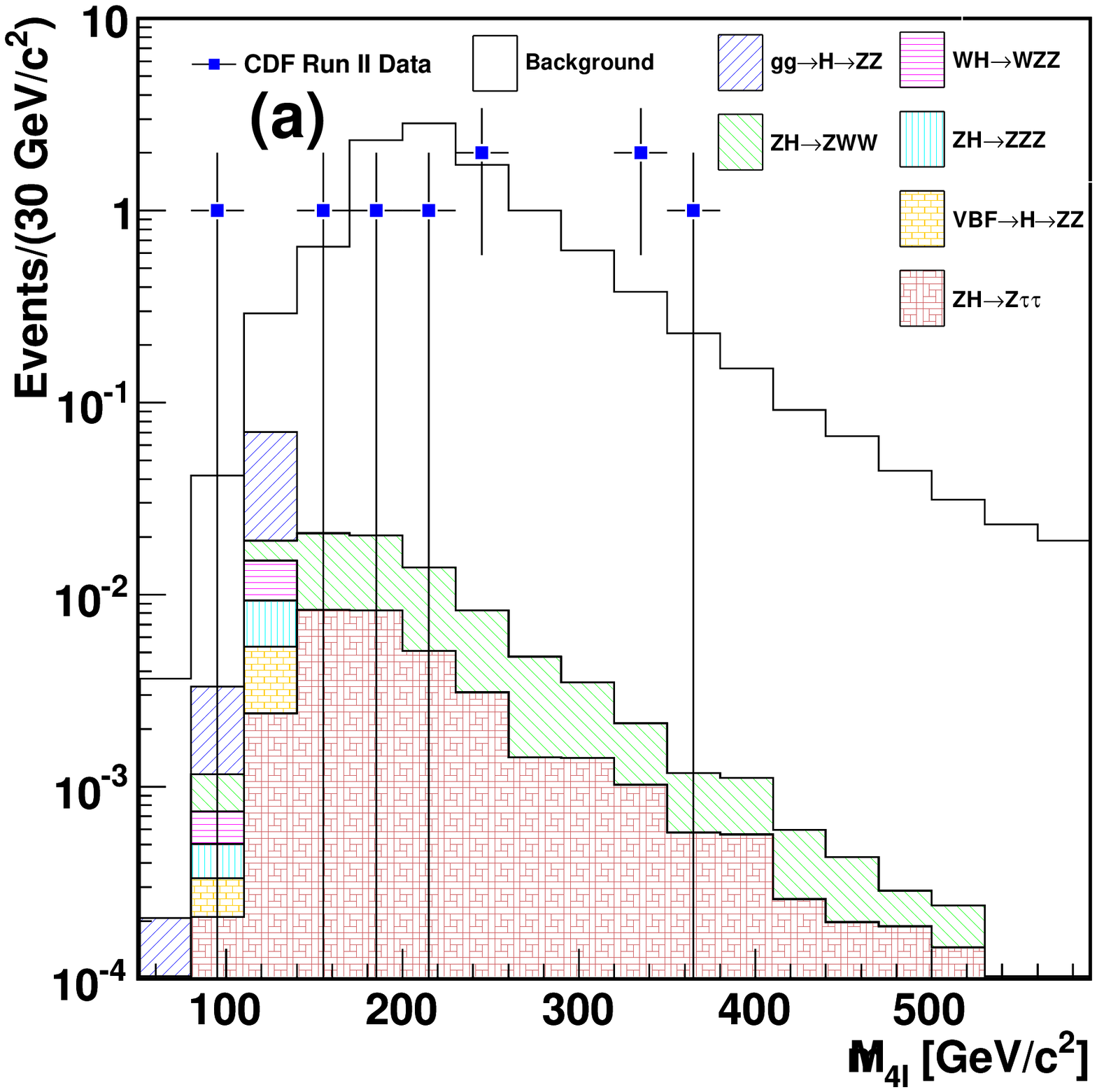} &
\includegraphics[height=5.5cm,width=0.49\textwidth]{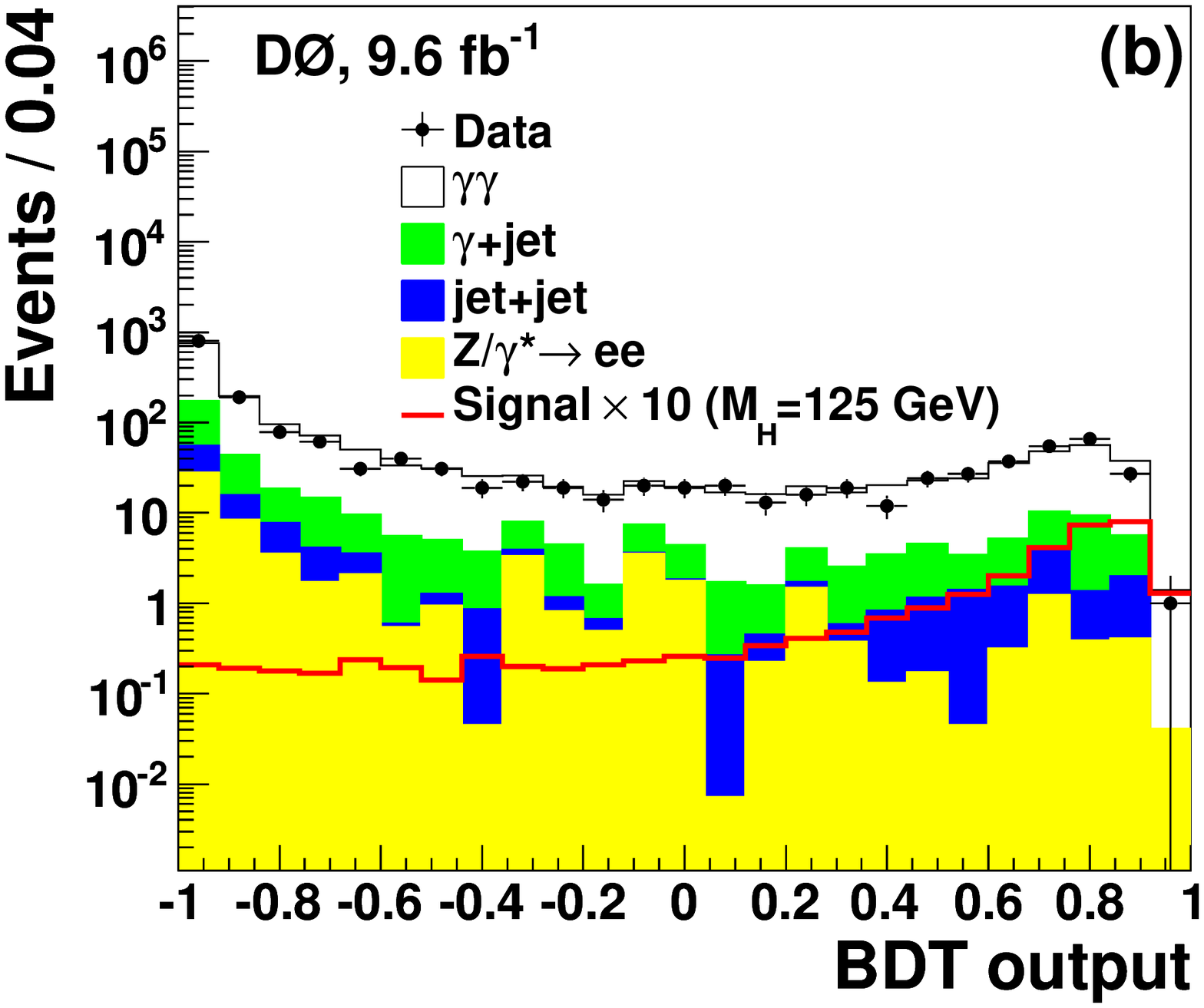} \\
\end{tabular}
\end{center}
\caption{(a)  Distribution of the four-lepton invariant mass used by the CDF $H\to ZZ^{(*)}\to 4\ell$ search.
The data (points with error bars) are compared to the total background prediction, shown with an open histogram.
Also shown are the expected contributions from a SM Higgs 
boson with  $m_H = 125$~GeV, broken down into several components, shown hatched and stacked. Adapted from Ref.~\citen{Aaltonen:2012ya}. 
(b) Distribution of the BDT distribution used by the D\O\ $H\to\gamma\gamma$ search in the photon-enriched sample.  
The data (points with error bars) are compared to the background prediction, 
broken down into its individual components. Also shown is the expected contribution from a SM Higgs 
boson with  $m_H = 125$~GeV scaled by a factor of 10.
From Ref.~\citen{Abazov:2013pci}.
\label{fig:hzz_hgamgam}}
\end{figure}

The CDF and D\O\ Collaborations have performed searches for
$H\to \gamma\gamma$ using the full Run II
dataset~\cite{Collaboration:2012pa,Abazov:2013pci}. The small
$H\to \gamma\gamma$ branching fraction in the SM makes these searches at the
Tevatron not promising in terms of sensitivity to the SM Higgs boson,
although the large enhancements possible to ${\cal B}(H\to\gamma\gamma)$ in beyond-SM scenarios open a window of
opportunity that makes them well justified.  These searches consider
the three main Higgs boson production modes, exploiting their
kinematic differences with respect to the main backgrounds, consisting
of non-resonant $\gamma\gamma$ production, followed by $\gamma$+jets
and QCD dijets with one or two jets misidentified as photons. The CDF
search selects photon candidates in both the central and forward
calorimeters, while the D\O\ analysis is restricted to photons in the
central calorimeter.  In both searches, events are classified in
categories with different signal-to-background ratio in order to optimize
the search sensitivity: the CDF analysis defines up to six different
categories (depending on pseudorapidity of the photons and whether or
not a photon candidate is identified as originating from a $\gamma \to e^+e^-$ conversion), while the
D\O\ analysis defines two categories (photon-enriched and
jet-enriched, depending on the output from an MVA used for photon
identification).  Both searches construct MVA discriminants exploiting
the diphoton mass as well as other kinematic variables to separate
the signal from the background. The CDF search applies MVA discriminants
only to the highest-sensitivity channels with two central photons,
using the diphoton mass in the rest of channels, whereas the D\O\ uses
the MVA discriminants in all analysis channels.
Figure~\ref{fig:hzz_hgamgam}(b) shows the MVA distribution used by
the D\O\ analysis in the photon-enriched region.  These searches have
a relatively constant sensitivity as a function of $m_H$ in the range
$100<m_H<140$ GeV.  The observed (expected) cross section limits at
$m_H=125$~GeV for the CDF and D\O\ analyses are 17.0 (9.9) and 
12.8 (8.7) times the SM prediction, respectively.

%%%%%%%%%%%%%%%%%%%%%
%%%%%%%%%%%%%%%%%%%%%

\subsection{Standard Model Higgs boson search results}
\label{sec:sec4_results}

Both the CDF and D\O\ Collaborations performed searches for the SM Higgs boson in the channels described in Secs.~\ref{sec:sec4_hbb}--\ref{sec:sec4_hgamgam}.
The interaction between the collaborations' analysis teams was minimal when the analyses were developed and
optimized.  As more data were collected, the results from each channel were combined together to form 
collaboration-wide results using the statistical
methods described in Sec.~\ref{sec:statistics}, and the communication between the experiments' analysis teams
increased in order to propagate the techniques that were found to be the most sensitive.  Each collaboration
prepared individual channel results as well as combined results, and the same techniques, described in 
Sec.~\ref{sec:statistics}, were used to combine CDF and D\O's results together to produce single results 
with the maximum sensitivity.

Producing combined results from CDF and D\O's searches required significant coordination between the
two collaborations, and thus the Tevatron New Physics and Higgs Working Group (TEVNPHWG) was formed.
The combinations needed to preserve all of the statistical power and systematic rigor
of the contributing analyses, and thus detailed exchange  of data distributions and predictions from each
signal and background process was performed.  All systematic uncertainties affecting
the rates and shapes of the predicted distributions were also exchanged and itemized by source.
The list of correlated systematic uncertainties was determined by the two collaborations, accounting
for cases where the same predictions and uncertainties were shared by both.  Recommendations from the
TEVNPHWG for central values and uncertainties for shared sources of systematic uncertainty were
propagated to the collaborations'
analysis teams to unify the treatment and to make the joint fits of data between the two
experiments consistent.  Frequently, combined results were required to be produced for the same conference
as the individual contributing results, and so the exchange formats, combination techniques, and
systematic uncertainty categories were formalized well in advance.  Combinations were
always performed twice, once by the CDF group members using the Bayesian method,
and once by the D\O\ group members using the Modified Frequentist method, and results were
approved only when consistency was achieved.

\subsubsection{Limits}

\begin{figure}[t]
\begin{centering}
\includegraphics[width=0.7\columnwidth]{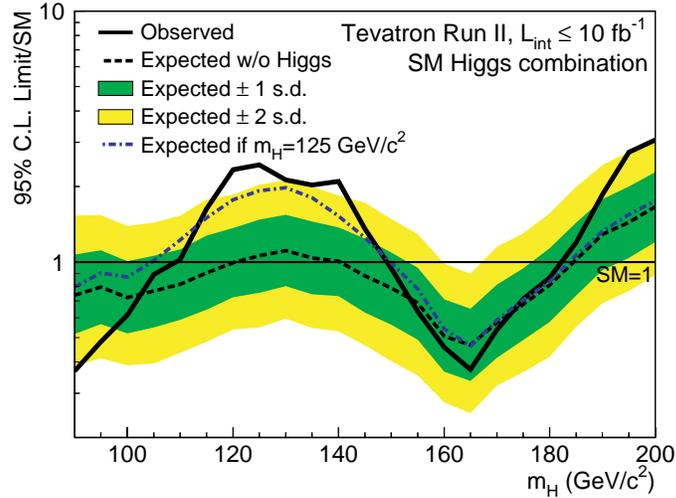}
\caption{
\label{fig:smlimits}
Observed and median expected (for the background-only 
hypothesis) 95\% C.L. Bayesian upper production limits expressed as 
multiples of the SM cross section as a function of Higgs boson mass 
for the combined CDF and D\O\ searches in all decay modes.
The dark- and light-shaded bands indicate, respectively, the one and 
two standard deviations (s.d.)~probability regions in which the limits are expected to fluctuate in the 
absence of signal.  The blue short-dashed line shows median expected 
limits assuming the SM Higgs boson is present at $m_H=125$~GeV.
From Ref.~\citen{Aaltonen:2013kxa}.}
\end{centering}
\end{figure}

During the first years of Run~II, the Tevatron experiments were not yet sensitive to the SM Higgs boson
at its predicted rate but could set limits on the signal strength modifier $\mu$.  Even though physics models do not
scale the five mechanisms $gg\rightarrow H$, $WH$, $ZH$, VBF, and $t{\bar{t}}H$ together, the expected
limit on a common scale factor defined the sensitivity of the searches.  If the observed limit on $\mu$ falls
below unity for a particular $m_H$, that value of the Higgs boson mass is excluded at the 95\% C.L.  Figure~\ref{fig:smlimits}
shows the observed and expected upper limits on $\mu$ (labeled ``95\% CL Limit/SM'') as a function of $m_H$ for the
full Run~II data sample.  Values of $m_H$ between 90~GeV and 109~GeV, and also between
149~GeV and 182~GeV, are excluded at the 95\% C.L.  The expected exclusion regions are
between 90~GeV and 120~GeV
and also between 140~GeV and 184~GeV, assuming no Higgs boson is present.
Excesses are seen in the
low-mass searches between $m_H$ values of 115~GeV and 135~GeV, as well as in the high-mass searches
(dominated by the $H\rightarrow WW^{(*)}$ searches) around $m_H=200$~GeV.   The excess at around $m_H=200$~GeV is in a region
where the sensitivity is not as strong as at lower masses, and where the mass resolution is quite poor.
Shown along with the expected limits
assuming no Higgs boson is present are the expected limits as a function of the test mass assuming a Higgs boson is
present at $m_H=125$~GeV.

\subsubsection{Significance}

\begin{figure}[t]
\begin{centering}
\includegraphics[width=0.47\columnwidth]{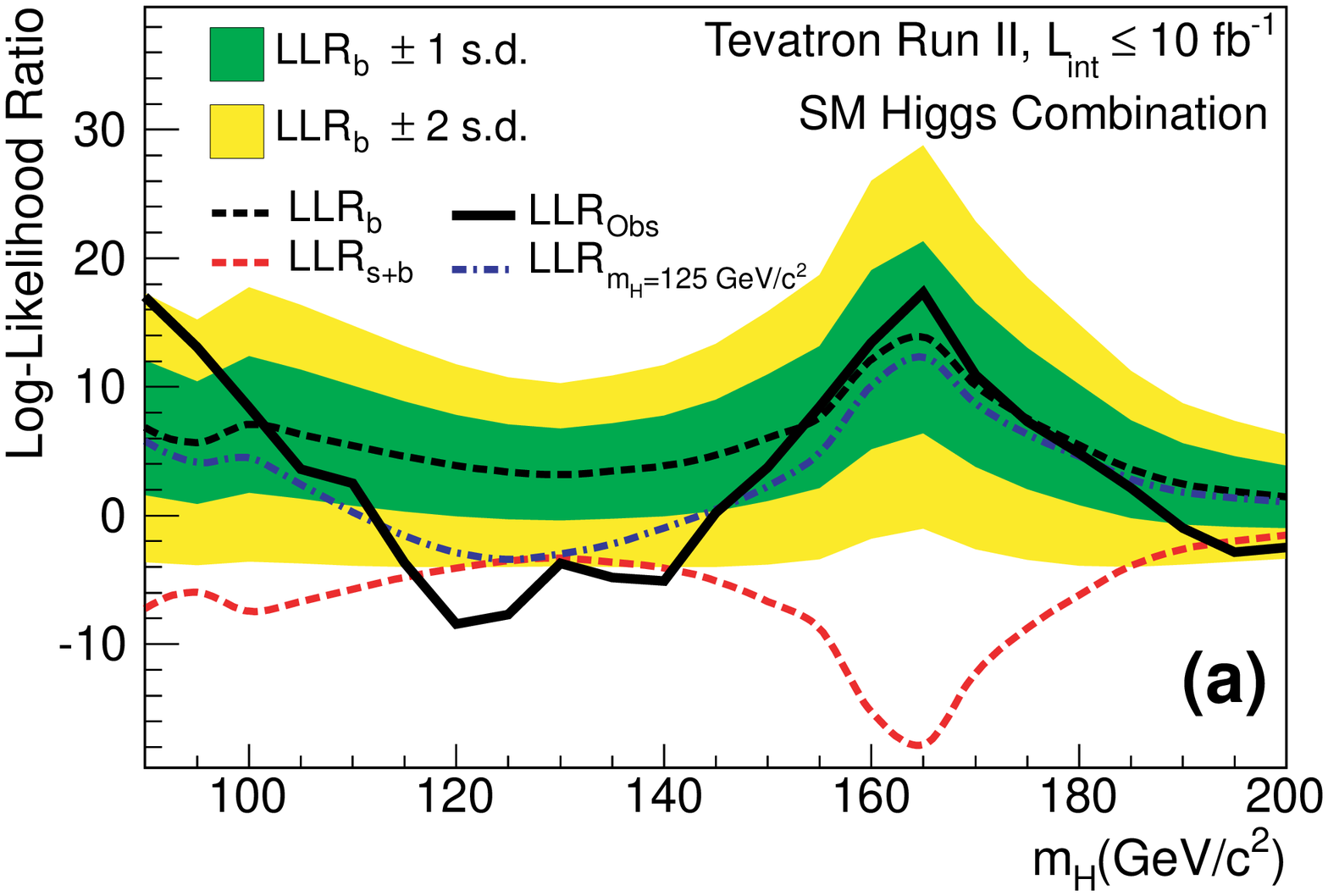}
\includegraphics[width=0.47\columnwidth]{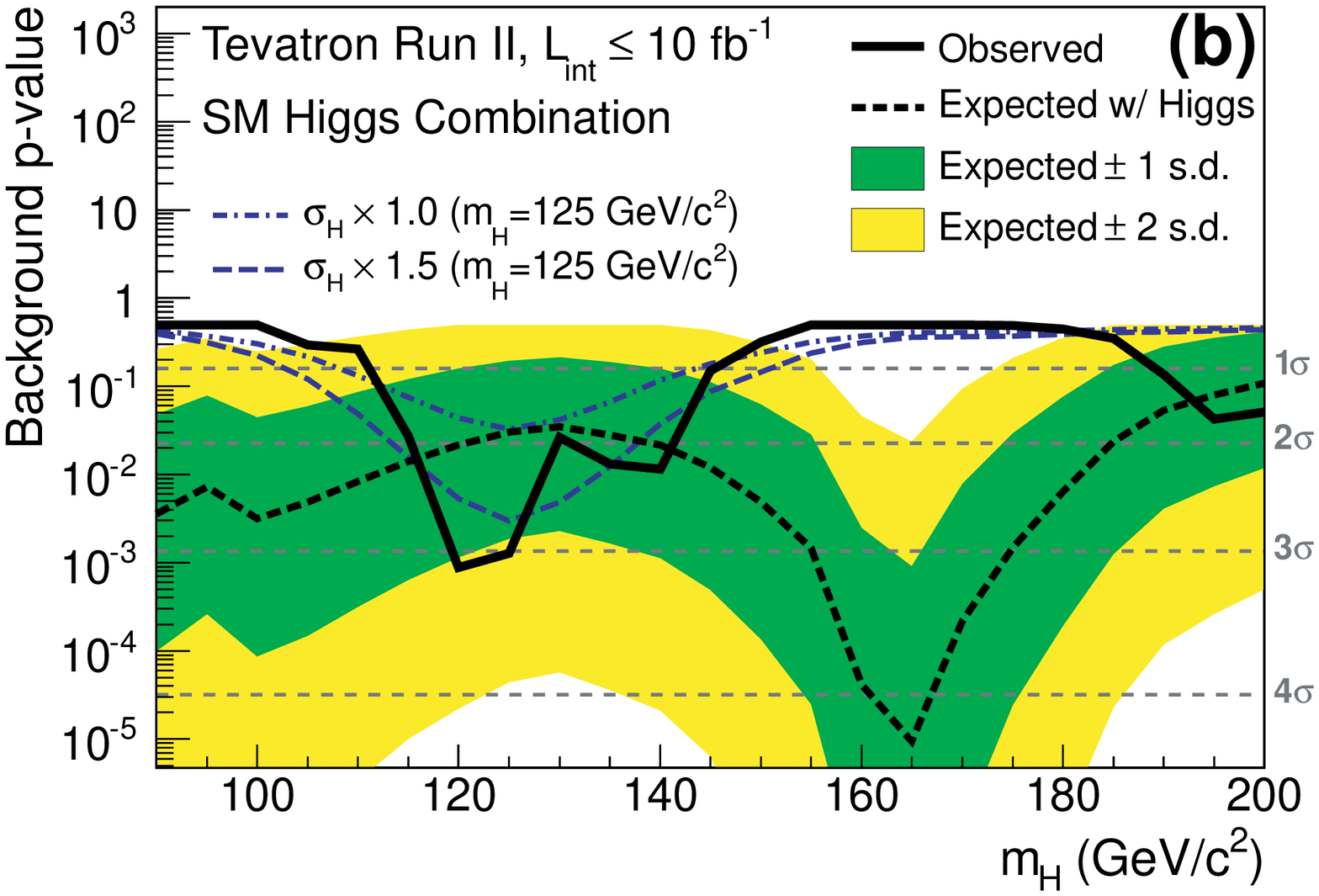} 
\caption{
\label{fig:combollr}
(a) Log-likelihood ratio (LLR) 
for the Tevatron's Higgs boson searches in all decay modes combined.  The solid line shows the 
observed LLR as a function of $m_H$. The median background-only expectation and its expected variation is
shown with a dark long-dashed line
and shaded bands.  The median expectation assuming the SM Higgs boson is present at $m_H=125$~GeV is shown
with a blue dashed line and the median expectation assuming a Higgs boson is present at each value of $m_H$ in turn is shown
with a red dashed line.  From Ref.~\citen{Aaltonen:2013kxa}.
(b) The background $p$-value as a function of $m_H$ is shown with a solid line.
The dotted black line shows the median expected values assuming a SM signal is present, 
evaluated separately at each $m_H$, and the shaded bands indicate the expected variations.
 The blue lines show the median expected 
$p$-values assuming the SM Higgs boson is present with $m_H$=125~GeV 
at signal strengths of 1.0 times (short-dashed) and 1.5 times (long-dashed) 
the SM prediction.  From Ref.~\citen{Aaltonen:2013kxa}.
}
\end{centering}
\end{figure}

To quantify the significance of excess data candidates compared with the background, the background-only
$p$-value 1-CL$_{\rm{b}}$ using LLR as the test statistic is computed.  The observed and expected values of LLR
are shown as functions of $m_H$ in Fig.~\ref{fig:combollr}(a).  The expected values are shown for the null hypothesis
(SM backgrounds but without a Higgs boson contribution) and the test hypothesis (the SM Higgs boson is present at
the $m_H$ being tested), and the  68\% and 95\% intervals around the null hypothesis's predictions are shown.  The
expected values assuming the SM Higgs boson is present at $m_H=125$~GeV are likewise shown.

The signal significance as a function of the tested $m_H$ is shown in Fig.~\ref{fig:combollr}(b).  It shows the probability
of obtaining an LLR value at least as signal-like as the observed value, as a function of $m_H$, as well as the median expected value
of this probability and its expected distribution owing to expected random outcomes
if no Higgs boson is present, if a Higgs boson is present at each value of $m_H$ tested, and if a Higgs boson is present
at $m_H=125$~GeV.  A local significance of 3.0~standard deviations (s.d.) is observed for $m_H=125$~GeV, and 1.9~s.d. 
are expected assuming the SM Higgs boson is present with the SM predicted rate~\cite{Aaltonen:2013kxa}.

\subsubsection{Cross section fits}

\begin{figure}[t]
\begin{centering}
\includegraphics[width=0.53\columnwidth]{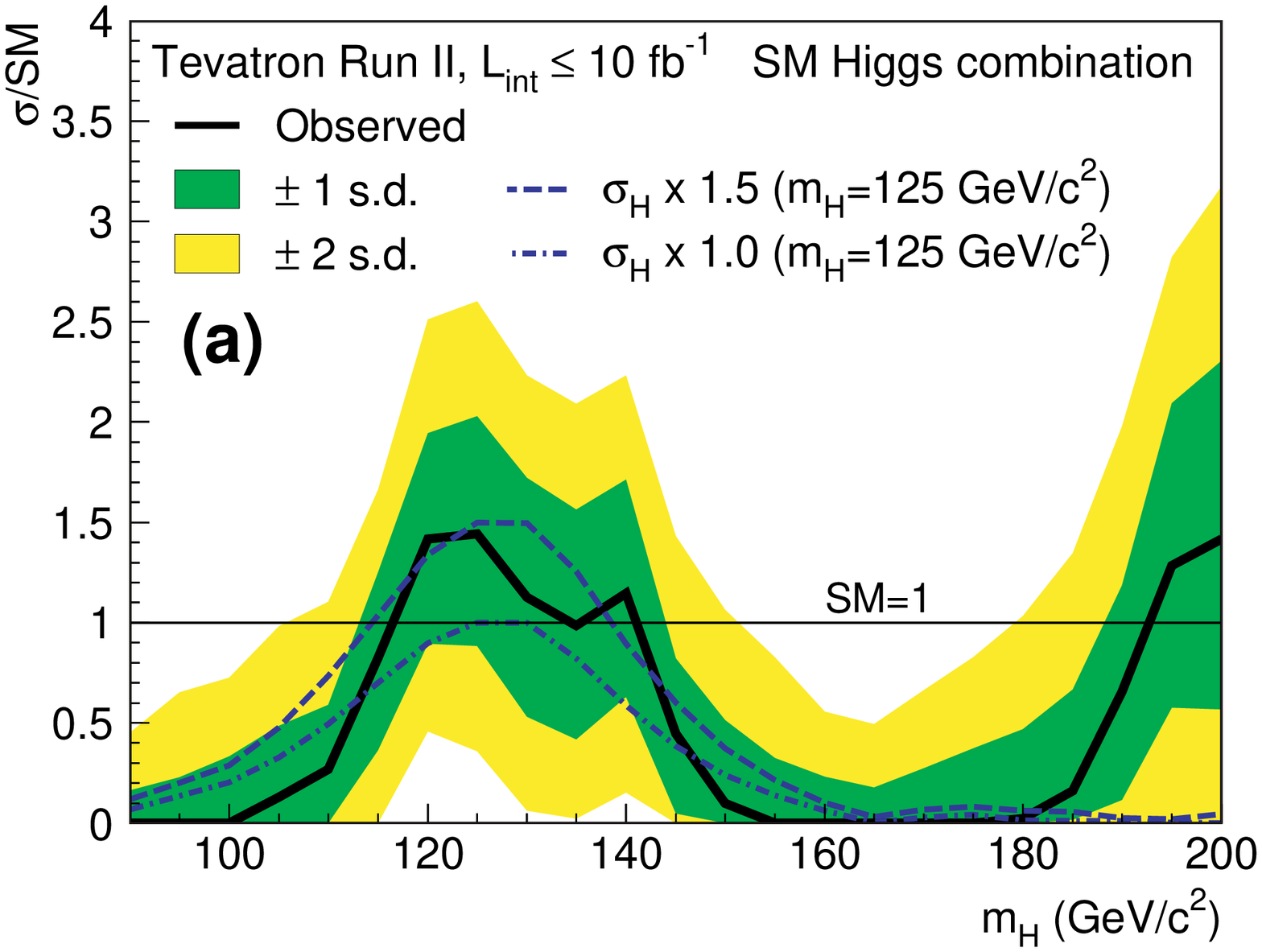}
\includegraphics[width=0.45\columnwidth]{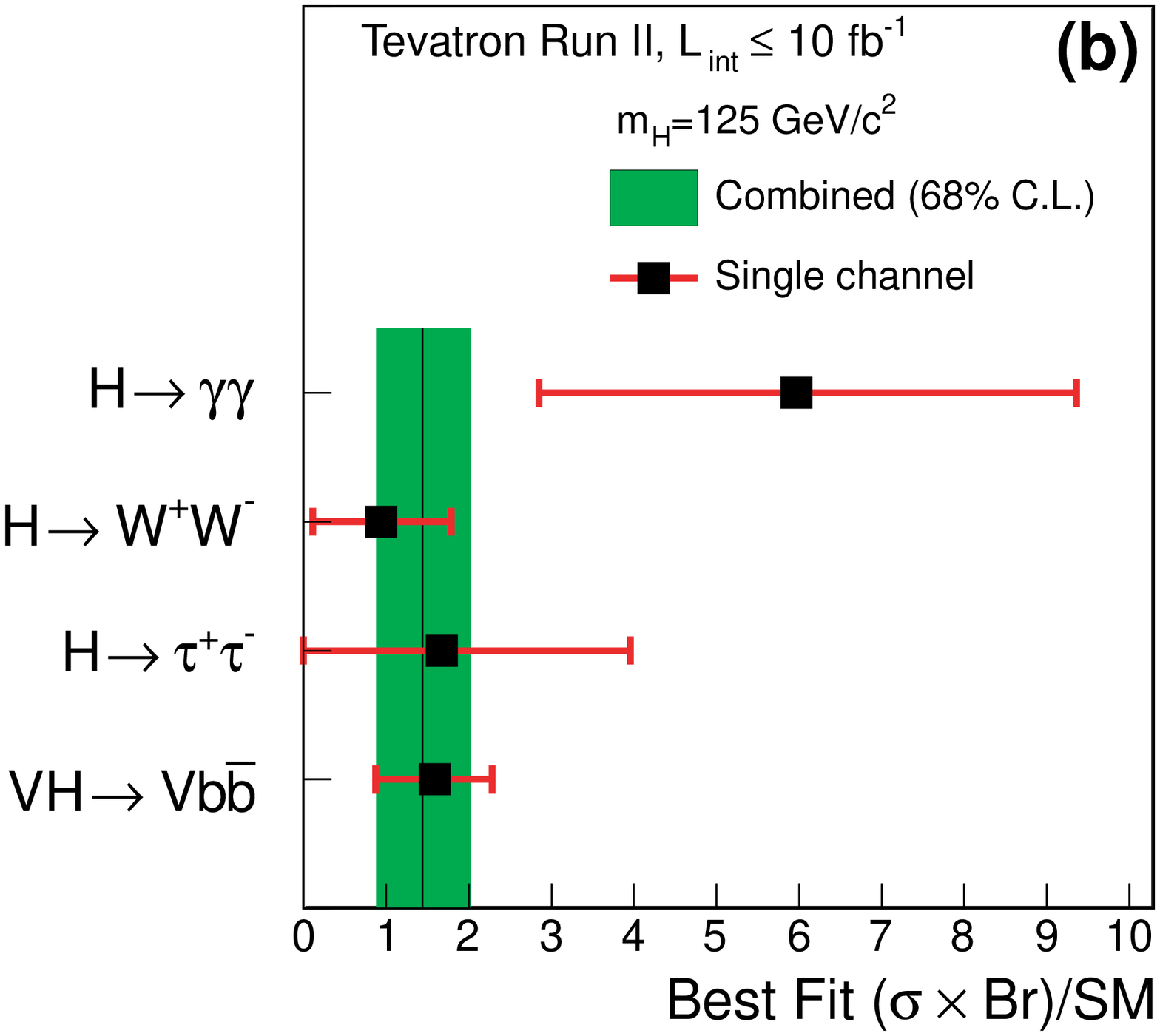}
\caption{
\label{fig:combBestFit}
(a) Best-fit signal cross section expressed as a ratio to 
the SM cross section as a function of Higgs boson mass for all of CDF and D\O's SM Higgs 
boson searches in all decay modes combined, assuming SM branching fractions.
The dark- and light-shaded bands show the one 
and two s.d.~uncertainty ranges on the fitted signal, respectively.  Also 
shown with blue lines are the median fitted cross sections expected for a 
SM Higgs boson with $m_H=125$~GeV at signal strengths of 1.0 times 
(short-dashed) and 1.5 times (long-dashed) the SM prediction. From Ref.~\citen{Aaltonen:2013kxa}.
(b) 
Best-fit values of $\mu=(\sigma\times{\cal B})$/SM using the Bayesian method 
for the combinations of CDF and D\O's Higgs boson search channels focusing on 
the $H\rightarrow b{\bar{b}}$,  $H\rightarrow\tau^+\tau^-$, $H\rightarrow WW^{(*)}$ and 
$H\rightarrow\gamma\gamma$ decay modes for a Higgs boson mass 
of 125~GeV.  The shaded band corresponds to the one s.d.~uncertainty
on the best-fit value of $\mu$ for all SM Higgs boson decay modes combined. From Ref.~\citen{Aaltonen:2013kxa}.
} 
\end{centering}
\end{figure}

The best-fit value of the
signal strength modifier $\mu$ is shown as a function of $m_H$ in Fig.~\ref{fig:combBestFit}(a), along
with the expectation assuming a Higgs boson is present at $m_H=125$~GeV, using the Bayesian method described
in Sec.~\ref{sec:statistics}.  The 68\% and 95\% intervals shown along with the cross section fit are the observed
credibility intervals and not the expected confidence intervals as shown in the LLR, limit, and $p$-value plots.

The properties of the excess of candidates seen by the Tevatron experiments are investigated first by measuring
the production cross section times the branching ratio in the several decay modes.  The searches performed are
typically sensitive to one decay mode each, although some searches have contributions from two or more, such as the
$H\rightarrow WW^{(*)}\rightarrow\ell^+\nu\ell^-{\bar{\nu}}$ searches, which have some acceptance for 
$H\rightarrow\tau^+\tau^-$ events in which the tau leptons decay leptonically.  All relevant channels are included
in the combination by decay mode in Fig.~\ref{fig:combBestFit}(b).  

\subsubsection{Coupling constraints}

\begin{figure}[t]
\begin{centering}
\includegraphics[width=0.47\columnwidth]{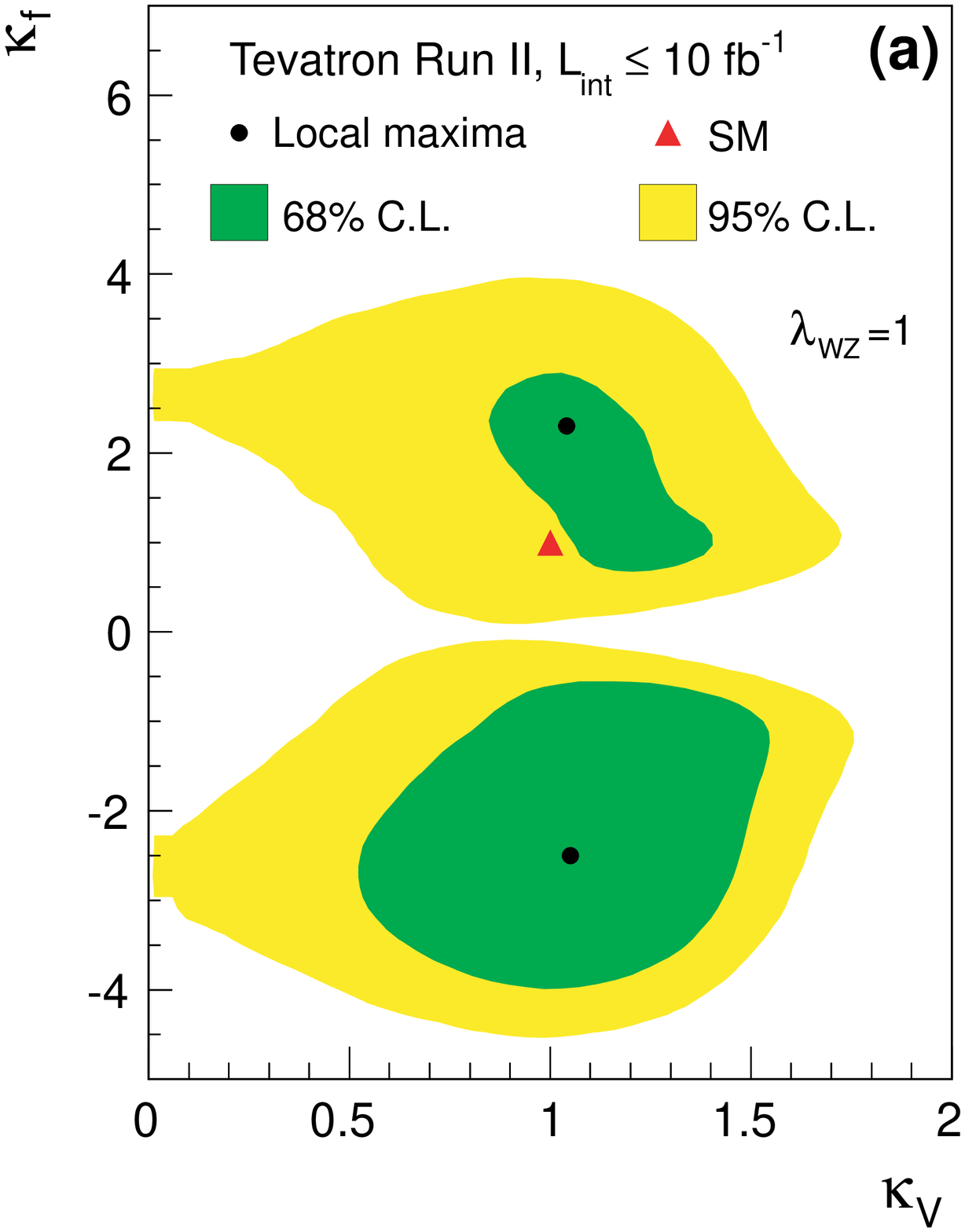}
\includegraphics[width=0.47\columnwidth]{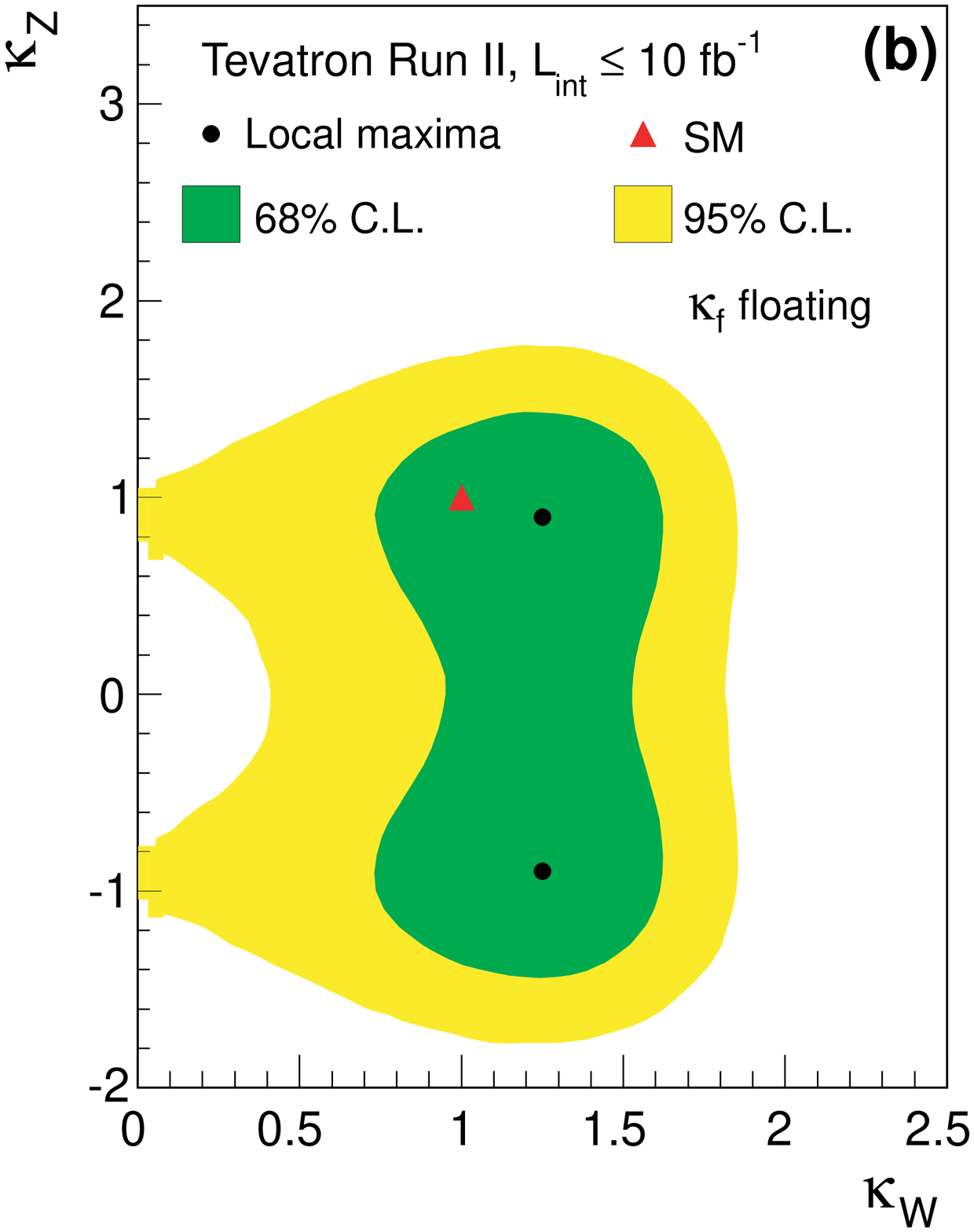}
\caption{
\label{fig:couplingplots}  Two-dimensional constraints in the $(\kappa_V,\kappa_f)$ 
plane (a), and the $(\kappa_W,\kappa_Z)$ plane (b), for the combined 
Tevatron searches for a SM-like Higgs boson with $m_H=125$~GeV.  The points that maximize the local posterior 
probability densities are marked with dots, and the 68\% and 95\% C.L. intervals are indicated 
with the dark- and light-shaded regions, respectively.  The SM prediction in each plane 
is marked with a triangle. From Ref.~\citen{Aaltonen:2013kxa}.
}
\end{centering}
\end{figure}

A further step in interpreting the excess in the Higgs boson searches at the Tevatron
is to test models in which the couplings of the $W$, $Z$, or fermions is modified relative to 
their SM predictions.  The prescription of Ref.~\citen{LHCHiggsCrossSectionWorkingGroup:2012nn} is followed, where
the Higgs boson's couplings to fermions are modified by a multiplicative factor 
$\kappa_f$, to the vector bosons $W$ and $Z$ by $\kappa_V$ when
tested together assuming custodial symmetry, and by $\kappa_W$ and $\kappa_Z$ when tested separately.
For each value of the $\kappa$ coupling modifiers, a new set of Higgs boson production cross sections and branching ratios
is computed starting with the SM predictions and modifying each component diagram by the relevant combination
of coupling modifiers.  For example, the $H\rightarrow\gamma\gamma$ width contains a contribution scaled by $\kappa_f$
due to the top-quark and $b$-quark loops, and a contribution with the opposite sign in the amplitude coming from the $W$-boson loop.
A uniform prior is assumed in the two-dimensional planes $(\kappa_V,\kappa_f)$ and $(\kappa_W,\kappa_Z)$, in which
the results are shown in Fig.~\ref{fig:couplingplots}.  No significant deviations from the SM predictions are seen.

\subsubsection{Tests of spin and parity}

Recent progress has been made at the Tevatron in testing the spin and parity of the Higgs boson using the model predictions
of Ref.~\citen{Ellis:2012xd}.  The threshold behavior of the associated production of a pseudoscalar ($J^P=0^-$) and a graviton-like
($J^P=2^+$) exotic higgs boson with a vector boson $V$ ($W$ or $Z$) differ markedly from those of the SM Higgs boson ($J^P=0^+$).
The SM Higgs associated production is an s-wave process and its cross section rises proportional to $\beta$ close to threshold,
where $\beta=2p/\sqrt{s}$, with $p$ being the magnitude of the three-momentum of the Higgs boson (or the vector boson) in the $VH$ rest frame,
and $\sqrt{s}$ being the total energy of the $VH$ system~\cite{Miller:2001bi}.  Associated production of a $0^-$ boson is a p-wave process
with a cross section that scales as $\beta^3$, and associated production of a graviton-like $2^+$ boson is a d-wave process
with a cross section that scales as $\beta^5$.  The distribution of the
invariant mass of the $VX$ system, where $X$ is either the SM Higgs boson
or one of the exotic Higgs-like particles proposed, is therefore quite different, with a much larger average value for the
$2^+$ particle than for the $0^-$ particle, with the smallest average value for the SM Higgs boson production.  The processes
$VX\rightarrow Vb{\bar{b}}$ are used to test for the presence of these exotic bosons, and the observable $m_{Vb{\bar{b}}}$ is a strong
discriminant among the possible signals and also the background processes.  Since there is no prediction for the cross section
of $VX$ production for the $J^P=0^-$ and $J^P=2^+$ particles, nor for the decay branching fractions,
the CDF and D\O\ Collaborations treated this search as a test for an exotic
new particle which may either replace the SM Higgs boson or be present along with it. 

Strong limits are obtained on the production cross sections times the decay branching ratios
$\sigma(VX)\times {\cal B}(X\rightarrow b{\bar{b}})$ for the $0^-$ and $2^+$ models.  D\O\ presents the limits in terms of the
fraction of the total Higgs boson production rate that could be from the exotic signal, while CDF sets limits on the
rate of exotic production.  Both collaborations compute $p$-values for excluding the exotic signals, assuming that the
production rates and decay branching fractions are the same for the exotic $X$ bosons as for the SM Higgs boson, obtaining
exclusions well in excess of the 95\% C.L. assuming this production rate~\cite{Abazov:2014doa,Aaltonen:2015tsa}.

The CDF and D\O\ Collaborations have combined the results of these analyses to produce the strongest constraints on these
models of exotic bosons~\cite{Aaltonen:2015mka}.  The observed (expected)
upper limit on $2^+$ boson production is 0.36 (0.33) times the rate predicted for SM Higgs 
boson production, and the upper limit on $0^-$ boson production is also 0.36 (0.32) times the SM Higgs boson rate,
both assuming that the SM Higgs boson is absent and is replaced with an exotic particle.

\begin{figure}[t]
\begin{centering}
\includegraphics[width=0.47\columnwidth]{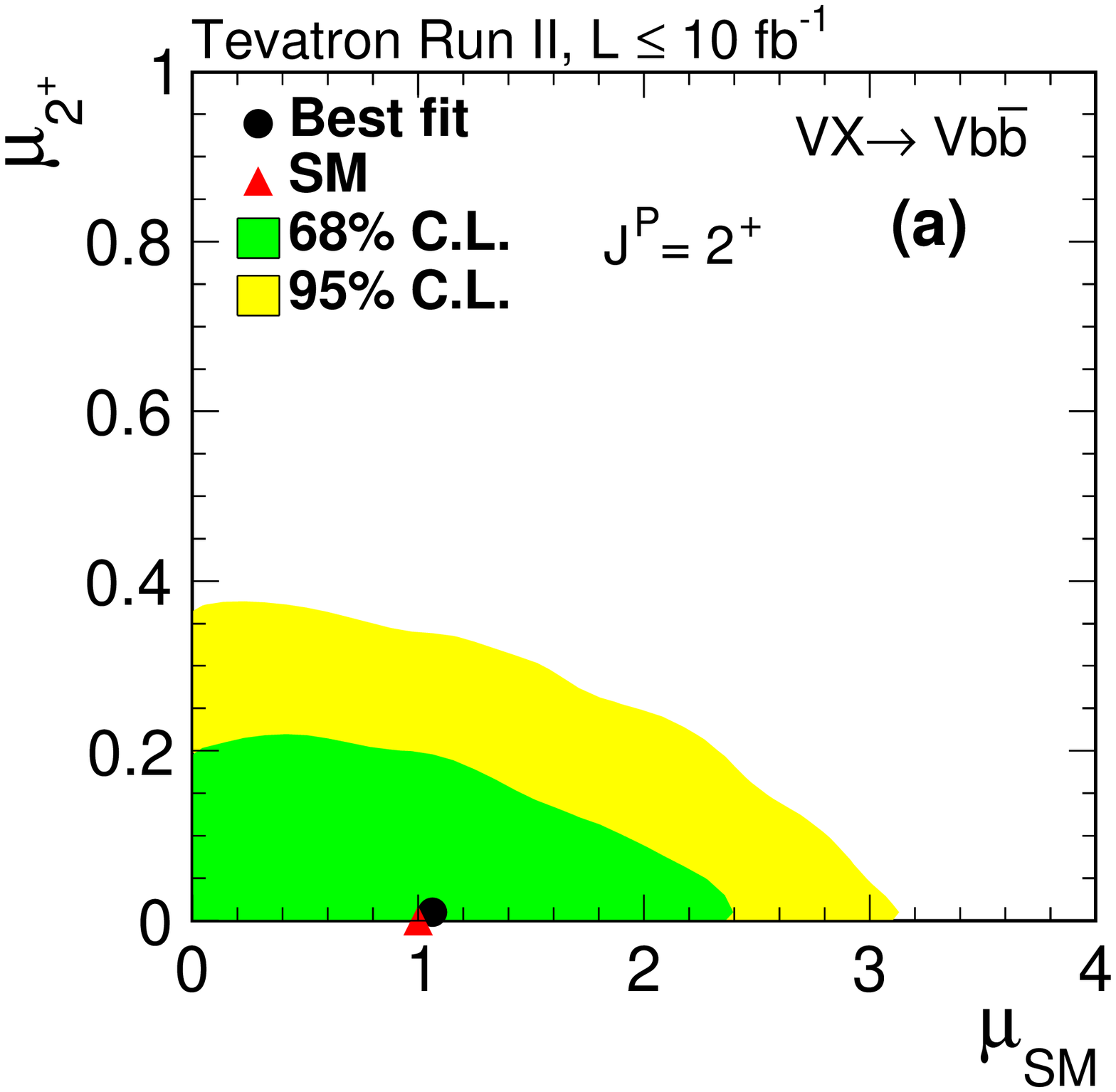}
\includegraphics[width=0.47\columnwidth]{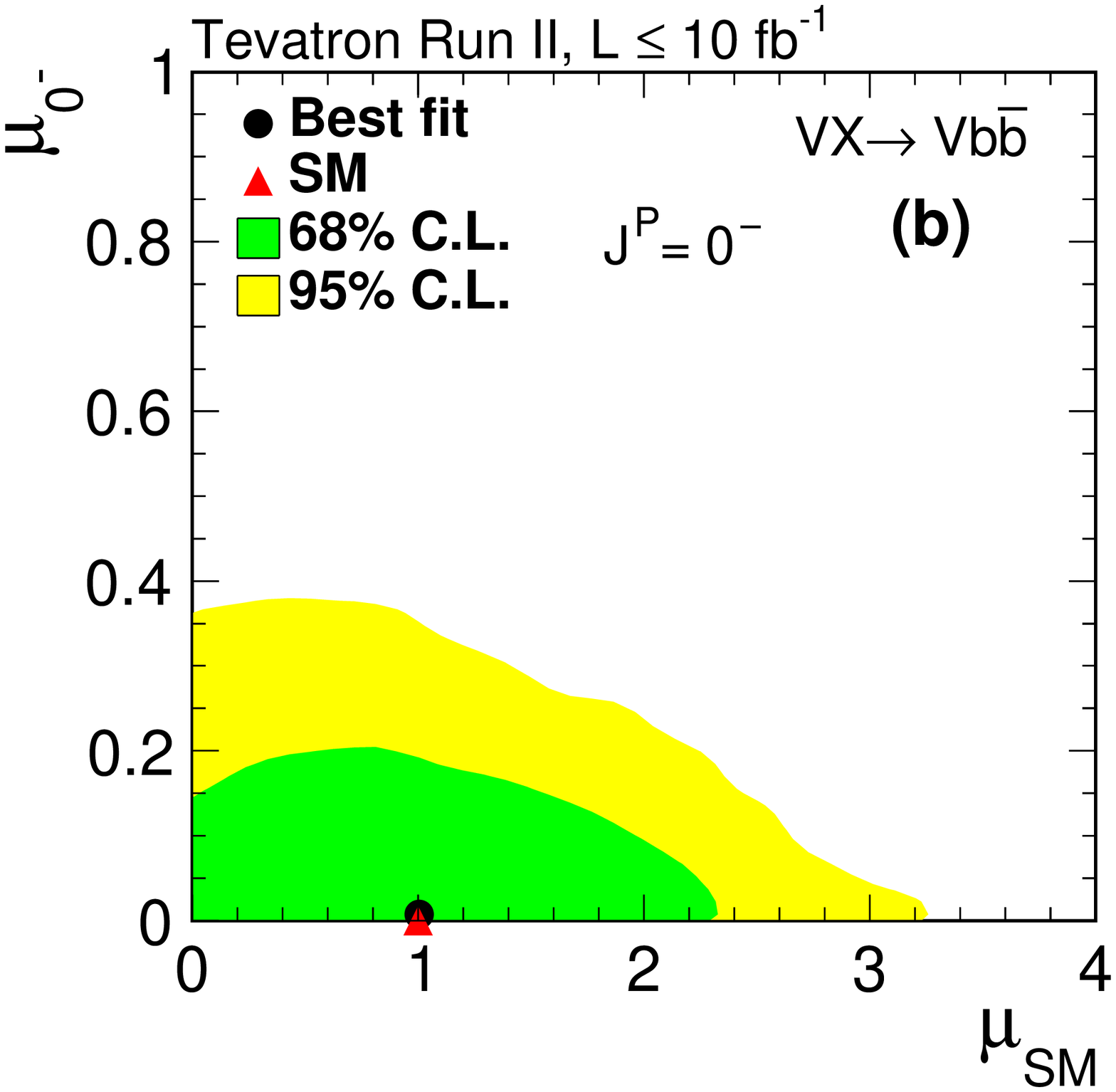}
\includegraphics[width=0.45\columnwidth]{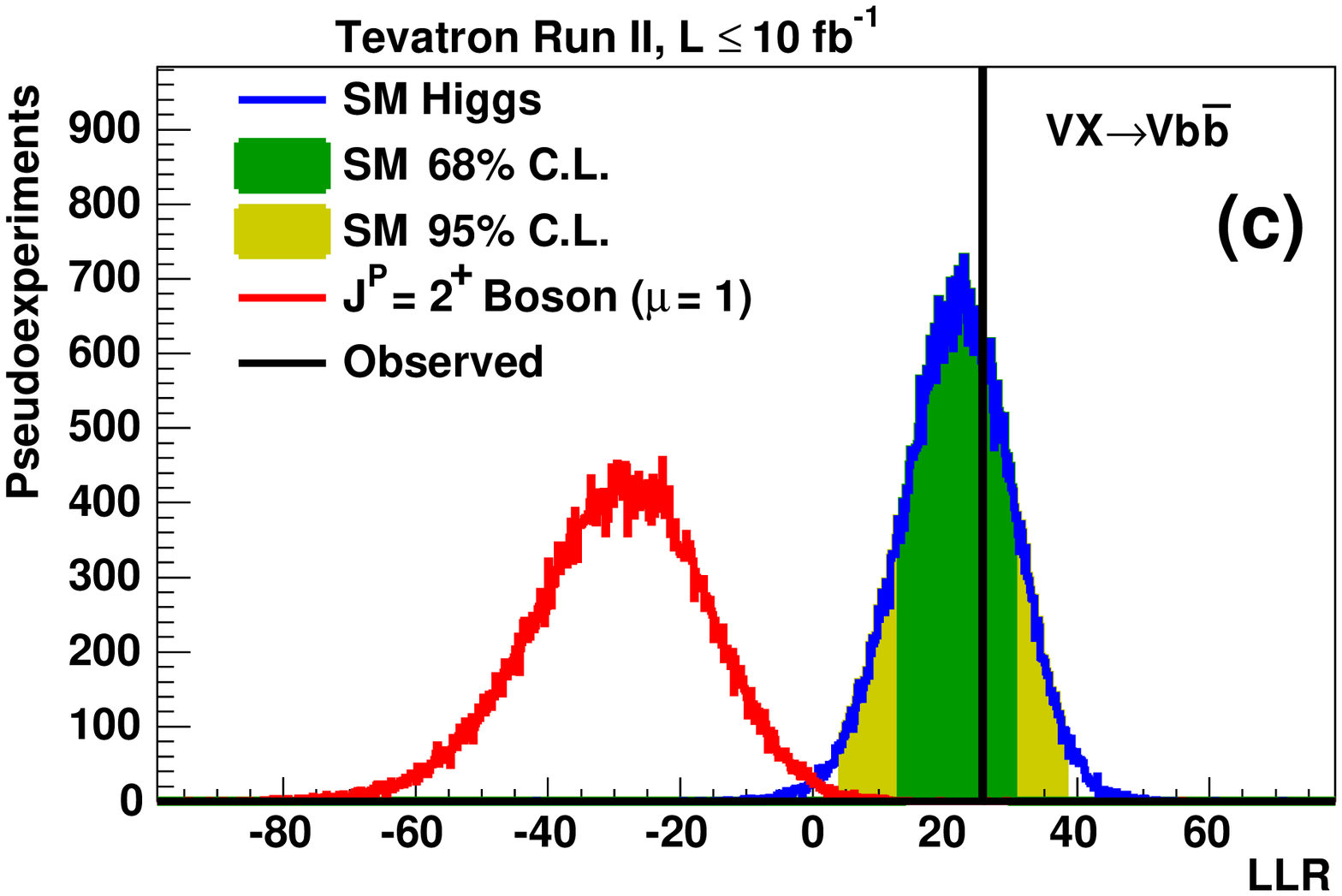}
\includegraphics[width=0.45\columnwidth]{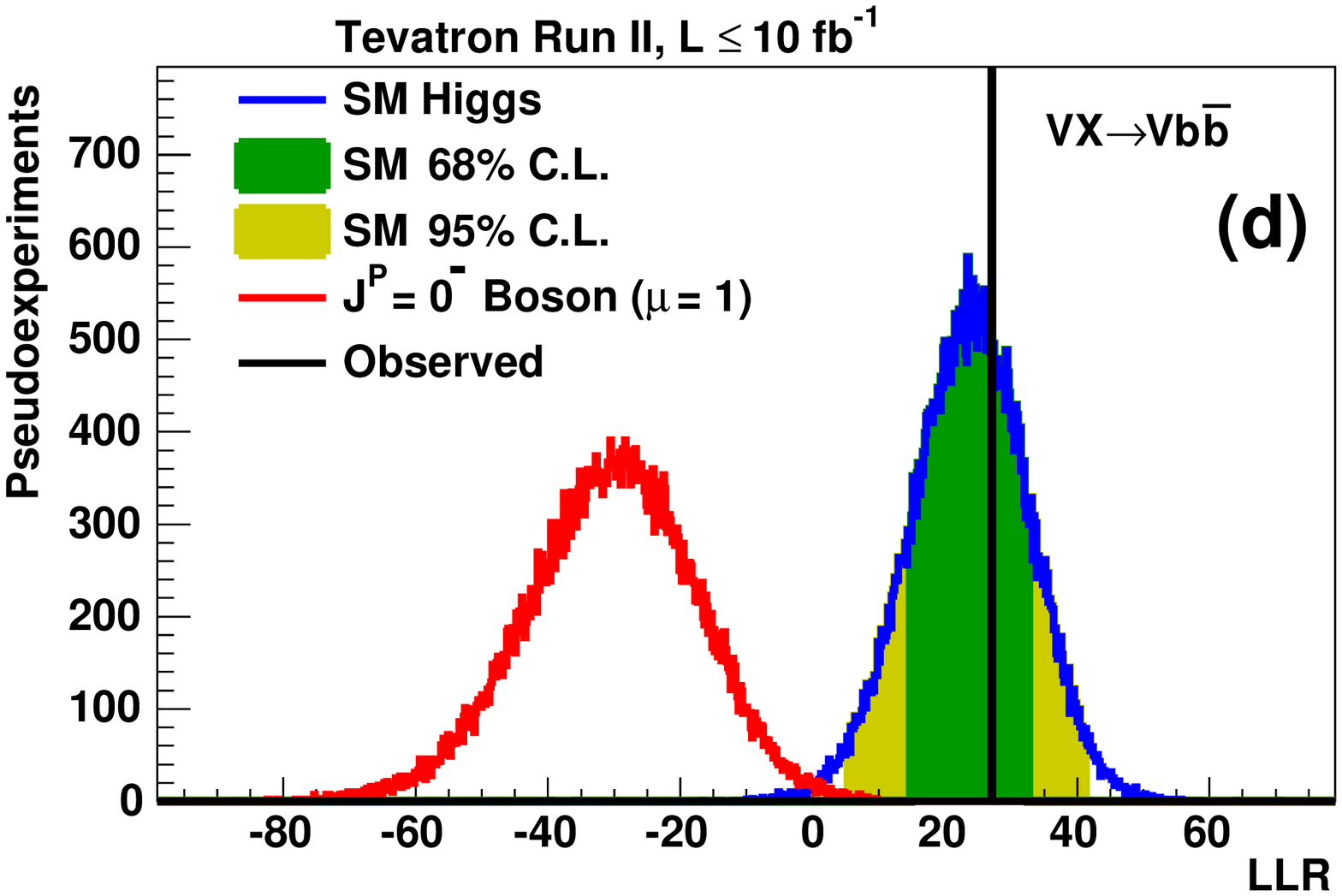}
\caption{
\label{fig:higgsjp}  Two-dimensional constraints in the $(\mu_{\rm{SM}},\mu_{\rm{exotic}})$ plane
for exotic models of Higgs boson production with $J^P=2^+$ (a) and $J^P=0^-$ (b).   Also shown are the
expected distributions of LLR comparing the data with the prediction assuming a SM ($0^+$) Higgs boson
against an alternative model with a $2^+$ boson produced with SM Higgs boson strength (c), and also for a $0^-$ boson
produced with SM Higgs boson strength (d).  From Ref.~\citen{Aaltonen:2015mka}.
}
\end{centering}
\end{figure}

Figure~\ref{fig:higgsjp} shows interpretations allowing for an arbitrary admixture of $0^+$ (SM) and exotic
Higgs bosons, separately for the combination of searches for the $2^+$ boson and the $0^-$ boson.  The
signal strength modifiers $\mu_{\rm{SM}}$, $\mu_{\rm{exotic}}$ are allowed to vary separatley and the Bayesian
posterior probability density is computed for both the $2^+$ and $0^-$ searches.  No evidence
is seen for either exotic particle, and the data are consistent with the presence of the SM Higgs boson in
both cases.  Figure~\ref{fig:higgsjp} also shows the distributions of LLR comparing the hypothesis that
the SM Higgs boson is present with its predicted strength and production and decay properties against the 
hypothesis that the boson is either a $J^P=2^+$ or $0^-$ particle, assuming SM Higgs boson production
strengths and decay branching ratio to $b{\bar{b}}$ for the exotic hypotheses.  These models are excluded
with CL$_{\rm{s}}$ values of $5.6\times 10^{-7}$ and $2.6\times 10^{-7}$, respectively.

These searches for exotic $J^P=2^+$ and $0^-$ bosons provide independent information about the spin and parity
of the Higgs boson from the constraints placed by the ATLAS~\cite{Aad:2013xqa,atlas_jp_prel} 
and CMS~\cite{Chatrchyan:2013mxa,Chatrchyan:2013iaa,Khachatryan:2014ira,Chatrchyan:2012jja,Khachatryan:2014kca}, since they test the
$X\rightarrow b{\bar{b}}$ decays instead of $X\rightarrow ZZ^{(*)}, WW^{(*)}, $ or $\gamma\gamma$ decays.

\section{Searches for Higgs Bosons Beyond the Standard Model}
\label{sec:sec5}

The phenomenology of Higgs boson production and decay relevant to searches for Higgs bosons in extensions of the SM is described
in Sec.~\ref{sec:bsmpheno}.
While the above extensions of the SM provide useful benchmarks, most searches are designed to be as model-independent as possible.

\subsection{Heavy neutral Higgs bosons decaying to vector bosons}

Searches for a non-SM heavy CP-even neutral Higgs boson decaying to vector bosons have been performed in the 
context of the SM Higgs boson searches discussed in  Secs.~\ref{sec:sec4_hww}--\ref{sec:sec4_hgamgam} 
(see Ref.~\citen{Aaltonen:2013kxa} for the combined results between CDF and D\O). 

\begin{figure}[t]
\begin{center}
\begin{tabular}{cc}
\includegraphics[height=5.3cm, width=0.49\textwidth]{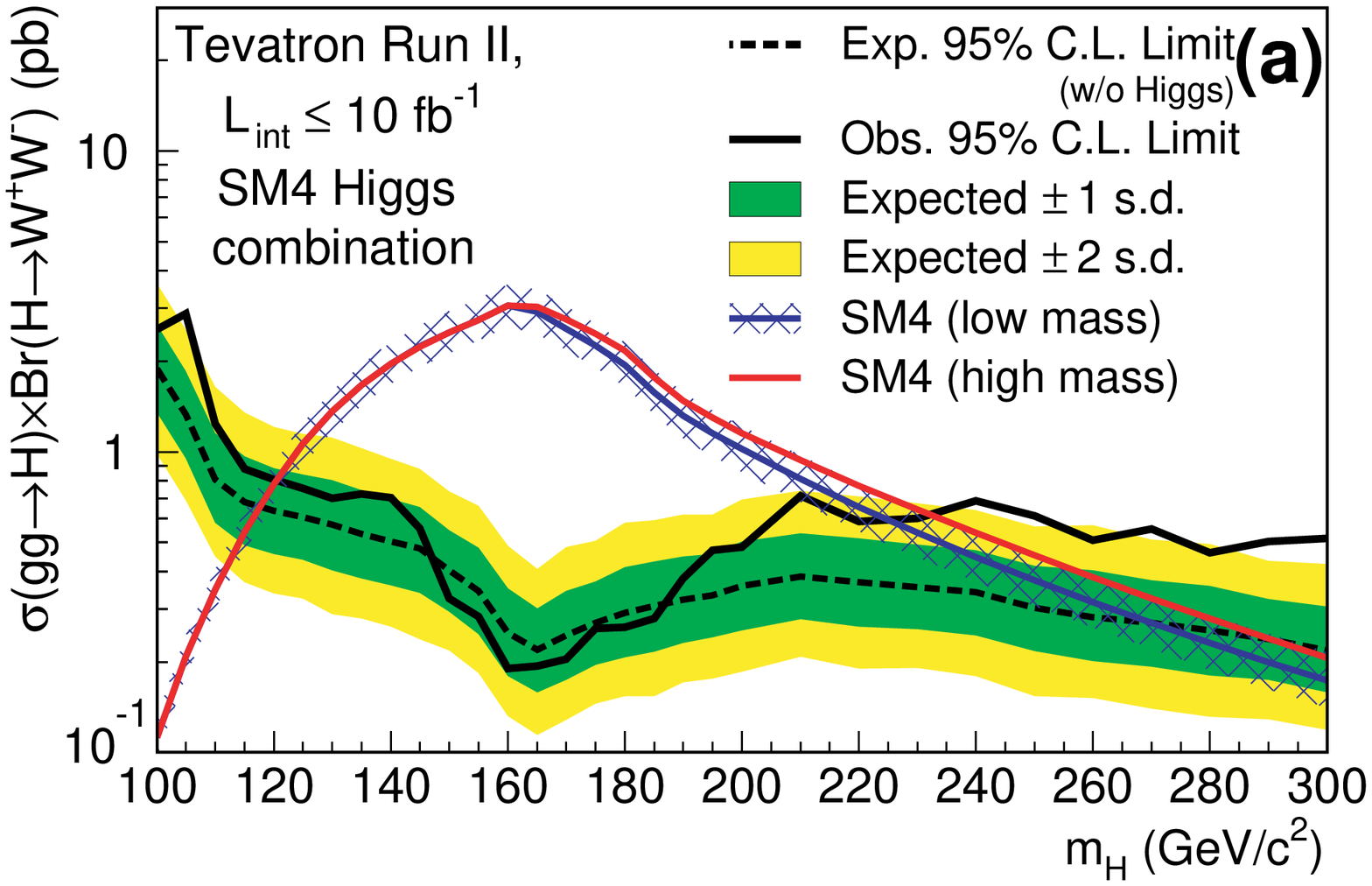} &
\includegraphics[height=5.5cm, width=0.49\textwidth]{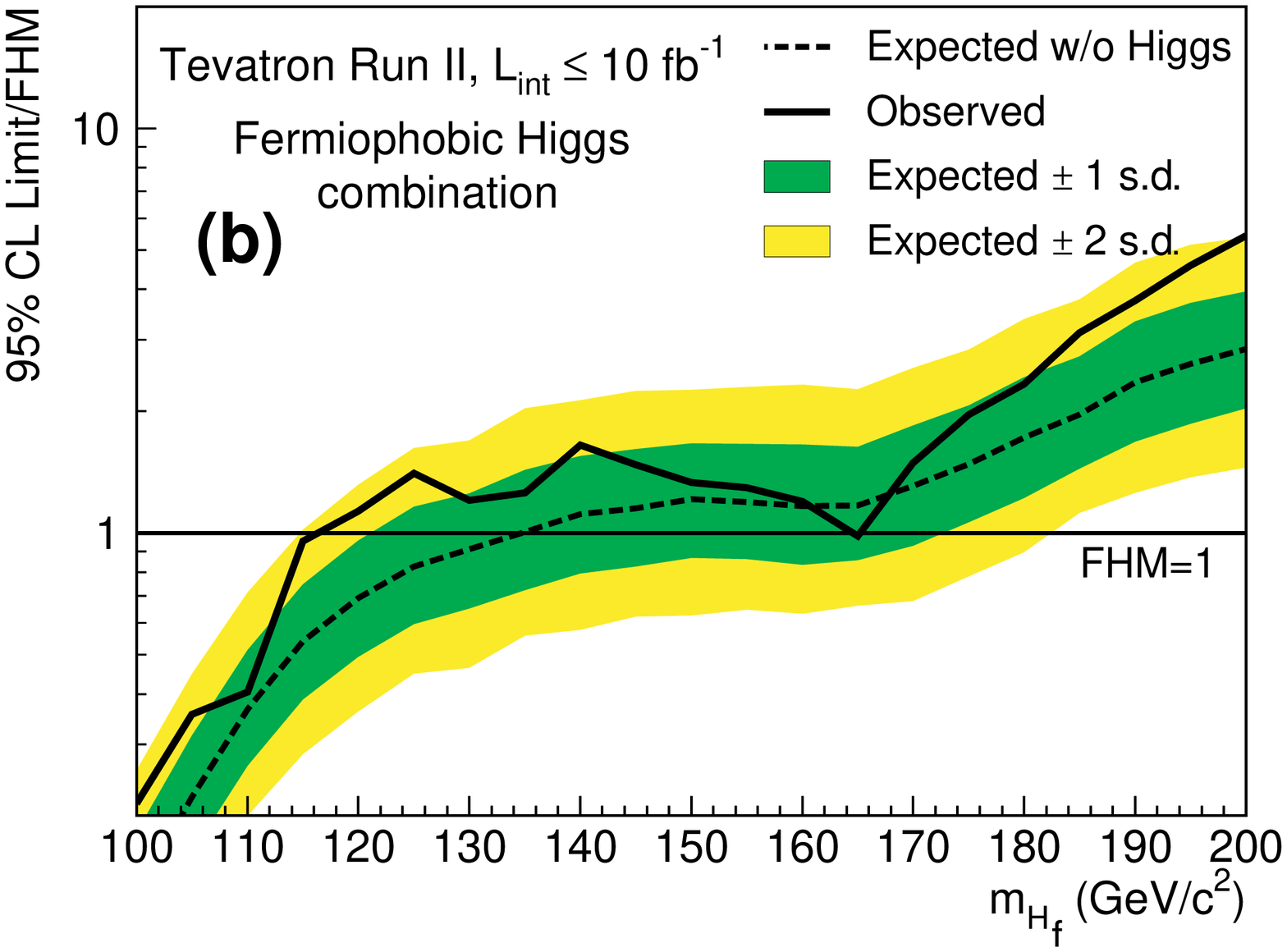} \\
\end{tabular}
\end{center}
\caption{(a) 95\% C.L. upper limits on $\sigma(gg\to H)\times  {\cal B}(H\to W^+W^-)$ as a function of $m_H$ from
the combination of CDF and D\O\ searches focused on this production and decay mode. Also shown are theoretical predictions for
the SM4 in the low- and high-mass scenarios (see text for details).
(b) 95\% C.L.  upper limits on the production cross section of a fermiophobic Higgs boson relative to the FHM prediction as a function of $m_{H_f}$
from the combination of CDF and D\O\ searches (see text for details).
From Ref.~\citen{Aaltonen:2013kxa}.
\label{fig:hbsm_1}}
\end{figure}

Searches for $H\to WW^{(*)}$ and $H\to ZZ^{(*)}$ have been performed in the
mass range of $m_H=$100--300 GeV. The MVA discriminants are
retrained at each value of $m_H$ considering only the $gg\to H$
production mode.  Searches for $gg\to H \to WW^{(*)}$ and $gg\to H \to ZZ^{(*)}$ 
are combined assuming the SM prediction for ${\cal B}(H\to WW^{(*)})/{\cal B}(H\to ZZ^{(*)})$ 
and 95\% C.L. upper limits are derived on $\sigma(H\to WW^{(*)})\times {\cal B}(H\to WW^{(*)})$ as a function of
$m_H$ (see Fig.~\ref{fig:hbsm_1}(a)). As shown in the same figure,
these results can be used to set constraints on models with a
sequential fourth generation of fermions (SM4) which, as described in Sec.~\ref{sec:bsmpheno},
leads to an enhancement in the $gg\rightarrow H$ production cross section by a factor of $\approx 9$.
This much larger production cross section provides a model that could be tested with
a smaller data sample, and with the complete Run II dataset, a much larger range in $m_H$ could 
be tested than in the case of the SM. The results are interpreted in the context of two
different SM4 scenarios, depending on the assumed masses of the
fourth-generation neutrino ($\nu4$) and charged lepton ($\ell 4$): a
``low mass" scenario with $(m_{\nu4},m_{\ell 4})=(80,100)$ GeV, such
that they maximally affect the $H\to WW^{(*)},ZZ^{(*)}$ branching ratios by
opening new decay modes for the Higgs boson, and a ``high-mass"
scenario with $m_{\nu4}=m_{\ell 4}=1$ TeV, where the $H\to WW^{(*)},ZZ^{(*)}$
branching ratios are unaffected. In the low-mass (high-mass) scenario
a Higgs boson with mass in the range 121--225 GeV (121--232 GeV) is
excluded at the 95\% C.L.

In addition, searches for $H\to \gamma\gamma$ and $H\to WW^{(*)}$ have
been performed to probe the existence of a fermiophobic Higgs boson
($H_f$), as predicted by e.g. Type-I 2HDMs. In the fermiophobic Higgs
model (FHM) considered by these searches, the Higgs boson does not
couple to fermions at tree level, which leads to substantial
modifications to the production cross sections and decay branching
ratios. On the one hand, the process $gg\to H_f$ is suppressed to a
negligible level whereas the $VH$ and VBF production modes remain nearly
unchanged relative to the corresponding processes in the SM. On the
other hand, direct decays to fermions are forbidden, resulting in a
large increase to the $H_f \to \gamma\gamma$ branching ratio at low
mass relative to the SM decay, while $H_f \to W^+ W^-$ dominates over
most of the mass range considered. The SM searches for $H\to
\gamma\gamma$ and $H\to WW^{(*)}$ have been reoptimized for the FHM
scenario by retraining the MVA discriminants after ignoring
the gluon-gluon fusion production mechanism, which significantly
affects the kinematic distributions of the Higgs boson compared to the
SM case. The combined limits from CDF and D\O\ on Higgs boson
production normalized to the FHM predictions are shown in
Fig.~\ref{fig:hbsm_1}(b) as a function of $m_{H_f}$. As a result,
fermiophobic Higgs bosons in the mass range 100--116 GeV are excluded
at the 95\% C.L., with an expected excluded mass range of 100-135 GeV.

%%%%%%%%%%%%%%%%%%%%%
%%%%%%%%%%%%%%%%%%%%%

\subsection{Heavy neutral Higgs bosons decaying to fermions}

As mentioned previously, the MSSM has five physical Higgs bosons:
three neutral ($h$, $H$ and $A$) and two charged ($H^\pm$).  At the
leading order, only two parameters are sufficient to describe the
Higgs sector, by convention taken to be the ratio of the two Higgs
doublets' vacuum expectation values, $\tan\beta$, and the mass of the
pseudoscalar boson, $m_A$.  Radiative corrections introduce additional
dependencies on other model parameters.  At large $\tan\beta$, one of
the CP-even Higgs bosons ($h$ or $H$) is approximately degenerate in
mass with the $A$ boson. These two almost-degenerate neutral states
are collectively referred to as $\phi$. In addition, the couplings to
the down-type fermions are enhanced by a factor of $\tan\beta$
relative to those in the SM. As a result, at high $\tan\beta$ the main decay modes are
$\phi\to b\bar{b}$ and $\phi \to \tau^+\tau^-$, with branching ratios
of approximately 90\% and 10\%, respectively. Also, the inclusive
$\phi$ production is dominated by gluon-gluon fusion ($gg \to \phi$,
with the $b$ quark playing a potentially important role in the loop) and
$b\bar{b}\to \phi$. The latter process may produce a $b$ quark in the
detector acceptance, via $g b \to \phi b$, which provides an important
experimental handle to suppress backgrounds.  The CDF and D\O\
Collaborations have searched for $\phi\to b\bar{b}$ and $\phi \to
\tau^+\tau^-$ in both inclusive and $b\phi$ production modes.  A
summary of the main features of these searches and their results is
provided below.

\subsubsection{$\phi \to b\bar{b}$}

\begin{figure}[t]
\begin{center}
\begin{tabular}{cc}
\includegraphics[height=5.1cm, width=0.49\textwidth]{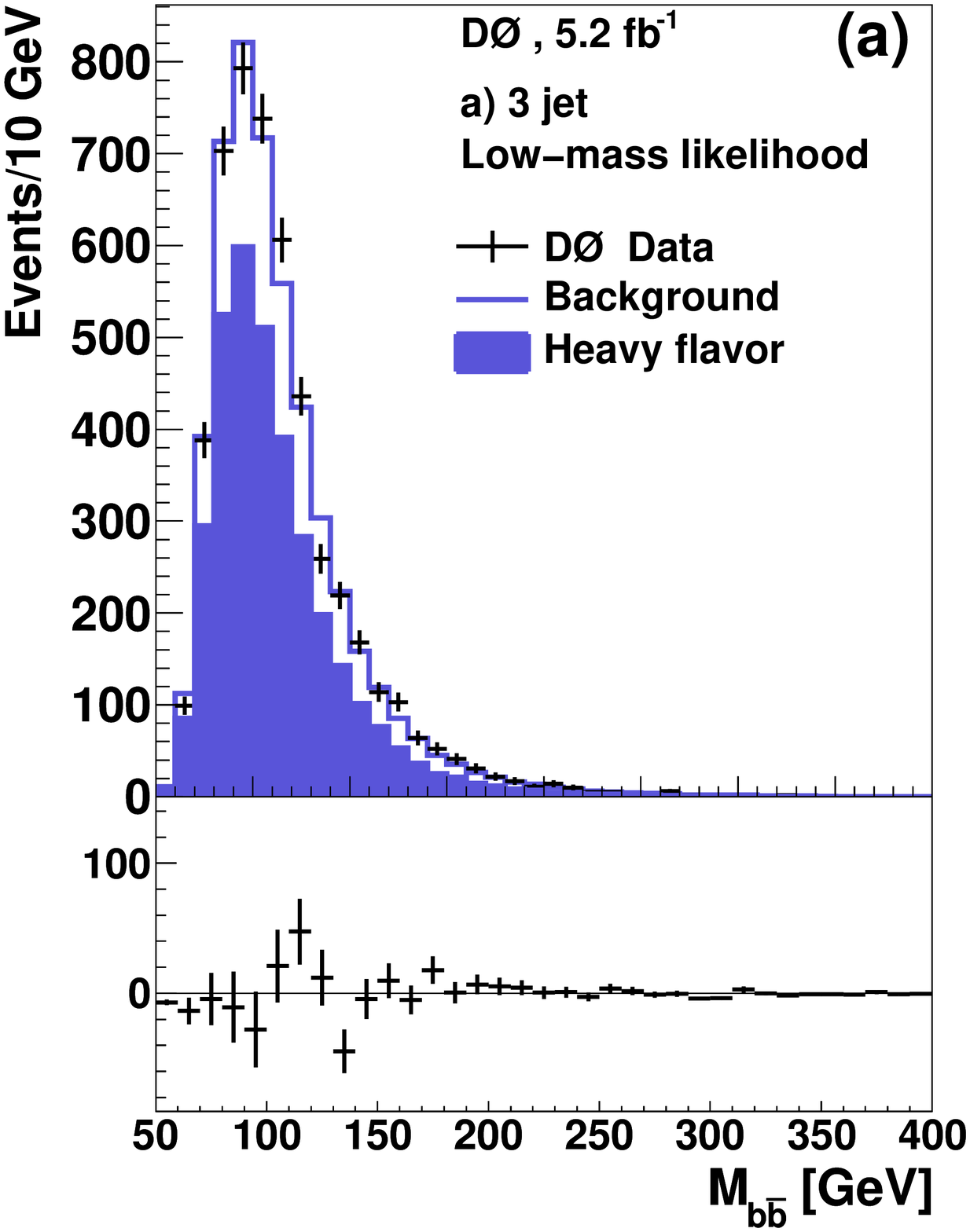} &
\includegraphics[height=5.5cm, width=0.49\textwidth]{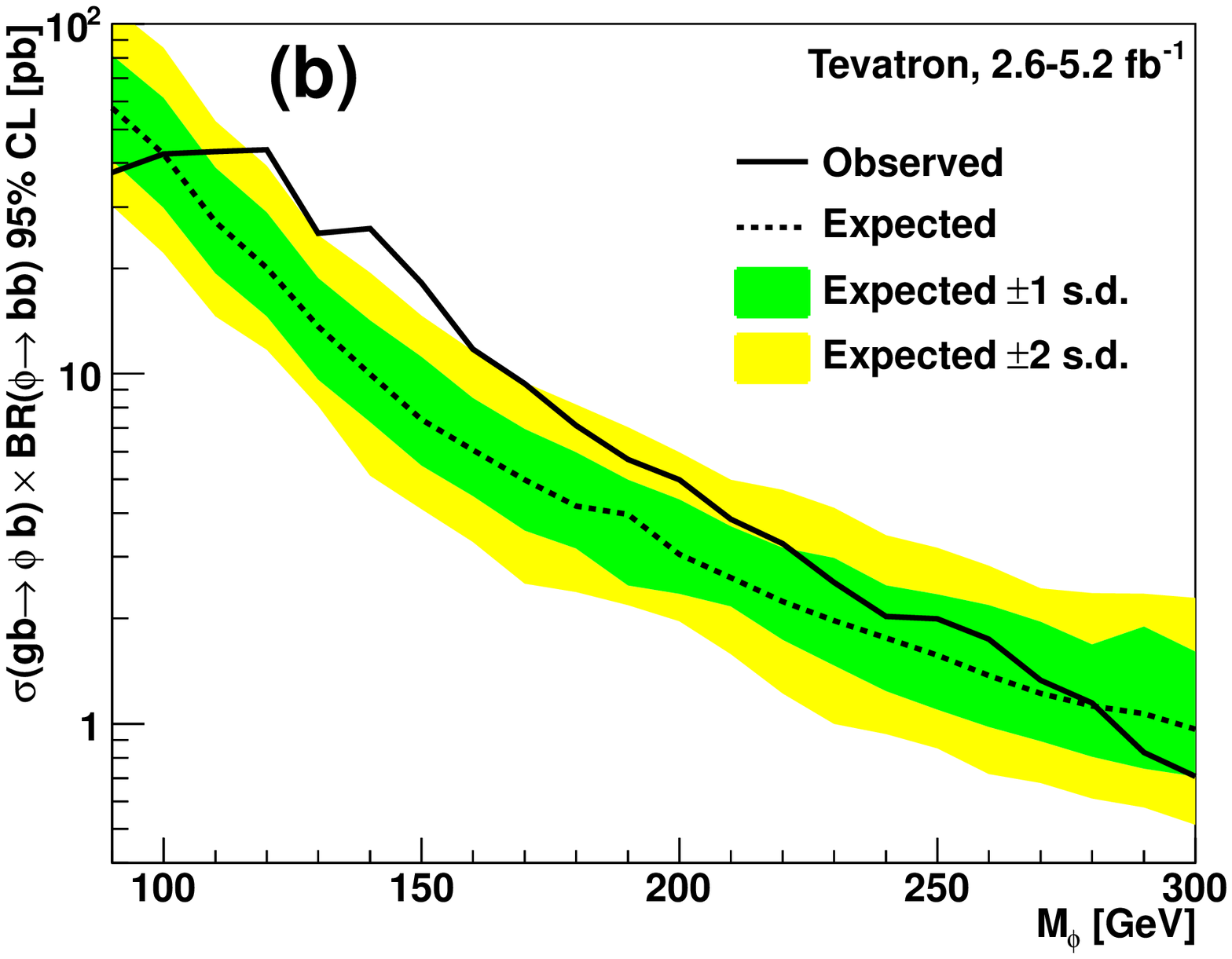} \\
\end{tabular}
\end{center} 
\caption{(a) Distribution of the reconstructed 2-jet invariant mass in the 3 $b$-tag exclusive channel after the low-mass likelihood
requirement in the D\O\ $b\phi \to bb\bar{b}$ search.
The data (points with error bars) are compared to the background prediction, for which the heavy-flavor component
($bbb$, $bbc$ and $bcc$) is shown as the shaded region.  The lower panel displays the difference between the data and the 
predicted background.
From Ref.~\citen{Abazov:2010ci}.
(b) Model-independent 95\% C.L. upper limits on $\sigma(gb\to \phi b)\times {\cal B}(\phi \to b\bar{b})$ from the combination 
of the CDF and D\O\ searches as a function of $m_\phi$ (see text for details). Two out of the three neutral Higgs bosons are assumed to be
degenerate in mass and to have a width significantly smaller than the experimental resolution.
From Ref.~\citen{Aaltonen:2012zh}.
\label{fig:hbsm_2}}
\end{figure}

Searches for $\phi \to b\bar{b}$ have been performed by the CDF and
D\O\ Collaborations, using 2.6 fb$^{-1}$ and 5.2 fb$^{-1}$ of Run II
data, respectively~\cite{Aaltonen:2011nh,Abazov:2010ci}. An inclusive
search for $\phi \to b\bar{b}$ would be extremely difficult due to the
overwhelming background from QCD $b\bar{b}$ production. Therefore, these searches
are performed in the associated production mode, $b\phi \to bb\bar{b}$, 
resulting in a signature with at least three $b$ jets in
the final state, with the third $b$ jet requirement providing
additional rejection against the QCD multijet background. Both CDF and
D\O\ searches employ multijet triggers including $b$-tagging
requirements. After selecting an offline sample with at least three
$b$-tagged jets, the final discriminating variable is the invariant
mass of the Higgs boson candidate, defined based on either the two
leading $b$-tagged jets (CDF) or the pairing that maximizes a
likelihood-ratio discriminant variable (D\O) (see
Fig.~\ref{fig:hbsm_2}(a)).  In the case of the D\O\ analysis, the
likelihood-ratio discriminant is also used to reject events for which
no pairing satisfies a given threshold value, a requirement that helps to 
further suppress the background.  The main challenge of the analysis
resides in the modeling of the QCD multijet background, for which no
reliable simulation exists.  Both analyses build a model of the
background in the 3 $b$-tag sample by using a large data sample
requiring exactly two $b$-tagged jets and applying suitable
corrections to account for the change in flavor composition and
possible kinematic distortions from the third $b$-tag requirement.

Different signal hypothesis are tested by varying the Higgs boson
mass, $m_\phi$, and the individual CDF and D\O\ analyses find local
excesses with significances of 2.8 s.d. at $m_\phi=180$ GeV and 2.5
s.d. at $m_\phi=120$ GeV, respectively. These excesses are not
significant after taking into account the LEE. Under the assumptions
that two out of the three neutral Higgs bosons are degenerate in mass,
and that the Higgs boson width is significantly smaller than the
experimental resolution, upper limits on the production cross section
times branching ratio, $\sigma(gb \to \phi b)\times {\cal B}(\phi \to b\bar{b})$ 
are set as a function of $m_\phi$. The cross section is
defined such that at least one $b$ quark not originating from the
$\phi$ decay has $\pt>20$~GeV and $|\eta|<2.5$.
Figure~\ref{fig:hbsm_2}(b) shows these upper limits for the
combination of the CDF and D\O\ analyses.  In addition, constraints
were placed in the ($\tan\beta$, $m_A$) plane for a particular MSSM
benchmark scenario, this time taking into account the Higgs boson
width effect.  It is worth noting that these constraints are strongly
dependent on higher-order radiative corrections, a feature not shared
by searches for $\phi \to \tau^+\tau^-$, owing to cancellations
between radiative corrections affecting the production and decay
processes. This makes both searches complementary since their
combination could shed light on the nature of a possible signal.

%%%%%%%%%%%%%%%%%%%%%
%%%%%%%%%%%%%%%%%%%%%

\subsubsection{$\phi \to \tau^+\tau^-$}

The first search for $\phi \to \tau^+\tau^-$ at a hadron collider was
performed by the CDF Collaboration using 86.3 pb$^{-1}$ of data at
$\sqrt{s}=1.8$ TeV collected during Run I~\cite{Acosta:2005bk}. 
This early search focused on events
with one tau decaying to an electron and neutrinos ($\tau \to e \nu_e
\nu_\tau$) and the other one decaying hadronically ($\tau_h$), and
demonstrated the feasibility to reconstruct the ditau invariant mass
when the tau candidates are not back-to-back. Much more sensitive
searches were carried out by the CDF and D\O\ Collaborations during
Run II, using up to 1.8 fb$^{-1}$ and 7.3 fb$^{-1}$ of data,
respectively~\cite{Abulencia:2005kq,Abazov:2011jh,Abazov:2011qz,Abazov:2011up}. Events
were selected requiring one or two tau candidates to decay
leptonically (excluding $ee$ and $\mu\mu$ final states, which suffer
from very large background from $Z/\gamma^*$ production), resulting in
final states denoted as $\tau_e\tau_h$, $\tau_\mu\tau_h$ and
$\tau_e\tau_\mu$. The CDF analysis~\cite{Abulencia:2005kq} and early
D\O\ analyses~\cite{Abazov:2011jh} considered the three channels,
while the final D\O\ analyses~\cite{Abazov:2011qz,Abazov:2011up} were
restricted to the $\tau_\mu\tau_h$ channel, which dominates the
sensitivity.

%%%%%%%%%%%%%%%%%%%%%
\begin{figure}[t]
\begin{center}
\begin{tabular}{cc}
\includegraphics[height=5.5cm, width=0.49\textwidth]{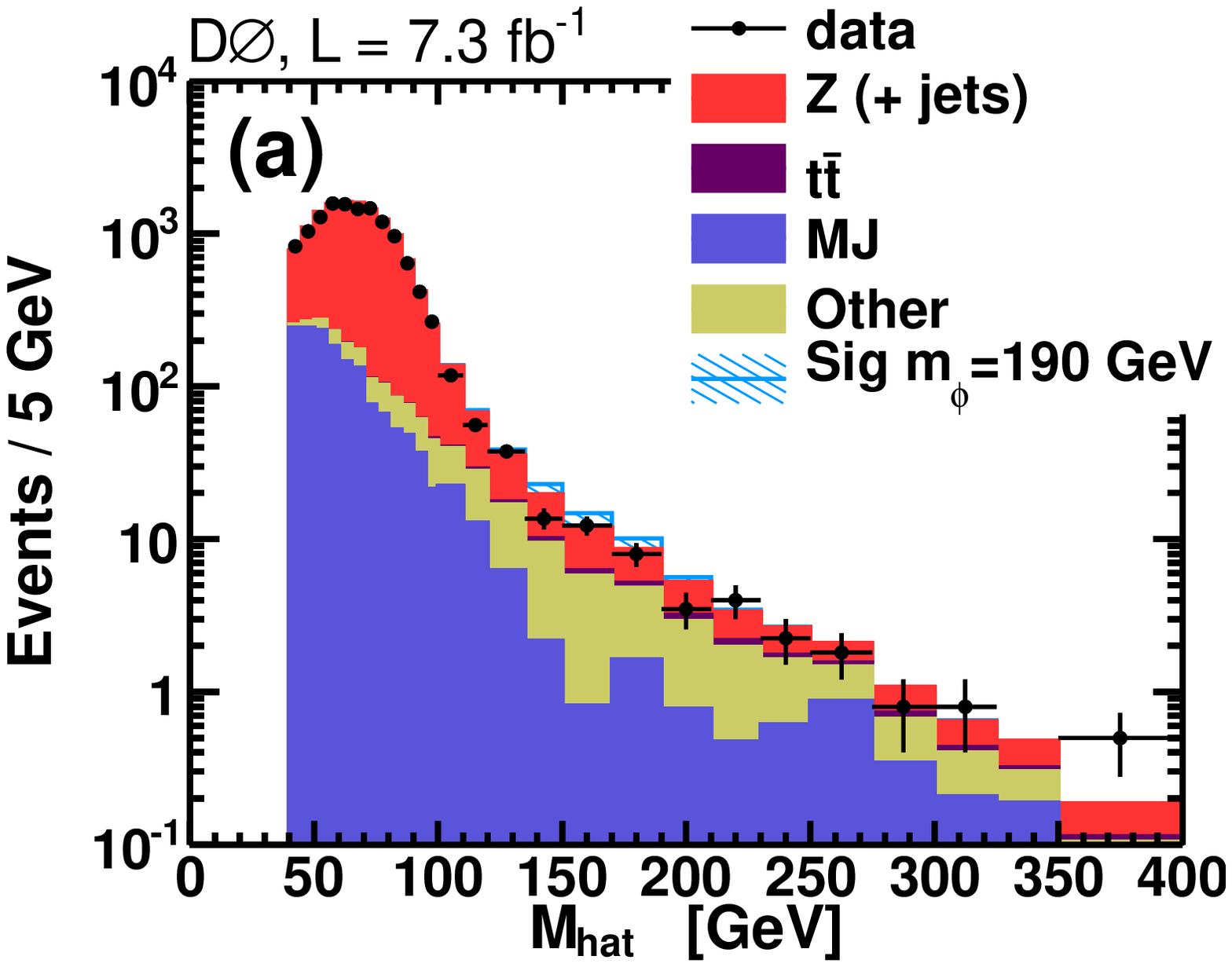} &
\includegraphics[height=5.5cm, width=0.49\textwidth]{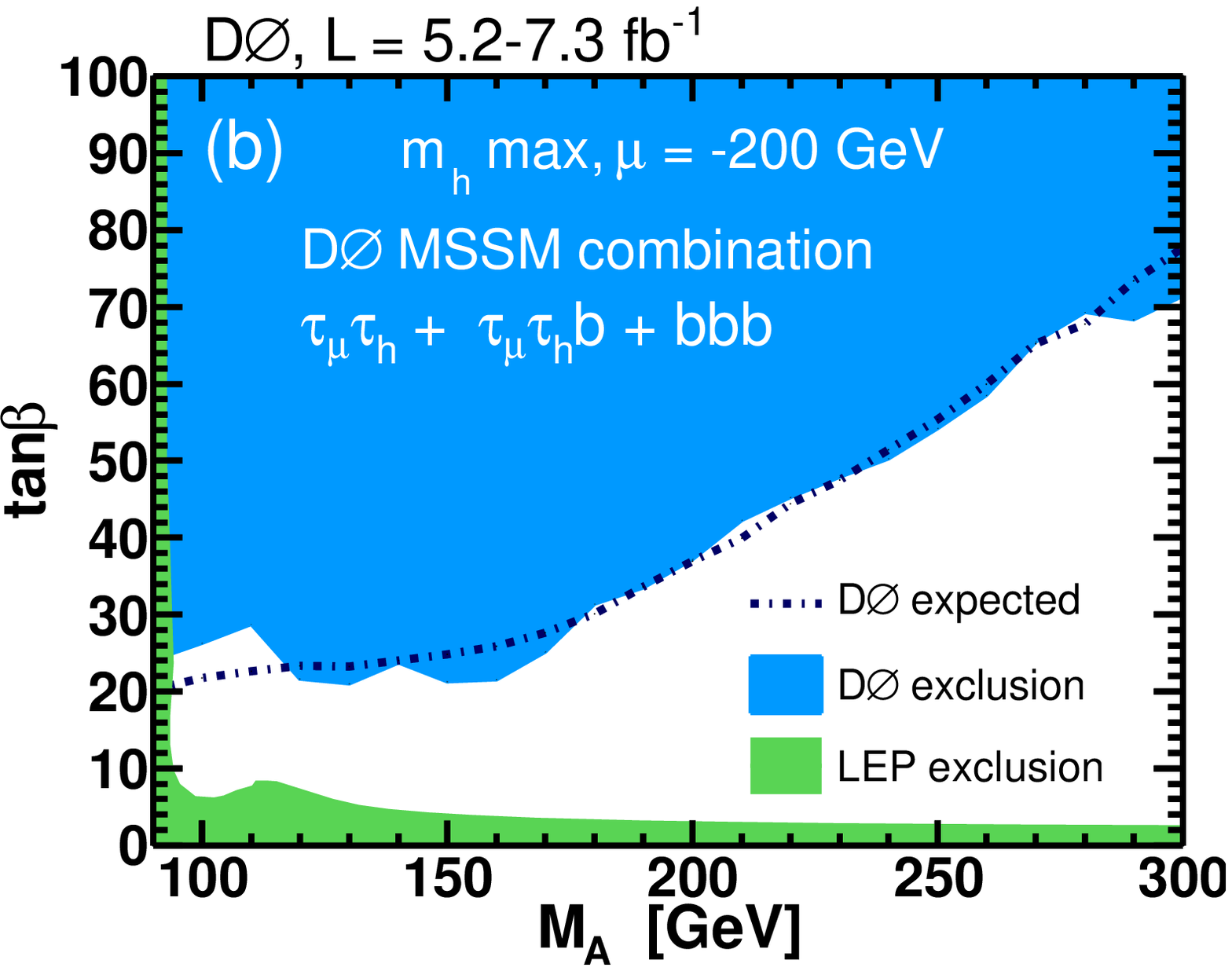} \\
\end{tabular}
\end{center}
\caption{(a) Distribution of $M_{\rm hat}$ in the inclusive $\tau_\mu\tau_h$ channel from the D\O\ search (see text for details).
The data (points with error bars) are compared to the background prediction, 
broken down into its individual components. Also shown is the expected contribution from a signal with $m_\phi = 190$~GeV. 
From Ref.~\citen{Abazov:2011up}.
(b) Constraints in the ($\tan\beta$,$m_A$) plane in a given MSSM benchmark scenario from the combination 
of the final $\phi \to \tau^+\tau^-$ and $\phi \to b\bar{b}$ D\O\ searches. 
From Ref.~\citen{Abazov:2011up}.
\label{fig:hbsm_3}}
\end{figure}
%%%%%%%%%%%%%%%%%%%%%

While all previous Tevatron searches were focused on the inclusive
production mode $gg,bb\to \phi$, the final D\O\ analysis considered
both the inclusive and $gb\to \phi b$ associated production modes, by
defining two non-overlapping analysis channels without and with the
requirement of an additional $b$-tagged jet, referred to as
$\tau_\mu\tau_h$ and $\tau_\mu\tau_h b$, respectively. In the case of
the $\tau_\mu\tau_h$ channel, the main background originates from
$Z/\gamma^*\to \tau^+\tau^-$, followed by QCD multijet and $W$+jets
production where one of the jets is misidentified as a hadronic
tau. The main discriminating variable used is the ditau invariant
mass, denoted as $M_{\rm hat}$, defined from the four-momenta of the
two leptons and the $\met$, in such a way that it represents the
minimum center-of-mass energy consistent with the decay of a ditau
resonance (see Fig.~\ref{fig:hbsm_3}(a)).  In the case of the
$\tau_\mu\tau_h b$ channel, there are large backgrounds from
$Z/\gamma^*$+jets, $t\bar{t}$ and QCD multijets. Dedicated
MVA discriminants are used to reject the $t\bar{t}$ and QCD
multijets backgrounds, as well as to discriminate signal from the
remaining total background. No significant excess above the SM
expectation is found in either of the channels, and constraints in
the ($\tan\beta$,$m_A$) plane in different MSSM benchmark scenarios
are derived. Both channels have comparable reach, with the
$\tau_\mu\tau_h b$ channel being somewhat more sensitive at lower
$m_A$, owing to the reduced background from $Z/\gamma^*\to
\tau^+\tau^-$ by the $b$-tagging
requirement. Figure~\ref{fig:hbsm_3}(b) shows the combination of
the D\O\ $\tau_\mu\tau_h$, $\tau_\mu\tau_h b$ channels and the $b\phi
\to bb\bar{b}$ search~\cite{Abazov:2010ci}, although the contribution
of the latter was very small and strongly dependent on the MSSM 
parameters assumed.

\subsection{Charged Higgs bosons}

At the Tevatron, charged Higgs bosons can be produced in different
modes depending on the value of their mass ($m_{H^\pm}$) compared to
the top quark mass. If $m_{H^+}<m_t-m_b$, the charged Higgs boson can
be produced in decays of the top quark, $t \to H^+ b$, competing with
the SM decay $t \to W^+ b$. Alternatively, if $m_{H^+}>m_t-m_b$, it 
can be produced through radiation from a third
generation quark or in association with a top quark, such as in the
process $q\bar{q}, gg \to t\bar{b}H^-$. Searches at the Tevatron have
been performed during Run I and Run II focused on the case of a light
charged Higgs boson appearing in top quark decays. At tree level, the
branching ratio ${\cal B}(t\to H^+b)$ is simply a function of
$m_{H^+}$ and $\tan\beta$. In the MSSM, additional dependencies on the
masses and couplings of other supersymmetric particles arise through
radiative corrections. The ${\cal B}(t\to H^+b)$ can typically be
sizable either at low $\tan\beta$ ($\lesssim 1$) or at high
$\tan\beta$ ($\gtrsim 15$). At low $\tan\beta$, $H^+$ decays
predominantly into $c\bar{s}$ for low $m_{H^+}$ ($\lesssim 130$~GeV)
and into $t^*\bar{b}$ ($\to W^{+(*)}b\bar{b}$) for higher
$m_{H^+}$. Instead, in the the high $\tan\beta$ regime, $H^+$ decays
into $\tau^+\nu$ almost 100\% of the time.

\begin{figure}[t]
\begin{center}
\begin{tabular}{cc}
\includegraphics[height=5.5cm, width=0.49\textwidth]{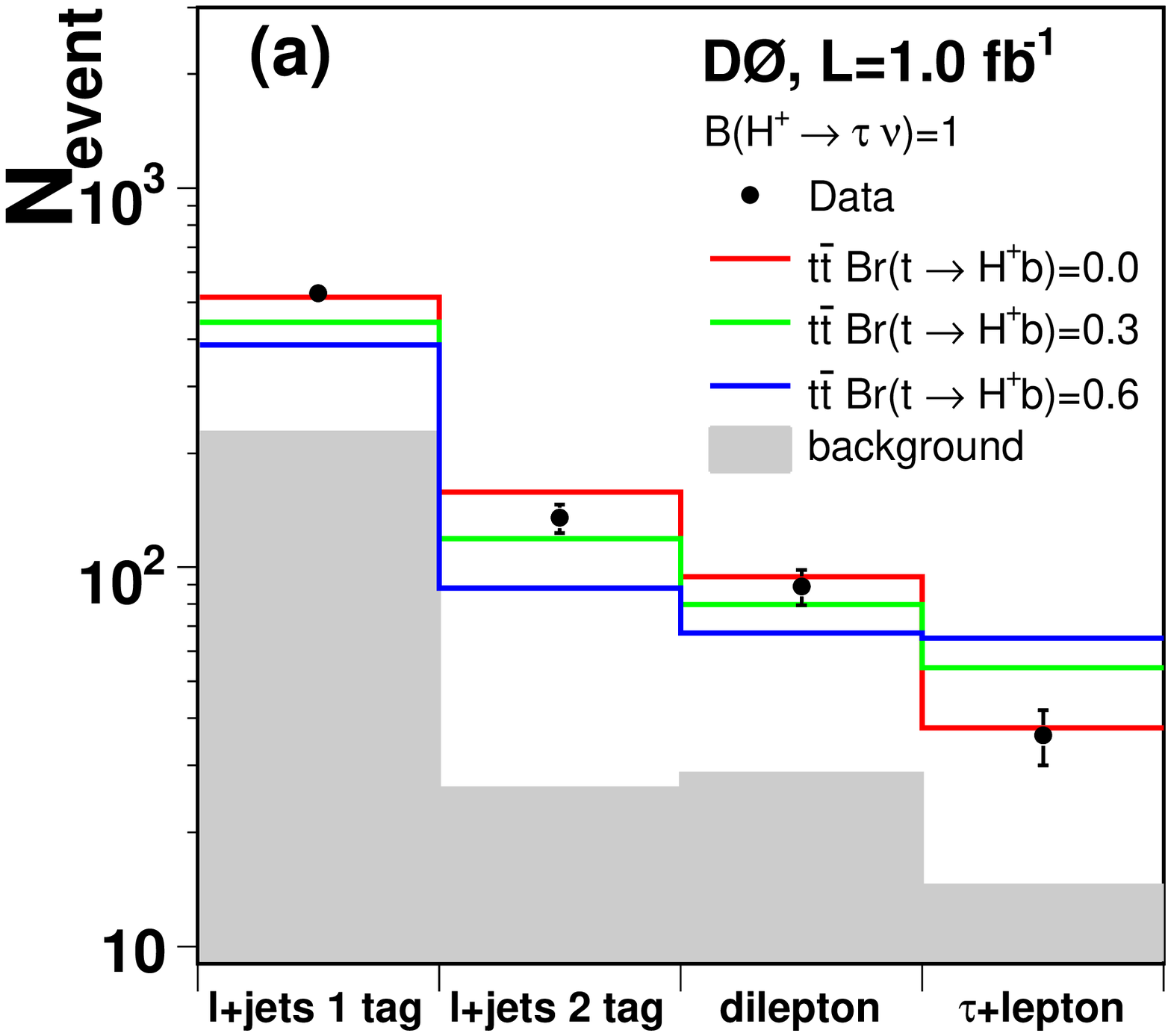} &
\includegraphics[height=5.5cm, width=0.49\textwidth]{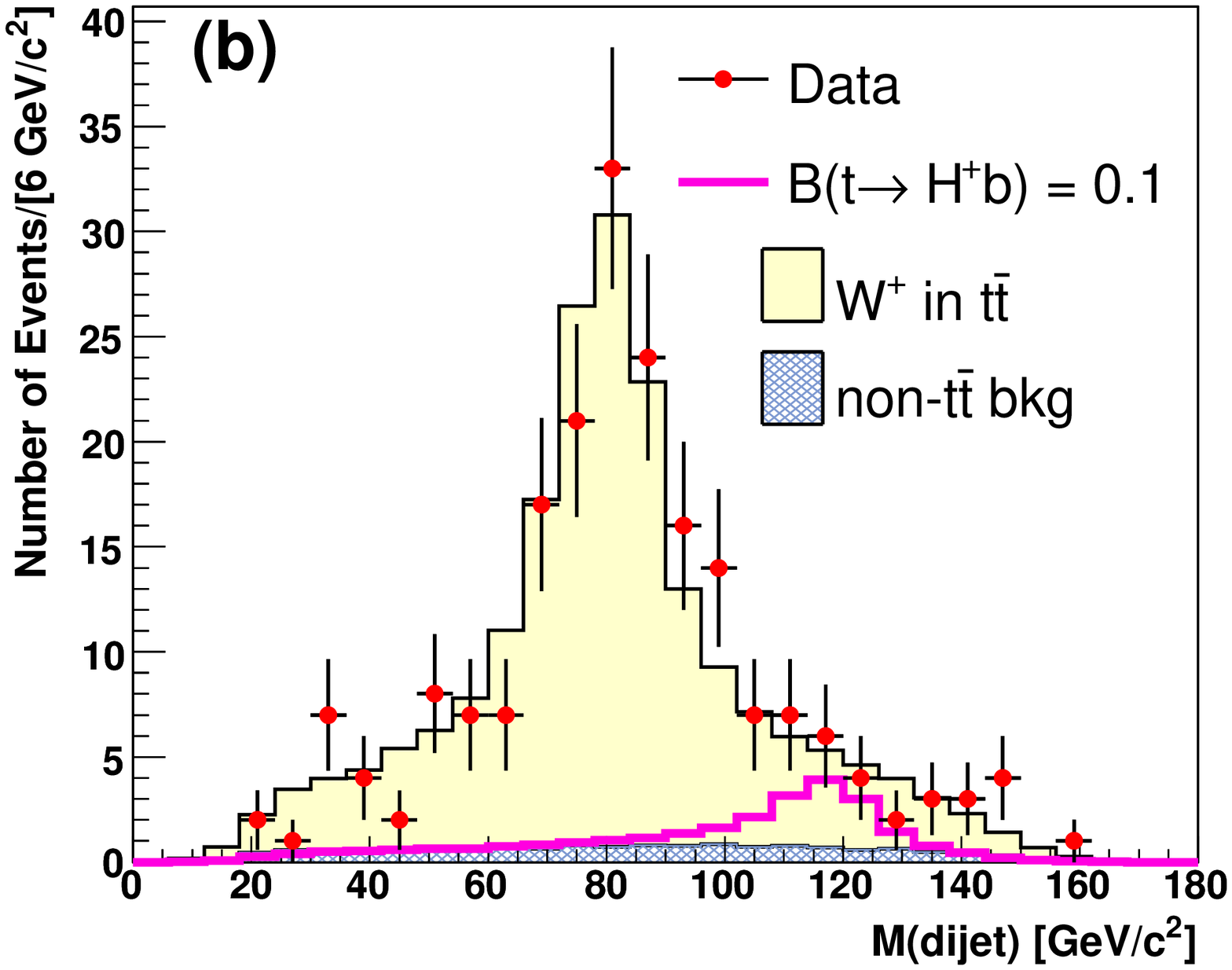} \\
\end{tabular}
\end{center}
\caption{(a) Distribution of the number of observed events in the different $t\bar{t}$ final states considered by the D\O\ analysis,
compared to the expectation assuming a charged Higgs boson with $m_{H^+}=80$~GeV decaying exclusively into $\tau^+\nu$ for
different values of ${\cal B}(t\to H^+b)$.
From Ref.~\citen{Abazov:2009aa}.
(b) Distribution of the dijet mass distribution from the CDF search for $H^+ \to c\bar{s}$ in top quark decays.
The data (points with error bars) are compared to the SM prediction and the expected signal contribution
from $H^+ \to c\bar{s}$ assuming $m_{H^+}=120$~GeV and ${\cal B}(t\to H^+b)$=0.1.
From Ref.~\citen{Aaltonen:2009ke}.
\label{fig:hbsm_4}}
\end{figure}

Searches for $t\to H^+b$ have been performed in $t\bar{t}$ final
states either by explicitly seeking an excess of a particular decay
mode, such as $t \to H^+ b \to \tau^+ \nu b$ or $t \to H^+ b \to c
\bar{s} b$, referred to as ``appearance'' or ``direct'' searches, or by
observing a deficit of a given final state owing to a reduction in its
branching ratio as a result of the competing effect from the $t \to H^+ b$ 
decay, referred to as ``disappearance'' or ``indirect''
searches. For instance, for non-zero branching ratio ${\cal B}(t \to H^+ b \to c \bar{s} b)$, 
the number of selected $t\bar{t}$ events is
expected to decrease in the $\ell$+jets ($\ell=e,\mu$),
$\ell\ell'$+jets, $\ell\tau_h$+$\met$+jets and $\tau_h$+$\met$+jets
final states. Instead, for non-zero branching ratio ${\cal B}(t \to
H^+ b \to \tau^+ \nu b)$, the number of selected $t\bar{t}$ events is
expected to increase in $\ell\tau_h$+$\met$+jets and
$\tau_h$+$\met$+jets final states, while decrease in the rest of final
states (see Fig.~\ref{fig:hbsm_4}(a)).

%%%%%%%%%%%%%%%%%%%%%
%%%%%%%%%%%%%%%%%%%%%

Early examples of both types of searches were carried out in Run~I. 
The CDF Collaboration performed an appearance search for $t \to H^+
b \to \tau^+ \nu b$ in the $\ell\tau_h$+$\met$+jets ($\ell=e,\mu$)
channel using 106 pb$^{-1}$ of Run I data~\cite{Affolder:1999au},
setting 95\% C.L. upper limits on ${\cal B}(t \to H^+ b)$ in the range
of 0.5--0.6 for $m_{H^+}$ in the range 60--160 GeV, assuming ${\cal B}(H^+ \to \tau^+ \nu)=1$. 
An appearance search was also performed by
the D\O\ Collaboration using 62 pb$^{-1}$ of Run I
data~\cite{Abazov:2001md}, this time in the more challenging
$\tau_h$+$\met$+jets final state. Finally, the D\O\ Collaboration also
performed a disappearance search by studying the $\ell$+jets channel
using 109 pb$^{-1}$ of Run I data~\cite{Abbott:1999eca}. In all cases,
constraints were set in the ($\tan\beta$, $m_{H^+}$) plane within the
MSSM at tree level.

During Run II, the CDF and D\O\ Collaborations have performed
significantly more sensitive searches by combining multiple analysis
channels with and without tau
leptons~\cite{Abulencia:2005jd,Abazov:2009aa}. The largest integrated
luminosity analyzed in this type of study is by the D\O\
Collaboration, corresponding to 1 fb$^{-1}$ of data, still only one
tenth of the total integrated luminosity recorded in Run II.  This
search analyzed simultaneously up to 14 channels covering the
$\ell$+jets ($\ell=e,\mu$), $\ell\ell'$+jets, and
$\ell\tau_h$+$\met$+jets final states.  The obtained 95\% C.L. upper
limits on ${\cal B}(t \to H^+ b)$ are in the range $\simeq 0.15$--0.2
depending on the assumed $m_{H^+}$ and scenario considered 
(${\cal B}(H^+ \to \tau^+ \nu)=1$ or ${\cal B}(H^+ \to c\bar{s})=1$). 
Constrains were also set in the ($\tan\beta$, $m_{H^+}$) plane for different 
benchmark scenarios.

The most restrictive direct search at the Tevatron for $t \to H^+ b \to \tau^+ \nu b$ 
was performed by the CDF Collaboration using 9
fb$^{-1}$ of Run II data~\cite{Aaltonen:2014hua}. This analysis is
focused on the $\ell\tau_h$+$\met$+jets ($\ell=e,\mu$) channel. A
novel feature of this search is the construction of a likelihood
discriminant that allows separating the single tau component from the
ditau component (where the charged lepton doesn't originate from the
$W$ decay but rather from a leptonic tau decay), yielding a direct
measurement of ${\cal B}(t \to \tau^+\nu b)=0.096 \pm 0.028$. Under the
assumption that ${\cal B}(H^+ \to \tau^+ \nu)=1$, this result excludes
${\cal B}(t \to H^+ b)>0.059$ at 95\% C.L. for $m_{H^+}$ in the range
80--140 GeV.

The CDF Collaboration also performed a direct search for 
$t \to H^+ b \to c \bar{s} b$ using 2.2 fb$^{-1}$ of Run II
data~\cite{Aaltonen:2009ke}. This analysis considers $t\bar{t}$
candidates in the $\ell$+jets final state and looks for evidence of 
the decay $t \to H^+ b \to c \bar{s} b$ by performing a kinematic reconstruction of
the final state and studying the dijet mass spectrum of the top quark
decaying hadronically, where the $H^+ \to c\bar{s}$ would appear as a
resonance above the $W$ mass peak (see
Fig.~\ref{fig:hbsm_4}(b)). No significance excess is found and
95\% C.L. upper limits on ${\cal B}(t \to H^+ b)$ of $\simeq 0.1$--0.3
are set for $m_{H^+}$ in the range of 60--150~GeV, assuming 
${\cal B}(H^+ \to c\bar{s})=1$.
 
\subsection{Light CP-odd Higgs bosons}

Although some of the original benchmarks for searches for light CP-odd
Higgs bosons ($a$) arising in singlet extensions of the Higgs sector have
changed following the discovery of a SM-like Higgs boson at the LHC,
which now could be the lightest or next-to-lightest CP-even state of
the extended Higgs sector, light pseudoscalars are still
interesting. On the one hand, a light CP-odd Higgs boson is a
potential axion candidate.  On the other hand, it can have
significant phenomenological implications in the study of an extended
Higgs sector: e.g. it can appear in the decay of the SM-like Higgs
boson ($h \to aa$), or become a dominant decay mode for a light
charged Higgs boson ($H^\pm \to W^{\pm (*)} a$).

The D\O\ Collaboration has performed searches for the SM Higgs boson
decaying to $h \to aa$ using 4.2 fb$^{-1}$ of Run~II
data~\cite{Abazov:2009yi}. Two different scenarios are considered,
depending on the assumed mass of the $a$ boson: (i) for $m_a < 2 m_\tau$,
both $a$ bosons are searched in decays to $\mu^+\mu^-$, giving
a signature with two pairs of collinear muons; (ii) for $2 m_\tau < m_a < 2 m_b$, 
one $a$ boson is required to decay to $\mu^+\mu^-$ and
the other to $\tau^+\tau^-$, giving a signature of one pair of
collinear muons and either large $\met$ and an additional (not
necessarily isolated) muon, or a loosely-isolated electron from 
$a \to \tau^+\tau^-$ opposite to the muon pair.  No significant excess above
the background prediction is found in either search and 95\% C.L. upper
limits on the production cross section times branching ratio are set as a
function of $m_a$, assuming $m_h=100$ GeV. In the case of the 
$h \to aa \to 4\mu$ search, the upper limits on 
$\sigma (h+X) \times {\cal B}(h \to aa \to 4\mu)$ are in the range of 10--5.6~fb for $m_a$ in the
range of 0.2143--3 GeV. In the case of the $h \to aa \to 2\mu 2\tau$
search, the upper limits on $\sigma (h+X) \times {\cal B}(h \to aa \to 2\mu 2\tau)$ 
are in the range of 19.1--33.7 fb for $m_a$ in the range
of 3.6--19~GeV. Assuming no significant difference in selection
efficiency between $m_h=100$ GeV and $m_h=125$ GeV, these upper limits
could be used to set constraints on ${\cal B}(h \to aa \to 4\mu)$ and
${\cal B}(h \to aa \to 2\mu 2\tau)$ for the SM Higgs boson discovered
by the ATLAS and CMS experiments at a mass of $\sim 125$ GeV.

The CDF Collaboration has searched for an $a$ boson using 2.7~fb$^{-1}$ of 
Run~II data~\cite{Aaltonen:2011aj} in the context of a
search for top quark decays to a charged Higgs boson, $t \to H^+ b$,
with subsequent decay $H^\pm \to W^{\pm (*)} a$ and $a \to \tau^+\tau^-$.  
In this case the decay products of the $a$ boson are
expected to have low momenta and the new decay mode for the $H^\pm$,
if dominant, would make the $H^\pm$ escape existing limits at the
Tevatron. The analysis selects $t\bar{t}$ candidates in the
$\ell$+jets final states, and searches for $a \to \tau^+\tau^-$ decays
by looking for at least one isolated track with $3 \leq \pt \leq 20$~GeV 
in the central detector. The main background to this search is
isolated tracks from the underlying event, which are modeled directly
from data.  By analyzing the $\pt$ spectrum of the isolated track and
under the assumptions that 
${\cal B}(H^\pm \to W^{\pm (*)} a) = {\cal B}(a \to \tau^+\tau^-)=1$, 
95\% C.L. upper limits on ${\cal B}(t \to H^+ b)<0.2$ are set for $m_{H^+}$ 
in the range of 90--160 GeV.

\subsection{Doubly-charged Higgs bosons}

Doubly-charged Higgs bosons ($H^{\pm\pm}$) arise in triplet extensions
of the Higgs sector and they couple directly to leptons, photons, $W$
and $Z$ bosons, and singly-charged Higgs bosons. The $H^{\pm\pm}_L$
and $H^{\pm\pm}_R$ bosons respectively couple to left- and
right-handed particles, and may have different fermionic couplings. At
the Tevatron, $H^{\pm\pm}$ would be dominantly produced in pairs
through the process $q\bar{q} \to Z/\gamma^* \to H^{++} H^{--}$, and
decays predominantly to charged leptons if $m_{H^{\pm\pm}}<2
m_{H^\pm}$ and $m_{H^{\pm\pm}} - m_{H^\pm} < M_W$. Searches at
the Tevatron have been performed both for lepton-flavor conserving as
well as lepton-flavor violating (LFV) decays, the latter having
potentially sizable branching ratios in particular models. These
analyses select events consistent with multilepton final states
and search for a resonance in the invariant mass of a SS dilepton pair.
They typically have very small background rates.

Regarding decays to light-flavor leptons, the CDF Collaboration has
searched for $p\bar{p} \to H^{++} H^{--}+X$, with $H^{\pm\pm} \to
e^\pm e^\pm, \mu^\pm \mu^\pm, e^\pm \mu^\pm$ using 240~pb$^{-1}$ of
Run~II data~\cite{Acosta:2004uj} and requiring only a pair of
SS leptons of either the same or different flavor. The
resulting 95\% C.L. lower limits on $m_{H^{\pm\pm}}$ are 133~GeV, 136~GeV and 
115~GeV, for exclusive $H^{\pm\pm}_L$ decays to $e^\pm e^\pm$,
$\mu^\pm \mu^\pm$, and $e^\pm \mu^\pm$, respectively, and 113~GeV for
exclusive $H^{\pm\pm}_R$ decays to $\mu^\pm \mu^\pm$. The D\O\
Collaboration has searched for $H^{\pm\pm} \to \mu^\pm \mu^\pm$ using
1.1~fb$^{-1}$ of Run~II data~\cite{Abazov:2008ab}. The resulting 95\%~C.L. 
lower limits on $m_{H^{\pm\pm}}$ improve to 150~GeV and 127~GeV for
$H^{\pm\pm}_L$ and $H^{\pm\pm}_R$, respectively, both exclusively
decaying to $\mu^\pm \mu^\pm$.

Searches for $H^{\pm\pm}$ decays involving hadronically-decaying tau
leptons have also been performed. The CDF Collaboration has searched
for LFV decays $H^{\pm\pm} \to \ell^\pm \tau^\pm$ ($\ell=e,\mu$) using
350~pb$^{-1}$ of Run~II data~\cite{Aaltonen:2008ip}, studying
separately events with exactly three or four leptons, where the
leading lepton was required to be an electron or muon and there had to
be at least one $\tau_h$ candidate. The resulting 95\%~C.L. lower limits
on $m_{H^{\pm\pm}}$ are 114~GeV and 112~GeV for exclusive
$H^{\pm\pm}_L$ decays to $e^\pm \tau^\pm$ and $\mu^\pm \tau^\pm$,
respectively. The D\O\ Collaboration has searched for $H^{\pm\pm} \to
\tau^\pm \tau^\pm, \mu^\pm \tau^\pm, \mu^\pm \mu^\pm$ using 7~fb$^{-1}$ of 
Run~II data~\cite{Abazov:2011xx} by selecting events with
at least one muon and at least two $\tau_h$ candidates. The resulting
95\%~C.L. lower limits on $m_{H^{\pm\pm}_L}$ are 128~GeV and 144~GeV for
exclusive decays to $\tau^\pm \tau^\pm$ and $\mu^\pm \tau^\pm$,
respectively, and 130 GeV for a model with equal branching ratios into
$\tau^\pm \tau^\pm$, $\mu^\pm \tau^\pm$ and $\mu^\pm \mu^\pm$.

Finally, the CDF Collaboration has also considered the scenario in which
the lifetime of the $H^{\pm\pm}$ boson is long ($c\tau>3$ m),
resulting in the $H^{\pm\pm}$ boson decaying outside the
detector. This search was performed using 292~pb$^{-1}$ of Run~II
data~\cite{Acosta:2005np}. The resulting signature is two isolated
central tracks leaving large ionization in the tracker and
calorimeters and having muon-like penetration properties due to their
large mass.  The resulting 95\%~C.L. lower limits on $m_{H^{\pm\pm}}$
are 133~GeV or 109~GeV if only $H^{\pm\pm}_L$ or $H^{\pm\pm}_R$ are
kinematically accessible, or 146~GeV if both are degenerate in mass.

\subsection{Higgs boson decays to hidden-sector particles}

The CDF and D\O\ Collaborations have performed searches for the SM Higgs
boson decaying into a pair of ``hidden valley" hadrons (HV), each of which in turn decays
into a $b\bar{b}$ pair, giving a striking experimental signature of
highly displaced secondary vertices with a very large number of tracks
attached from the $b$-quark
decays~\cite{Abazov:2009ik,Aaltonen:2011rja}. The CDF and D\O\ analyses 
use 3.2~fb$^{-1}$ and 3.6~fb$^{-1}$ of Run~II data,
respectively. The CDF analysis searches for a pair of jets, where each of
them contains a reconstructed secondary vertex, and both jets appear
to originate from a common displaced point in space where the HV
hadron decayed. The D\O\ analysis instead requires at least two jets
in the event and at least two secondary vertices, and applies
stringent requirements on the secondary vertex properties to suppress
backgrounds, which are eventually dominated by interactions of
particles with the tracker material. In both searches, backgrounds are
estimated directly from data.  No evidence of a signal is found in
either search, and limits on the production cross section of a SM
Higgs boson times the branching ratio for $H\to HV \overline{HV} \to
b\bar{b} b\bar{b}$ are set, for different values of the Higgs boson
mass, the HV mass and its lifetime.

The CDF Collaboration has also performed a generic search for
anomalous production of multiple leptons produced in association with
$W$ and $Z$ bosons using 5.1~fb$^{-1}$ of Run II
data~\cite{Aaltonen:2012pu}. This search is sensitive to a wide range
of scenarios that predict multiple electrons and muons, including
clusters of leptons that are produced spatially close to each other,
often referred to as
``lepton-jets"~\cite{Ruderman:2009tj,Chen:2009iua,Mardon:2009gw}. No
significant excess is observed above the SM background expectation and
a 95\%~C.L. upper limit on the production cross section is set for a
benchmark model in which the Higgs boson decays mainly to a pair of
the lightest supersymmetric neutralinos, which in turn decay through a
dark sector to lepton-jets~\cite{Falkowski:2010cm,Falkowski:2010gv}.

\section{Summary and Conclusions}

The CDF and D\O\ Collaborations vigorously sought the Higgs boson
predicted by the SM and have produced evidence consistent with
such a particle and inconsistent with the background-only prediction
with a significance level of 3.1 standard deviations in the
$H\rightarrow b{\bar{b}}$ searches in July 2012.  As of this writing, the sensitivity of the
combined Tevatron analyses 
remains competitive with results from ATLAS~\cite{Aad:2014xzb} and CMS~\cite{Chatrchyan:2014vua}
in the $H\rightarrow b{\bar{b}}$ decay mode, even though the LHC's integrated luminosity is higher per experiment
than the Tevatron total, and the center of mass
energy is roughly a factor of four higher at the LHC.  The fact that the Tevatron was a $p{\bar{p}}$ collider while
the LHC is a $pp$ collider makes the Tevatron results complementary to those of the LHC.
Measurements at the Tevatron of the production
rates times the decay branching fractions in the $H\rightarrow b{\bar{b}}$, 
$H\rightarrow WW^{(*)}$, $H\rightarrow\tau^+\tau^-$,
$H\rightarrow\gamma\gamma$, and $H\rightarrow ZZ^{(*)}$ searches are
consistent with the predictions for the SM Higgs boson
with a mass of approximately 125~GeV, which is the mass of the
Higgs boson observed by ATLAS and CMS. 
Constraints from CDF and D\O\ on the couplings and the spin and parity
likewise are consistent with the presence of the SM Higgs
boson and disfavor exotic interpretations, as well as admixtures of
signals from exotic particles and the SM Higgs boson.

Searches for Higgs bosons predicted by extensions of the SM, such as
the additional neutral and charged Higgs bosons of the MSSM and other
two Higgs doublet models do not find evidence for any new particles
beyond those predicted by the SM.

The searches for Higgs bosons at the Tevatron have been an excellent
proving ground for new techniques used to collect, validate, simulate, and analyze hadron
collider data, where the expected signal yields are small and the
background rates are large and highly uncertain.  The use of machine
learning techniques and the splitting of data samples into multiple
categories with different sensitivities improved the statistical power of the searches.  The
impact of systematic uncertainties on the results yielded by these new
methods was evaluated with techniques common in high-energy physics
experiments -- validation using control samples and sidebands, as well
as propagation of uncertainties in all predictions through the
multivariate techniques while handling correlations and uncertainties in the
distributions of observable variables.  Standard statistical
techniques were used to extract final results.  These same techniques
have been adopted in searches for many new phenomena at the Tevatron and
the LHC, as well as measurements of newly established phenomena, such
as single top quark production, diboson production, and Higgs boson
production.

\section*{Acknowledgments}
We thank Gregorio Bernardi, Craig Group, and Ken Herner for useful comments and discussions. 

We thank the Fermilab staff and technical staffs of
the participating institutions for their vital contributions.
We acknowledge support from the DOE and NSF
(USA), ARC (Australia), CNPq, FAPERJ, FAPESP
and FUNDUNESP (Brazil), NSERC (Canada), NSC,
CAS and CNSF (China), Colciencias (Colombia), MSMT
and GACR (Czech Republic), the Academy of Finland,
CEA and CNRS/IN2P3 (France), BMBF and DFG (Germany),
DAE and DST (India), SFI (Ireland), INFN
(Italy), MEXT (Japan), the KoreanWorld Class University
Program and NRF (Korea), CONACyT (Mexico),
FOM (Netherlands), MON, NRC KI and RFBR (Russia),
the Slovak R\&D Agency, the Ministerio de Ciencia
e Innovaciå«on, and Programa Consolider--Ingenio 2010
(Spain), The Swedish Research Council (Sweden), SNSF
(Switzerland), STFC and the Royal Society (United
Kingdom), the A.P. Sloan Foundation (USA), and the
EU community Marie Curie Fellowship contract 302103.

\appendix

\bibliographystyle{ws-ijmpa}
\bibliography{tevhiggs}

\end{document}